\documentclass[12pt]{article}

\usepackage[dvipdfmx]{graphicx}

\usepackage{fancybox}
\usepackage{float}
\usepackage{amsfonts}
\usepackage{amsmath}
\usepackage{amsbsy}
\usepackage{amssymb}
\usepackage{amsthm}
\usepackage{bm}
\usepackage{epsfig}
\usepackage{latexsym}
\usepackage{pdflscape}
\usepackage{color}
\usepackage{here}
\numberwithin{equation}{section}

\allowdisplaybreaks

\DeclareMathOperator{\Tr}{Tr}
\DeclareMathOperator{\Det}{Det}

\newcommand{\llangle}{\langle\!\langle}
\newcommand{\rrangle}{\rangle\!\rangle}

\newcounter{aff}

\begin{document}

\begin{titlepage}

\begin{flushright}
{\footnotesize OCU-PHYS 459, TU-1042, KIAS-P17019}
\end{flushright}
\begin{center}
{\Large\bf 
Instanton Effects in Rank Deformed Superconformal\\[6pt]
Chern-Simons Theories from Topological Strings
}\\
\bigskip\bigskip
{\large 
Sanefumi Moriyama\,\footnote{\tt moriyama@sci.osaka-cu.ac.jp},
\quad
Shota Nakayama\,\footnote{\tt nakayama@tuhep.phys.tohoku.ac.jp},
\quad
Tomoki Nosaka\,\footnote{\tt nosaka@yukawa.kyoto-u.ac.jp}
}\\
\bigskip\bigskip
${}^{*\dagger}$\,{\it Department of Physics, Graduate School of Science,
Osaka City University}\\
${}^*$\,{\it Osaka City University 
Advanced Mathematical Institute (OCAMI)}\\
{\it Sumiyoshi-ku, Osaka 558-8585, Japan}\\[6pt]
${}^\dagger$\,{\it Department of Physics, Graduate School of Science,
Tohoku University}\\
{\it Aoba-ku, Sendai 980-8578, Japan}\\[6pt]
${}^\ddagger$\,{\it School of Physics, Korea Institute for Advanced Study}\\
{\it Dongdaemun-gu, Seoul 02455, Korea}
\end{center}

\begin{abstract}
In the so-called $(2,2)$ theory, which is the U$(N)^4$ circular quiver superconformal Chern-Simons theory with levels $(k,0,-k,0)$, it was known that the instanton effects are described by the free energy of topological strings whose Gopakumar-Vafa invariants coincide with those of the local $D_5$ del Pezzo geometry.
By considering two types of one-parameter rank deformations U$(N)\times $U$(N+M)\times $U$(N+2M)\times $U$(N+M)$ and U$(N+M)\times $U$(N)\times $U$(N+M)\times $U$(N)$, we classify the known diagonal BPS indices by degrees.
Together with other two types of one-parameter deformations, we further propose the topological string expression when both of the above two deformations are turned on.

\centering\includegraphics[scale=0.45,angle=-90]{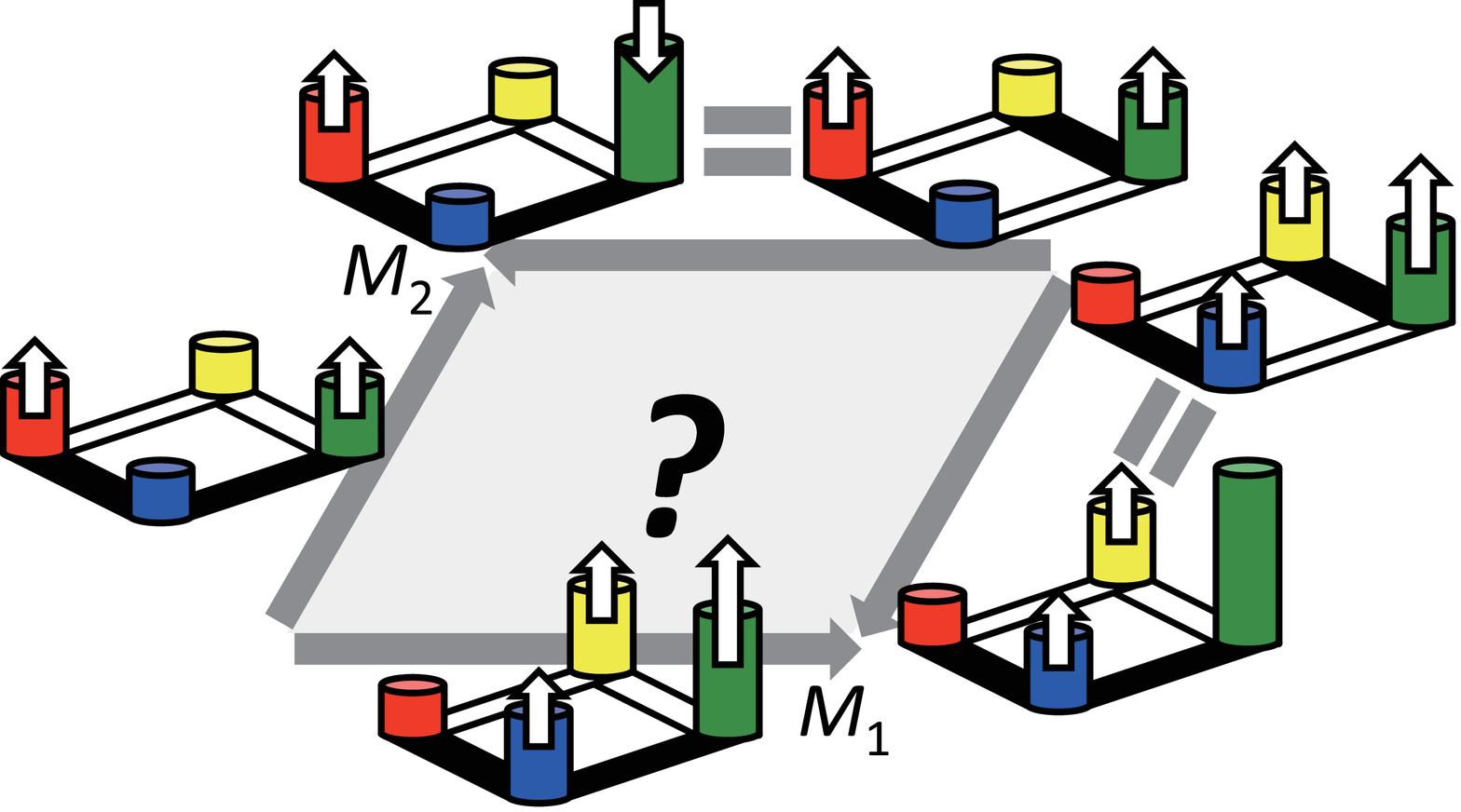}
\end{abstract}

\end{titlepage}
\setcounter{footnote}{0}

\tableofcontents

\section{Introduction}
\label{introduction}

M-theory is a mysterious theory.
From the AdS/CFT correspondence, it was conjectured \cite{KT} from the gravity side that the free energy of M2-branes, the degrees of freedom of fundamental excitations in M-theory, is $N^{3/2}$ in the large $N$ limit.
After the proposal \cite{ABJM,HLLLP2,ABJ} that the ${\mathcal N}=6$ superconformal Chern-Simons theory with gauge group U$(N_1)_k\times$U$(N_2)_{-k}$ and two pairs of bifundamental matters describes $\min(N_1,N_2)$ M2-branes and $|N_2-N_1|$ fractional M2-branes on the target geometry ${\mathbb C}^4/{\mathbb Z}_k$, this conjecture was confirmed from the gauge theory side.
Namely, after using the localization technique \cite{KWY}, the partition function of the ABJM theory on $S^3$, which is originally defined by the infinite-dimensional path integral, is reduced to a finite-dimensional matrix integration.
Then, the 't Hooft expansion\footnote{The $N^{3/2}$ scaling was also derived in the direct M-theory limit $N\to\infty$ with $k$ kept finite in \cite{HKPT} for a wide class of theories dual to the tri-Sasaki Einstein geometries \cite{MS}.} of the matrix model is applicable \cite{DMP1,DMP2,FHM}.
Due to the peculiar power ${3/2}$, it is natural to ask what the corrections in the large $N$ limit are.
The perturbative corrections were found to be summed up to the Airy function \cite{FHM}, which turned out to be just a starting point of the full exploration.

One of the celebrating results in this matrix model is the correspondence to the topological string theory.
After the discovery of the Fermi gas formalism \cite{MP} which rewrites the partition function into that of a Fermi gas system with $N$ particles, more information on the matrix model was obtained.
Besides the 't Hooft expansion, we can perform the WKB small $k$ expansion \cite{MP,CM} or study the numerical fitting from the exact values of the partition function \cite{HMO1,PY,HMO2,HMO3}.
Finally, the non-perturbative effects of the matrix model in the grand canonical ensemble are given explicitly by the free energy of the topological string theory on the local ${\mathbb P}^1\times{\mathbb P}^1$ geometry \cite{DMP1,DMP2,HMMO}.
Although this result was originally found for the case of equal ranks $N_2=N_1$, it was soon generalized into the case of different ranks $N_2\ne N_1$ \cite{MM,HO}.

Generally it is interesting to ask whether we can classify/engineer the geometries through the matrix models.
To understand this we have to consider an example where we have more degrees of freedom on both the gauge theory side and the geometry side.
It is already a non-trivial question whether any quiver superconformal Chern-Simons theory corresponds to some Calabi-Yau geometry or not.
In the subsequent works \cite{HM,MN1,MN2,MN3,HHO}, the investigation started with a special class of ${\cal N}=3$ theories: the circular quiver superconformal Chern-Simons theories with unitary gauge groups\footnote{Instead of the circular quiver or the $\widehat A$ quiver, the $\widehat D$ quiver was also investigated carefully in \cite{GAH,CHJ,ADF,MN4,CJ}.}.
It was shown in \cite{GW,HLLLP1,IK1,TY,IK2} that, for the circular quiver, the theory enjoys the supersymmetry ${\mathcal N}=4$ when the Chern-Simons level $k_a$ associated to each vertex is expressed as $k_a=k(s_a-s_{a-1})/2$ with $s_a=\pm 1$. 
Namely, the full information of the ${\mathcal N}=4$ theories is encoded in the list of $s_a=\pm 1$, and we may refer to the theory with
\begin{align}
\{s_a\}=\{\underbrace{+1,+1,\cdots,+1}_{p_1},\underbrace{-1,-1,\cdots,-1}_{q_1},
\underbrace{+1,+1,\cdots,+1}_{p_2},\underbrace{-1,-1,\cdots,-1}_{q_2},\cdots\},
\end{align}
as $(p_1,q_1,p_2,q_2,\cdots)$ model \cite{MN1}.
For example, the theory with alternating $s_a$ is denoted as the $(1,1,\cdots,1)$ model.
The physical interpretation of this theory is the orbifold and its grand potential was studied in \cite{HM}.
Another limiting case $\{s_a\}=\{(+1)^p,(-1)^q\}$, or the $(p,q)$ model \cite{MN1}, which is expected to be a building block of all other ${\mathcal N}=4$ theories, was also studied carefully in \cite{MN1,MN2,MN3,HHO}.

If we consider general ranks, this class turns out to have a rich structure.
Namely, the theories with different sequences of $\{s_a\}$ are expected to be connected and to form a non-trivial moduli space as a whole.
In fact, since the ordering of $s_a=\pm 1$ corresponds to the ordering of the quantum operators \cite{MN1} in the one-particle density matrix in the Fermi gas formalism, the theories which are different only in the ordering are equivalent in the classical limit.
This may suggest that these theories correspond to the topological string theory on the same Calabi-Yau manifold \cite{GHM1}.
Furthermore, the ${\cal N}=4$ theory with the circular quiver can be translated to a type IIB brane system by replacing the edges $s_a=\pm 1$ with $(1,k)5$-/NS5-branes and the vertices with $N$ D3-branes stretching between two 5-branes. 
In this context the ordering of $\{s_a\}$ is the ordering of the $(1,k)$5-/NS5-branes.
This suggests that the theories with different ordering indeed should be unified if we consider general ranks, as a pair of 5-branes can be exchanged at the expense of creation/annihilation of the D3-branes, which is known as Hanany-Witten transitions \cite{HW}.

In this paper we consider the $(2,2)$ model with general ranks.
In fact, the $(2,2)$ model is the simplest one which is essentially distinctive from the ABJM theory U$(N_1)_k\times$U$(N_2)_{-k}$ due to the following observations.
Firstly, without rank deformations, the grand potential of the $(2,2)$ model was found to have the structure of the free energy of topological strings \cite{MN3}.
We actually found that the diagonal Gopakumar-Vafa invariants of the $(2,2)$ model match completely with those of the local $D_5$ del Pezzo geometry (table 6 in \cite{KKV}).
Secondly, the $(2,2)$ model allows a rank deformation with maximally three parameters (besides the overall rank $N$), while the local $D_5$ del Pezzo geometry has $5$ degrees of freedom for the K\"ahler parameters (besides the overall scaling), which implies a richer structure between the ranks and the K\"ahler parameters.
Thirdly, the $(2,2)$ model is the theory with the smallest number of 5-branes which has a non-trivial dependence on the ordering.
Therefore, although so far a direct relation to the topological string theory was not found for any other ${\cal N}=4$ theories, it is worthwhile to study the rank deformations of the $(2,2)$ model thoroughly to understand the relation between the ${\cal N}=4$ theories and the topological string theories.

Among the general rank deformations, we find that the following two one-parameter rank deformations are particularly tractable: U$(N)_k\times$U$(N+M)_0\times$U$(N+2M)_{-k}\times$U$(N+M)_0$ and U$(N+M)_k\times$U$(N)_0\times$U$(N+M)_{-k}\times$U$(N)_0$.
Unlike the situation with the general R charges \cite{N}, the FI terms completely cancel in these deformations.
By using the technique \cite{MM} called the open string formalism, we can factorize the grand canonical partition function of these models as the product of the grand partition function of $M=0$ and the determinant of a matrix whose matrix elements are expressed with a generalization of the two functions $\phi_m(q)$ and $\psi_m(q)$ introduced in \cite{MN3}.
Although these functions were originally introduced from a technical reason to compute the grand canonical partition function of $M=0$, it is interesting to find that these functions appear naturally in the rank deformations as well.
Moreover, from the computation we can give a clear reason for the appearance of the two functions: $\phi_m(q)$ and $\psi_m(q)$ are associated to the vertices with non-zero levels and zero levels respectively.

The result again has the structure of the free energy of topological strings.
By matching the table of the BPS indices of the local $D_5$ del Pezzo geometry (the tables in section 5.4 of \cite{HKP}) with the instanton coefficients obtained from the numerical studies of these two rank deformations, we can classify the BPS indices by various degrees.
Interestingly, we have obtained different classifications of the BPS indices for the two one-parameter rank deformations, where the BPS indices in the latter case are a further classification of those in the former case.

After studying the two one-parameter rank deformations of the $(2,2)$ model, we propose a unification of them by a two-parameter deformation.
As a non-trivial check of our unification and our classification of the BPS indices, we turn to another ${\cal N}=4$ theory, the $(1,1,1,1)$ model, whose rank deformations are connected to those of the $(2,2)$ model.
We discuss the Hanany-Witten transition to see how these two models are related and study the rank deformations of the $(1,1,1,1)$ model as well.
We have found that all of our classifications of the BPS indices are consistent with the rank deformations on the $(1,1,1,1)$ side.

This paper is organized as follows.
In the next section, we shall review the $(2,2)$ model without rank deformations.
As we see in section \ref{22revisit}, already by revisiting the result we find a lot of information to split the BPS indices, if we assume the existence of two K\"ahler parameters and compare the BPS indices with the table in \cite{HKP}.
In section \ref{22:0121}, we study the first deformation U$(N)_k\times$U$(N+M)_0\times$U$(N+2M)_{-k}\times$U$(N+M)_0$.
Though the open string formalism resembles to that of the ABJM case, it also contains a new indefinite feature, which requires a careful regularization.
After this regularization, we are able to give a clear reason for the appearance of the two functions $\phi_m(q)$ and $\psi_m(q)$.
In section \ref{22:1010}, we study the second deformation U$(N+M)_k\times$U$(N)_0\times$U$(N+M)_{-k}\times$U$(N)_0$.
The non-perturbative effects are much complicated and we explain that the result can be given by the free energy of topological strings if we introduce six K\"ahler parameters.
After that, in section \ref{two1111} we study other two rank deformations by utilizing the Hanany-Witten transition and propose to unify the four one-parameter rank deformations with a two-parameter deformation.
Finally in section \ref{conclusion}, we summarize and discuss future directions.
The paper also contains four appendices.
Appendix \ref{notations} and appendix \ref{detformula} provide some useful formulas for the open string formalism.
The exact values of the partition function and the grand potential for the four one-parameter rank deformations are summarized in appendix \ref{valuesofZandJandJtilde}.
These results are crucial in the determination of the K\"{a}hler parameters and the BPS indices.
In appendix \ref{22closed} we provide the closed string formalism for U$(N+M)_k\times$U$(N)_0\times$U$(N+M)_{-k}\times$U$(N)_0$, where the duality to the $(1,1,1,1)$ model at $M=k/2$ is manifest.

\section{ABJM theory and $(2,2)$ model}
\label{sec22model}

Since we will discuss the topological string description of the non-perturbative effects in the $(2,2)$ model with rank deformations, we shall start by reviewing the Fermi gas formalism and the non-perturbative effects for the standard ABJM theory with the rank deformation and the $(2,2)$ model without rank deformations.
We shall see that the topological string description of the non-perturbative effects in the ABJM theory is successful, though there is room for improvement for that in the $(2,2)$ model.
Most of the contents in this section are reviews, though in section \ref{not_trivial1} and section \ref{22revisit} we will also raise some questions and clarify some points which we believe are not so trivial even to the experts but important for our later analysis.
The main references of the review part are \cite{HMMO,MM,MN3} (see also \cite{PTEP,Marino}).

\subsection{ABJM theory}

\subsubsection{Open string formalism}
\label{ABJMFG}

After applying the localization technique \cite{KWY}, the partition function of the ABJM theory on $S^3$ reduces to a matrix model
\begin{align}
Z^\text{ABJM}_k(N_1,N_2)
=\int\frac{D^{N_1}\mu}{N_1!}\frac{D^{N_2}\nu}{N_2!}
\frac{\prod_{m\ne m'}^{N_1}2\sinh\frac{\mu_m-\mu_{m'}}{2}
\prod_{n\ne n'}^{N_2}2\sinh\frac{\nu_n-\nu_{n'}}{2}}
{\prod_{m=1}^{N_1}\prod_{n=1}^{N_2}(2\cosh\frac{\mu_m-\nu_n}{2})^2},
\label{ZABJM}
\end{align}
with the Fresnel integrations $(k>0)$
\begin{align}
D\mu=\frac{d\mu}{2\pi}e^{\frac{ik}{4\pi}\mu^2},\quad
D\nu=\frac{d\nu}{2\pi}e^{-\frac{ik}{4\pi}\nu^2}.
\label{dmudnu}
\end{align}
Without loss of generality, we often set $M=N_2-N_1\ge 0$ for the ABJM matrix model.
Otherwise, we can apply the complex conjugation.
Hereafter we set $(N_1,N_2)=(N,N+M)$.

One of the standard techniques to study this ABJM model with generally non-equal ranks $M\ge 0$ is to utilize the following two essentially same determinant formulas \cite{MM}
\begin{align}
&\frac{\prod_{m<m'}^{N}2\sinh\frac{\mu_m-\mu_{m'}}{2}
\prod_{n<n'}^{N+M}2\sinh\frac{\nu_n-\nu_{n'}}{2}}
{\prod_{m=1}^{N}\prod_{n=1}^{N+M}2\cosh\frac{\mu_m-\nu_n}{2}}\nonumber\\
&\quad=(-1)^{MN}e^{\frac{M}{2}\sum_m\mu_m}\det\begin{pmatrix}
\bigl[Q(\mu_m,\nu_n)\bigr]
_{\begin{subarray}{c}1\le m\le N\\1\le n\le N+M\end{subarray}}\\
\bigl[E_{l_p}(\nu_n)\bigr]
_{\begin{subarray}{c}1\le p\le M\\1\le n\le N+M\end{subarray}}
\end{pmatrix}e^{-\frac{M}{2}\sum_n\nu_n},\nonumber\\
&\frac{\prod_{n>n'}^{N+M}2\sinh\frac{\nu_n-\nu_{n'}}{2}
\prod_{m>m'}^{N}2\sinh\frac{\mu_m-\mu_{m'}}{2}}
{\prod_{n=1}^{N+M}\prod_{m=1}^{N}2\cosh\frac{\nu_n-\mu_m}{2}}\nonumber\\
&\quad=(-1)^{MN}e^{\frac{M}{2}\sum_n\nu_n}\det\begin{pmatrix}
\bigl[P(\nu_n,\mu_m)\bigr]
_{\begin{subarray}{c}1\le n\le N+M\\1\le m\le N\end{subarray}}&
\bigl[E_{a_q}(\nu_n)\bigr]
_{\begin{subarray}{c}1\le n\le N+M\\1\le q\le M\end{subarray}}
\end{pmatrix}e^{-\frac{M}{2}\sum_m\mu_m},
\label{vdmcauchy}
\end{align}
with $\{l_p\}_{p=1}^M=\{M-\frac{1}{2},M-\frac{3}{2},\cdots,\frac{1}{2}\}$, $\{a_q\}_{q=1}^M=\{-M+\frac{1}{2},-M+\frac{3}{2},\cdots,-\frac{1}{2}\}$ and
\begin{align}
Q(\mu,\nu)=\frac{1}{2\cosh\frac{\mu-\nu}{2}},\quad
E_l(\nu)=e^{l\nu},\quad
P(\nu,\mu)=\frac{1}{2\cosh\frac{\nu-\mu}{2}},\quad
E_a(\nu)=e^{a\nu},
\end{align}
which are obtained by combining the Vandermonde determinant and the Cauchy determinant.
By multiplying these two determinant formulas, the extra exponential factors $e^{\pm\frac{M}{2}\sum_m\mu_m}$ and $e^{\pm\frac{M}{2}\sum_n\nu_n}$ cancel among themselves and we can perform the $\nu$ integrations by using a determinant formula in \cite{MM} (which is further generalized in appendix \ref{detformula})
\begin{align}
Z^\text{ABJM}_k(N,N+M)=\int\frac{D^{N}\mu}{N!}\det
\begin{pmatrix}\bigl[(QP)(\mu_m,\mu_{m'})\bigr]_{N\times N}&
\bigl[(QE_a)(\mu_m)\bigr]_{N\times M}\\
\bigl[(E_lP)(\mu_{m'})\bigr]_{M\times N}&
\bigl[(E_lE_a)\bigr]_{M\times M}&
\end{pmatrix}.
\end{align}
Here we regard $Q(\mu,\nu)$, $P(\nu,\mu)$ and $E_l(\nu)$, $E_a(\nu)$ respectively as matrices and vectors with continuous indices and the multiplications are given by contracting these continuous indices with the integrations in \eqref{dmudnu}.

Furthermore if we define the grand canonical partition function as
\begin{align}
\Xi^\text{ABJM}_{k,M}(z)=\sum_{N=0}^\infty z^NZ^\text{ABJM}_k(N,N+M),
\label{Xinoabs}
\end{align}
and apply another determinant formula in \cite{MM}, we arrive at the expression
\begin{align}
\frac{\Xi^\text{ABJM}_{k,M}(z)}{\Xi^\text{ABJM}_{k,0}(z)}
=\det\bigl(H_{l,a}(z)\bigr),
\end{align}
with
\begin{align}
\Xi^\text{ABJM}_{k,0}(z)=\Det(1+zPQ),\quad
H_{l,a}(z)=E_l(1+zPQ)^{-1}E_a.
\end{align}
Here both the Fredholm determinant $\Det=\exp\Tr\log$ and the matrix element $H_{l,a}(z)$ are defined in the small $z$ expansion, consisting of traces and matrix elements of the powers of $PQ$.

To evaluate the exact values of the partition function, it is convenient to introduce the coordinate operator $\widehat q$ and the momentum operator $\widehat p$ satisfying the canonical commutation relation $[\widehat q,\widehat p]=i\hbar$ with the identification $\hbar=2\pi k$.
We also introduce the coordinate eigenstate $|q\rangle$ and the momentum eigenstate $|p\rrangle$ as well as summarize some useful formulas in appendix \ref{notations}.
In fact, after the Fourier transformation and the similarity transformation, $\Xi_{k,0}^{\text{ABJM}}(z)$ simply becomes
\begin{align}
\Xi^\text{ABJM}_{k,0}(z)=\Det(1+z\widehat\rho),\quad
\widehat\rho=\frac{1}{\sqrt{2\cosh\frac{\widehat q}{2}}}
\frac{1}{2\cosh\frac{\widehat p}{2}}
\frac{1}{\sqrt{2\cosh\frac{\widehat q}{2}}}.
\end{align}
Notice that the density matrix ${\widehat\rho}$ has the following structure \cite{TW,PY}
\begin{align}
[{\widehat M},{\widehat\rho}]={\widehat E}|0\rrangle\llangle 0|{\widehat E},
\end{align}
with
\begin{align}
{\widehat E}=E({\widehat q})
=\frac{e^{\frac{\widehat q}{2k}}}{\sqrt{2\cosh\frac{\widehat q}{2}}},\quad
{\widehat M}=M({\widehat q})=e^{\frac{\widehat q}{k}},
\end{align}
This structure allows us to decompose the powers $\langle q_1|\widehat\rho^n|q_2\rangle$ as
\begin{align}
\langle q_1|{\widehat\rho}^n|q_2\rangle
=\frac{E(q_1)E(q_2)}{M(q_1)+(-1)^{n-1}M(q_2)}
\sum_{m=0}^{n-1}(-1)^m\phi_m(q_1)\phi_{n-1-m}(q_2),
\end{align}
with $\phi_m(q)=\langle q|{\widehat E}^{-1}\widehat\rho^m{\widehat E}|0\rrangle$, which can be computed recursively as
\begin{align}
\phi_{m+1}(q)=\int\frac{dq'}{2\pi}
\langle q|{\widehat E}^{-1}\widehat\rho{\widehat E}|q'\rangle\phi_m(q').
\label{recurrence}
\end{align}
This expression reduces the computation of $\langle q_1|{\widehat\rho}^n|q_2\rangle$ to a much simpler computation of multiplying ${\widehat\rho}$ to $\widehat E|0\rrangle$ subsequently.
By rewriting $E_l$ and $E_a$ with the momentum eigenstates $\llangle 2\pi il|$ and $|{-2\pi ia}\rrangle$ we can also reduce the computation of $H_{l,a}(z)$ to the computation of the functions obeying the same recurrence relation \eqref{recurrence} as that of $\phi_m(q)$.
Indeed these functions are a straightforward generalization of $\phi_m(q)$ by replacing the initial state $|0\rrangle$ with $|{-2\pi i(a-\frac{1}{2})}\rrangle$.

\subsubsection{Instanton effects}
\label{abjmnp}

The main strategy in studying the instanton effects in the ABJM matrix model with rank deformations is to evaluate the exact values of the partition function, read off the instanton coefficients numerically and interpolate each instanton coefficient to a function of $(k,M)$.

If we define the grand potential $\overline J^\text{ABJM}_{k,M}(\mu)$ as
\begin{align}
e^{\overline J^\text{ABJM}_{k,M}(\mu)}
=\sum_{N=0}^\infty e^{N\mu}|Z^\text{ABJM}_k(N,N+M)|,
\label{Xi}
\end{align}
apparently $\overline J^\text{ABJM}_{k,M}(\mu)$ is invariant under the $2\pi i$ shift of $\mu$.
Due to this $2\pi i$ shift symmetry, $\overline J^\text{ABJM}_{k,M}(\mu)$ acquires an oscillating behavior with respect to $\mu$.
To get rid of the oscillation, it is convenient to define the {\it reduced} grand potential $J^\text{ABJM}_{k,M}(\mu)$ as
\begin{align}
\sum_{n=-\infty}^\infty e^{J^\text{ABJM}_{k,M}(\mu+2\pi i n)}
=\sum_{N=0}^\infty e^{N\mu}|Z^\text{ABJM}_k(N,N+M)|,
\label{Jreduced}
\end{align}
by adding infinite numbers of replicas from the beginning.
Though it is ambiguous to decide which one in the replicas to be the reduced grand potential, we can avoid the ambiguity in the definition by choosing the real one satisfying $(J^\text{ABJM}_{k,M}(\mu))^*=J^\text{ABJM}_{k,M}(\mu^*)$.
The concept of introducing the reduced grand potential appears repetitively in the following for various theories other than the ABJM theory, though we always define the reduced grand potential in the same way as \eqref{Jreduced}.

Then, the reduced grand potential is expressed in terms of the free energy of the topological string theory on the local ${\mathbb P}^1\times{\mathbb P}^1$ geometry.
First if we separate the grand potential into the perturbative part and the non-perturbative part in the large $\mu$ limit, $J^\text{ABJM}_{k,M}(\mu)=J^\text{ABJM,pert}_{k,M}(\mu)+J^\text{ABJM,np}_{k,M}(\mu)$, the perturbative part is given by
\begin{align}
J^\text{ABJM,pert}_{k,M}(\mu)
=\frac{C^\text{ABJM}_k}{3}\mu^3+B^\text{ABJM}_{k,M}\mu+A^\text{ABJM}_k,
\end{align}
with the coefficients $C^\text{ABJM}_k$, $B^\text{ABJM}_{k,M}$
\begin{align}
&C^\text{ABJM}_k=\frac{2}{\pi^2k},\quad
B^\text{ABJM}_{k,M}=\frac{1}{3k}-\frac{k}{12}
+\frac{k}{8}\biggl(1-\frac{2M}{k}\biggr)^2,
\end{align}
and $A^\text{ABJM}_k$ \cite{KEK,HaOk} $(g_m(k)=(k+(-1)^m(2m-k))/4)$
\begin{align}
&A^{\text{ABJM}}_k
=\begin{cases}
\displaystyle -\frac{\zeta(3)}{\pi^2k}-\frac{2}{k}\sum_{m=1}^{\frac{k}{2}-1}m\Bigl(\frac{k}{2}-m\Bigr)\log 2\sin\frac{2\pi m}{k},&k:\text{even},\\
\displaystyle -\frac{\zeta(3)}{8\pi^2k}+\frac{k}{4}\log 2-\frac{1}{k}\sum_{m=1}^{k-1}
g_m(k)(k-g_m(k))\log 2\sin\frac{\pi m}{k},&k:\text{odd},
\end{cases}
\label{AABJM}
\end{align}
for integral $k$.
For the non-perturbative part there appear bound states of worldsheet instantons $e^{-\frac{4\mu}{k}}$ and membrane instantons $e^{-2\mu}$.
It turns out that if we re-expand $J^{\text{ABJM}}_{k,M}(\mu)$ by the effective chemical potential $\mu_\text{eff}$ which is defined by
\begin{align}
\mu_\text{eff}=\begin{cases}
\displaystyle\mu-2(-1)^{\frac{k}{2}-M}e^{-2\mu}{}_4F_3
\biggl(1,1,\frac{3}{2},\frac{3}{2};2,2,2;16(-1)^{\frac{k}{2}-M}e^{-2\mu}\biggr),
&k:\text{even},\\
\displaystyle\mu+e^{-4\mu}{}_4F_3
\biggl(1,1,\frac{3}{2},\frac{3}{2};2,2,2;-16e^{-4\mu}\biggr),
&k:\text{odd},
\end{cases}
\end{align}
the bound states are completely taken care of by the worldsheet instantons and the non-perturbative part consists only of pure worldsheet instantons and pure membrane instantons
\begin{align}
J^\text{ABJM}_{k,M}(\mu)&=J^\text{ABJM,pert}_{k,M}(\mu_\text{eff})
+\widetilde J^\text{ABJM,np}_{k,M}(\mu_\text{eff}),\nonumber\\
\widetilde J^\text{ABJM,np}_{k,M}(\mu_\text{eff})
&=J^\text{ABJM,WS}_{k,M}(\mu_\text{eff})
+\widetilde J^\text{ABJM,MB}_{k,M}(\mu_\text{eff}),
\end{align}
with ($s_L=2j_L+1$, $s_R=2j_R+1$)
\begin{align}
J^\text{ABJM,WS}_{k,M}(\mu_\text{eff})
&=\sum_{j_L,j_R}\sum_{\bm d}N_{j_L,j_R}^{\bm d}
\sum_{n=1}^\infty
\frac{s_R\sin 2\pi g_sns_L}
{n(2\sin\pi g_sn)^2\sin 2\pi g_sn}e^{-n{\bm d}\cdot{\bm T}},
\nonumber\\
\widetilde J^\text{ABJM,MB}_{k,M}(\mu_\text{eff})
&=\sum_{j_L,j_R}\sum_{\bm d}N_{j_L,j_R}^{\bm d}
\sum_{n=1}^\infty
\frac{\partial}{\partial g_s}
\biggl[
g_s\frac{-\sin\frac{\pi n}{g_s}s_L\sin\frac{\pi n}{g_s}s_R}
{4\pi n^2(\sin\frac{\pi n}{g_s})^3}e^{-n\frac{{\bm d}\cdot{\bm T}}{g_s}}
\biggr].
\label{HMMOtop}
\end{align}
Here the two K\"ahler parameters ${\bm T}=(T^+;T^-)$ and the string coupling constant $g_s$ are
\begin{align}
T^{\pm}
=\frac{4\mu_\text{eff}}{k}\pm\pi i\biggl(1-\frac{2M}{k}\biggr),\quad
g_s=\frac{2}{k},
\label{ABJMid}
\end{align}
while $N_{j_L,j_R}^{\bm d}$ are the BPS indices of the local $\mathbb{P}^1\times\mathbb{P}^1$ geometry.
In this way the degrees of freedom for the rank deformations of two gauge groups reproduce the two degrees of freedom for the K\"{a}hler parameters of local $\mathbb{P}^1\times\mathbb{P}^1$.

Above we have introduced two similar quantities $e^{\overline J_{k,M}^{\text{ABJM}}}$ \eqref{Xi} and $\Xi_{k,M}^\text{ABJM}(z)$ \eqref{Xinoabs} with or without taking the absolute values.
Although they are same if there are no rank deformations since the values of the partition function are real positive, they play respective roles for general ranks.
$\Xi(z)_{k,M}^{\text{ABJM}}$ plays the role of a generating function in the computation of exact values of the partition function.
If we use $\Xi_{k,M}^{\text{ABJM}}(z)$ to define the reduced grand potential $J_{k,M}^{\text{ABJM}}(\mu)$, however, in general we cannot impose the condition $(J_{k,M}^{\text{ABJM}}(\mu))^*=J_{k,M}^{\text{ABJM}}(\mu^*)$ to simplify the large $\mu$ expansion.
Therefore, after computing the partition functions through $\Xi_{k,M}^{\text{ABJM}}(z)$ we need to switch to $e^{\overline J_{k,M}^{\text{ABJM}}(\mu)}$ by taking the absolute values.

\subsubsection{Comment on definition of grand potential}
\label{not_trivial1}

Although we have noted that without loss of generality we can consider $M=N_2-N_1\ge 0$, it is interesting to ask how the grand potential changes across $M=0$.
From the inverse transformation of \eqref{Jreduced}, we find
\begin{align}
|Z^\text{ABJM}_k(N,N+M)|
=\int_{-\infty i}^{\infty i}\frac{d\mu}{2\pi i}e^{J^\text{ABJM}_{k,M}(\mu)-N\mu}.
\end{align}
Suppose that this relation is valid for $M<0$, in the sense that the reduced grand potential $J^\text{ABJM}_{k,M}(\mu)$ is given by the expressions in section \ref{abjmnp} with a naive substitution of $M\,(<0)$.
Let us change the rank variables into positive ones $\bar N=N+M$, $\bar N+\bar M=N$,
\begin{align}
|Z^\text{ABJM}_k(\bar N+\bar M,\bar N)|
=\int_{-\infty i}^{\infty i}\frac{d\mu}{2\pi i}
e^{J^\text{ABJM}_{k,-\bar M}(\mu)-(\bar N+\bar M)\mu}.
\end{align}
From this we expect that if we define the grand potential as
\begin{align}
\sum_{n=-\infty}^\infty e^{J^\text{ABJM}_{k,-\bar M}(\mu+2\pi in)}
=\sum_{\bar N=0}^\infty e^{(\bar N+\bar M)\mu}
|Z^\text{ABJM}_k(\bar N+\bar M,\bar N)|,
\label{invZ}
\end{align}
the grand potential coincides with the original one with negative $M$.
In fact, we can check that $J^\text{ABJM}_{k,\bar M}(\mu)+\bar M\mu$ with $\bar M\ge 0$, when expressed with the same effective chemical potential $\mu_\text{eff}$, coincides with the free energy of the topological string theory with the K\"ahler parameters $T^\pm=4\mu_{\text{eff}}/k\pm \pi i(1+2{\bar M}/k)$.
Note that in \eqref{invZ} the power of $e^{\mu}$ has to match with the same argument of the partition function regardless of the relative magnitude of the ranks, though the lower bound of the sum is not vey important.
This observation is crucial later in our discussion of the rank deformation with multiple parameters.

\subsection{$(2,2)$ model without rank deformations}
\label{22_0000}

Now let us turn to the review of the $(2,2)$ model without rank deformations.
It turns out that most of the techniques and the results of the ABJM matrix model apply similarly to the $(2,2)$ model.

\subsubsection{Fermi gas formalism}

With the localization technique, the partition function of the $(2,2)$ model on $S^3$ reduces to the following matrix model
\begin{align}
&Z_k(N)=\int\frac{D^N\mu D^N\kappa D^N\nu D^N\lambda}{(N!)^4}\nonumber\\
&\times
\frac{\prod_{m\ne m'}^N2\sinh\frac{\mu_m-\mu_{m'}}{2}
\prod_{k\ne k'}^N2\sinh\frac{\kappa_k-\kappa_{k'}}{2}
\prod_{n\ne n'}^N2\sinh\frac{\nu_n-\nu_{n'}}{2}
\prod_{l\ne l'}^N2\sinh\frac{\lambda_l-\lambda_{l'}}{2}}
{\prod_{m,k'}^N2\cosh\frac{\mu_m-\kappa_{k'}}{2}
\prod_{k,n'}^N2\cosh\frac{\kappa_k-\nu_{n'}}{2}
\prod_{n,l'}^N2\cosh\frac{\nu_n-\lambda_{l'}}{2}
\prod_{l,m'}^N2\cosh\frac{\lambda_l-\mu_{m'}}{2}},
\end{align}
with
\begin{align}
D\mu=\frac{d\mu}{2\pi}e^{\frac{ik}{4\pi}\mu^2},\quad
D\kappa=\frac{d\kappa}{2\pi},\quad
D\nu=\frac{d\nu}{2\pi}e^{-\frac{ik}{4\pi}\nu^2},\quad
D\lambda=\frac{d\lambda}{2\pi}.
\label{integrations}
\end{align}

As in the case of the ABJM theory \cite{MP}, it was found that the grand canonical partition function $\Xi_k(z)=\sum_{N=0}^\infty z^NZ_k(N)$ is rearranged into a Fredholm determinant
\begin{align}
\Xi_k(z)=\Det(1+z{\widehat\rho}),\quad
\widehat\rho=\frac{1}{2\cosh\frac{\widehat q}{2}}
\frac{1}{(2\cosh\frac{\widehat p}{2})^2}
\frac{1}{2\cosh\frac{\widehat q}{2}}.
\label{22nodeform}
\end{align}
A systematic method to compute these traces is established again by noticing the following structure of the matrix element of ${\widehat\rho}$ \cite{MN3}
\begin{align}
[{\widehat M},{\widehat\rho}]
={\widehat q}{\widehat E}|0\rrangle\llangle 0|{\widehat E}
-{\widehat E}|0\rrangle\llangle 0|{\widehat E}{\widehat q},
\label{TWPYstructure}
\end{align}
with
\begin{align}
{\widehat E}=E({\widehat q})
=\frac{e^{\frac{\widehat q}{2k}}}{2\cosh\frac{\widehat q}{2}},\quad
{\widehat M}=M({\widehat q})=2\pi ke^{\frac{\widehat q}{k}},
\end{align}
(see \eqref{pket_qket_rotate}).
From \eqref{TWPYstructure} we obtain
\begin{align}
\langle q_1|{\widehat\rho}^n|q_2\rangle
=\frac{E(q_1)E(q_2)}{M(q_1)-M(q_2)}\sum_{m=0}^{n-1}
\bigl[\psi_m(q_1)\phi_{n-1-m}(q_2)-\phi_m(q_1)\psi_{n-1-m}(q_2)\bigr],
\end{align}
with
\begin{align}
\phi_m(q)=\langle q|{\widehat E}^{-1}{\widehat\rho}^m{\widehat E}|0\rrangle,\quad
\psi_m(q)=\langle q|{\widehat E}^{-1}{\widehat\rho}^m
{\widehat E}{\widehat q}|0\rrangle,
\label{phipsi22}
\end{align}
which we can compute recursively in $m$ \cite{MN3}.
With this method we can compute many exact values for the $(2,2)$ model as well to study the non-perturbative effects.
Interestingly, compared with the ABJM theory, here we need to introduce two functions \eqref{phipsi22} to accomplish this analysis.
In section \ref{FGforM0_regularization} we will comment on the origin of the two functions, which becomes apparent after we introduce the rank deformations.

\subsubsection{Instanton effects}
\label{22np}

To summarize the result for the reduced grand potential \eqref{Jreduced} of the $(2,2)$ model, it is again convenient to introduce the effective chemical potential which is now given as
\begin{align}
\mu_{\text{eff}}
=\mu+4e^{-\mu}\,_4F_3\biggl(1,1,\frac{3}{2},\frac{3}{2};2,2,2;-16e^{-\mu}\biggr).
\label{mueff}
\end{align}
Then, $J_k(\mu)$ is split into the perturbative part and the non-perturbative part
\begin{align}
J_k(\mu)=J_k^{\text{pert}}(\mu_\text{eff})
+\widetilde J_k^{\text{np}}(\mu_\text{eff}).
\end{align}
Here $J_k^{\text{pert}}(\mu_\text{eff})$ is the perturbative part in the large $\mu_\text{eff}$ expansion
\begin{align}
J_k^{\text{pert}}(\mu_\text{eff})
=\frac{C_k}{3}\mu_\text{eff}^3+B_k\mu_\text{eff}+A_k,
\end{align}
where $C_k$, $B_k$ and $A_k$ are given as \cite{MN1}
\begin{align}
C_k=\frac{1}{2\pi^2k},\quad
B_k=-\frac{1}{6k}+\frac{k}{6},\quad
A_k=4A^{\text{ABJM}}_{2k},
\end{align}
with $A^{\text{ABJM}}_k$ given previously in \eqref{AABJM}.
As in the ABJM case, due to the redefinition of the chemical potential \eqref{mueff}, significant simplifications happen in the non-perturbative effects.
Namely, the bound states of the worldsheet instantons $e^{-\frac{\mu}{k}}$ and the membrane instantons $e^{-\mu}$ are taken care of by the pure worldsheet instantons and the non-perturbative part is simply given as
\begin{align}
{\widetilde J}_k^{\text{np}}(\mu_{\text{eff}})
=J_k^\text{WS}(\mu_{\text{eff}})+\widetilde J_k^\text{MB}(\mu_{\text{eff}}),
\end{align}
with the pure worldsheet instantons
\begin{align}
J_k^\text{WS}(\mu_{\text{eff}})
=\sum_{m=1}^\infty d_m(k)e^{-m\frac{\mu_{\text{eff}}}{k}},
\label{pureWS}
\end{align}
and the pure membrane instantons
\begin{align}
\widetilde J_k^\text{MB}(\mu_{\text{eff}})
=\sum_{\ell=1}^\infty({\widetilde b}_\ell(k)\mu_{\text{eff}}
+{\widetilde c}_\ell(k))e^{-\ell\mu_{\text{eff}}},\quad
{\widetilde c}_\ell(k)=-k^2\frac{d}{dk}\frac{{\widetilde b}_\ell(k)}{\ell k}.
\end{align}
The instanton coefficients simplify in the following multi-covering structure
\begin{align}
d_m(k)=\sum_{n|m}\frac{1}{n}\delta_{\frac{m}{n}}\Bigl(\frac{k}{n}\Bigr),\quad
\widetilde b_\ell(k)=\sum_{n|\ell}\frac{1}{n}\widetilde\beta_{\frac{\ell}{n}}(nk),
\label{22multicover}
\end{align}
where the first few coefficients are given by
\begin{align}
&\delta_1(k)=\frac{4}{\sin^2\frac{\pi}{k}},\quad
\delta_2(k)=\frac{-5}{\sin^2\frac{\pi}{k}},\quad
\delta_3(k)=\frac{12}{\sin^2\frac{\pi}{k}},\nonumber\\
&\delta_4(k)=\frac{-48}{\sin^2\frac{\pi}{k}}+5,\quad
\delta_5(k)=\frac{240}{\sin^2\frac{\pi}{k}}-96,
\end{align}
and
\begin{align}
&\widetilde\beta_1(k)=-\frac{2\sin 2\pi k}{\pi\sin^2\pi k},\quad
\widetilde\beta_2(k)=\frac{8\sin 2\pi k+\sin 4\pi k}{2\pi\sin^2\pi k},\quad
\widetilde\beta_3(k)=-\frac{6\sin 2\pi k+6\sin 4\pi k}{\pi\sin^2\pi k},\nonumber\\
&\widetilde\beta_4(k)=\frac{9\sin 2\pi k+30\sin 4\pi k+9\sin 6\pi k}
{\pi\sin^2\pi k},\nonumber\\
&\widetilde\beta_5(k)
=-\frac{20\sin 2\pi k+100\sin 4\pi k+100\sin 6\pi k+20\sin 8\pi k}
{\pi\sin^2\pi k}.
\label{22beta}
\end{align}

In the previous paper \cite{MN3} it was found that, as in the ABJM model,
${\widetilde J}_k^{\text{np}}(\mu_{\text{eff}})$ of the $(2,2)$ model is consistent with the expression of the free energy of topological strings \eqref{HMMOtop},
if we identify the K\"ahler parameter and the string coupling as
\begin{align}
T_\text{eff}=\frac{\mu_\text{eff}}{k},\quad g_s=\frac{1}{k},
\label{oldid}
\end{align}
since the worldsheet and membrane instanton effects are respectively given by $e^{-\frac{\mu_\text{eff}}{k}}$ and $e^{-\mu_\text{eff}}$.
Here by ``consistent'' we mean that a set of the numbers $N_{j_L,j_R}^d$ exists (not necessarily uniquely), with which the expression \eqref{HMMOtop} reproduces the instanton coefficients $\delta_d(k)$ and $\widetilde\beta_d(k)$ determined by fitting.
Note that in \cite{MN3} we assumed boldly the absence of the discrete $B$-field effect from the simple multi-covering structure without sign factors \eqref{22multicover} (unlike the ABJM case \eqref{ABJMid} with the effect of discrete $B$-field).
Otherwise, the determination of the BPS indices was impossible with too many unknowns.

Though the identification of the (diagonal) BPS indices\footnote{Here and in the following, the norm $|{\bm d}|$ stands for the sum of all components of ${\bm d}=(d_1,d_2,\cdots)$, $|{\bm d}|=\sum_id_i$, and the summation $\sum_{|{\bm d}|=d}$ stands for that under this constraint $\sum_{\{(d_1,d_2,\cdots)|d_i\in{\mathbb Z}_{\ge 0},\sum_id_i=d\}}$.
Unless mentioned explicitly, we often abbreviate $|{\bm d}|$ simply as $d$.}
$N_{j_L,j_R}^d=\sum_{|{\bm d}|=d}N_{j_L,j_R}^{{\bm d}}$ is ambiguous, the following particular set of linear combinations $n_g^d$
\begin{align}
\sum_{g=0}^\infty n_g^d(2\sin\pi g_s)^{2g}
=\sum_{j_L,j_R}N_{j_L,j_R}^d\frac{s_R\sin 2\pi g_s s_L}{\sin 2\pi g_s},
\end{align}
called the (diagonal) Gopakumar-Vafa invariants, can be straightforwardly read off from the worldsheet instanton coefficients.
As a result we found that they completely coincide with the Gopakumar-Vafa invariants of the local $D_5$ del Pezzo geometry (table 6 in \cite{KKV}).
After noticing that the inverse of the density matrix ${\widehat \rho}^{-1}$ in the classical limit $[{\widehat q},{\widehat p}]\rightarrow 0$, $\rho^{-1}=\sum_{i=1}^9 e^{m_iq+n_ip}$, gives the Newton polygon $\{(m_i,n_i)\}_{i=1}^9=\{(0,0),(0,\pm 1),(\pm 1,0),(\pm 1,\pm 1)\}$ of the local $D_5$ del Pezzo geometry (see polygon 15 in \cite{HKP}), this coincidence is probably not so surprising.
Rather it can be regarded as a non-trivial check for the conjecture on the quantization of spectral curves\footnote{We thank Alba Grassi for valuable discussions.} \cite{GHM1,WZH}.

We also found, however, that the membrane instantons are inconsistent with the topological string expression \eqref{HMMOtop} with the BPS indices on the local $D_5$ del Pezzo geometry (tables in section 5.4 of \cite{HKP}).
Since the match of the Gopakumar-Vafa invariants is highly non-trivial, we are strongly confident of the relation with the local $D_5$ del Pezzo geometry.
Although originally we adopt the ansatz \eqref{oldid} to determine the BPS indices to reduce the unknowns, after being confident with the relation to the local $D_5$ del Pezzo geometry, we can borrow the diagonal BPS indices directly from the table in \cite{HKP} and ask what is the correct identification of the K\"ahler parameters and the BPS indices.
Below, we shall start from the identification.

\subsubsection{Splitting of K\"{a}hler parameters and BPS indices}
\label{22revisit}

Here let us see how to resolve the discrepancy in the BPS indices.
We note that already at this point without rank deformations we have much information about the BPS indices.
First we introduce a minor change in \eqref{HMMOtop}.
In \cite{HMO2,HMMO} it was noticed that for the BPS indices on the local ${\mathbb P}^1\times{\mathbb P}^1$ geometry satisfying $2j_L+2j_R-1\in 2\mathbb{Z}$, the divergences in the instanton coefficients $\widetilde J^\text{np}_{k}(\mu_\text{eff})=J^\text{WS}_{k}(\mu_\text{eff})+\widetilde J^\text{MB}_{k}(\mu_\text{eff})$ are cancelled among the worldsheet instantons and the membrane instantons.
When the constraint is violated, the cancellation of the divergences does not work for \eqref{HMMOtop} any more.
Indeed the BPS indices listed in section 5.4 of \cite{HKP} do not satisfy the constraint $2j_L+2j_R-1\in 2\mathbb{Z}$.

To restore the cancellation of the divergence, we need to modify the expression of the worldsheet instantons slightly while keeping the expression of the membrane instantons,
\begin{align}
J^\text{WS}_{k}(\mu_\text{eff})
&=\sum_{j_L,j_R}\sum_{\bm d}N_{j_L,j_R}^{\bm d}
\sum_{n=1}^\infty
\frac{(-1)^{(2j_L+2j_R-1)n}s_R\sin 2\pi g_sns_L}
{n(2\sin\pi g_sn)^2\sin 2\pi g_sn}e^{-n{\bm d}\cdot{\bm T}},
\nonumber\\
\widetilde J^\text{MB}_{k}(\mu_\text{eff})
&=\sum_{j_L,j_R}\sum_{\bm d}N_{j_L,j_R}^{\bm d}
\sum_{n=1}^\infty
\frac{\partial}{\partial g_s}
\biggl[
g_s\frac{-\sin\frac{\pi n}{g_s}s_L\sin\frac{\pi n}{g_s}s_R}
{4\pi n^2(\sin\frac{\pi n}{g_s})^3}e^{-n\frac{{\bm d}\cdot{\bm T}}{g_s}}
\biggr].
\label{top}
\end{align}
Indeed, with \eqref{top}
we can show by an explicit regularization $k\rightarrow k+\epsilon$ that the divergent coefficients at each $e^{-n{\bm d}\cdot {\bm T}}$ add up to the following finite quadratic polynomial in the K\"{a}hler parameters
\begin{align}
&J^\text{WS}_{k}(\mu_\text{eff})\Big|_{j_L,j_R,{\bm d},kn}
+{\widetilde J}^{\text{MB}}_k(\mu_{\text{eff}})\Bigr|_{j_L,j_R,{\bm d},n}
=N_{j_L,j_R}^{{\bm d}}(-1)^{kn(s_L+s_R+1)}\nonumber\\
&\times\biggl[\frac{s_Ls_R}{4\pi^2kn^3}
\Bigl(\frac{(kn{\bm d}\cdot{\bm T})^2}{2}+kn{\bm d}\cdot{\bm T}+1\Bigr)
+\frac{s_Ls_R}{24n}\Bigl(\frac{6-4s_L^2}{k}+k(3-s_L^2-s_R^2)\Bigr)\biggr]
e^{-kn{\bm d}\cdot{\bm T}}.
\label{cancellation}
\end{align}
Here $|_{j_L,j_R,{\bm d},n}$ stands for the summand in \eqref{top} with the indicated quantum numbers.

Although the pole cancellation works by the above change, this may jeopardize the multi-covering structure \eqref{22multicover} of the worldsheet instanton in the Gopakumar-Vafa interpretation.
Looking more closely at the table of the BPS index in \cite{HKP}, however, we find the BPS indices still satisfy a modified constraint $2j_L+2j_R-1-d\in 2\mathbb{Z}$.
The modified constraint allows us to replace the sign factor $(-1)^{(2j_L+2j_R-1)n}$ with $(-1)^{dn}$, which can be compensated if we define
the K\"{a}hler parameters as
\begin{align}
T^i=\frac{\mu_{\text{eff}}}{k}+\pi i(\text{odd integer}).
\end{align}
Then, the multi-covering structure is preserved by defining
\begin{align}
\delta_d(k)=\sum_{j_L,j_R}\sum_{|{\bm d}|=d}N^{\bm d}_{j_L,j_R}
\frac{s_R\sin\frac{2\pi}{k}s_L}{(2\sin\frac{\pi}{k})^2\sin\frac{2\pi}{k}}.
\end{align}

These modifications keep the worldsheet instantons, while change the membrane instantons.
For simplicity, here let us assume that there are two K\"ahler parameters, each of which is shifted by $\pm\pi i$,
\begin{align}
T^\pm=\frac{\mu_\text{eff}}{k}\pm\pi i,\quad
g_s=\frac{1}{k},
\label{parameters}
\end{align}
as in the case of the ABJM theory \cite{HO}.
For the membrane instanton the multi-covering structure \eqref{22multicover} is kept unchanged, while each component $\widetilde\beta_d(k)$ is modified as (${\bm d}=(d^+;d^-)$)
\begin{align}
\widetilde\beta_d(k)=\sum_{j_L,j_R}\sum_{d^++d^-=d}N^{(d^+;d^-)}_{j_L,j_R}
e^{-\pi ik(d^+-d^-)}\frac{-d\sin\pi ks_L\sin\pi ks_R}
{4\pi(\sin\pi k)^3}.
\label{betad}
\end{align}

Now we try to find the BPS indices $N_{j_L,j_R}^{(d^+;d^-)}$ which are consistent with the actual membrane instanton coefficients \eqref{22beta}.
Since we are already confident of the relation to the $D_5$ del Pezzo geometry, we start with the assumption that the diagonal BPS indices are given by the table in section 5.4 of \cite{HKP}.
Then, assuming the positivity of the BPS indices $(-1)^{d^++d^--1}N_{j_L,j_R}^{(d^+;d^-)}\ge 0$ the problem reduces to how to partition these diagonal values\footnote{Note that in the current notation the BPS indices in the table of \cite{HKP} should be interpreted as $(-1)^{d-1}N_{j_L,j_R}^d$.}
\begin{align}
N_{j_L,j_R}^d=\sum_{d^++d^-=d}N_{j_L,j_R}^{(d^+;d^-)}.
\end{align}
Interestingly we find that \eqref{22beta} are consistent with \eqref{betad} even under these assumptions.
Actually the constraints completely fixes the BPS indices for small $d$.
The results for $1\le d\le 5$ are listed in table \ref{BPS}.
Due to the reality of the grand potential, the BPS index associated to the degree $(d^+;d^-)$ and that associated to the conjugate degree $(d^-;d^+)$ are the same,
\begin{align}
N^{(d^+;d^-)}_{j_L,j_R}=N^{(d^-;d^+)}_{j_L,j_R}.
\end{align}
In table \ref{BPS} we have denoted these two conjugate degrees in different columns.
\begin{table}
\begin{center}
\begin{tabular}{|c||c|c||c|}
\hline
$|{\bm d}|$&
\multicolumn{2}{|c||}{$\;{\bm d}=(d^+;d^-)\;$}
&$(-1)^{|{\bm d}|-1}\sum_{j_L,j_R}
N^{\bm d}_{j_L,j_R}(j_L,j_R)$\\
\hline\hline
$1$&$(1;0)$&$(0;1)$&$8(0,0)$\\
\hline
$2$&$(2;0)$&$(0;2)$&$(0,\frac{1}{2})$\\
\cline{2-4}
&\multicolumn{2}{|c||}{$(1;1)$}&$8(0,\frac{1}{2})$\\
\hline
$3$&$(2;1)$&$(1;2)$&$8(0,1)$\\
\hline
$4$&$(3;1)$&$(1;3)$&$8(0,\frac{3}{2})$\\
\cline{2-4}
&\multicolumn{2}{|c||}{$(2;2)$}&$(0,\frac{1}{2})+29(0,\frac{3}{2})+(\frac{1}{2},2)$\\
\hline
$5$&$(4;1)$&$(1;4)$&$8(0,2)$\\
\cline{2-4}
&$(3;2)$&$(2;3)$&$8(0,1)+64(0,2)+8(\frac{1}{2},\frac{5}{2})$\\
\hline
\end{tabular}
\caption{The BPS indices $N^{\bm d}_{j_L,j_R}$ identified for the $(2,2)$ model under the assumption of two K\"ahler parameters with the opposite $B$-field effect \eqref{parameters}.}
\label{BPS}
\end{center}
\end{table}
For example, the second membrane instanton in \eqref{22beta} can be decomposed as
\begin{align}
\widetilde\beta_2(k)=\frac{8\sin 2\pi k+\sin 4\pi k}{2\pi\sin^2\pi k}
=(e^{-2\pi ik}+e^{2\pi ik}+8)\frac{\sin 2\pi k}{2\pi \sin^2\pi k},
\end{align}
where each term in the first factor can be interpreted as the contribution from $(d^+;d^-)=(2;0)$, $(0;2)$ and $(1;1)$ respectively.

\section{Two rank deformations}
\label{22_1010_0121}

In the previous section we have solved the mismatch of the BPS indices by introducing two K\"ahler parameters \eqref{parameters} and classifying the known diagonal BPS indices.
Of course it is reasonable to doubt whether this ad hoc classification is meaningful.
In the following we shall consider the partition function of the $(2,2)$ model with the rank deformation U$(N_1)_k\times$U$(N_2)_0\times$U$(N_3)_{-k}\times$U$(N_4)_0$,
\begin{align}
&{Z_k(N_1,N_2,N_3,N_4)}
=\int\frac{D^{N_1}\mu}{N_1!}\frac{D^{N_2}\kappa}{N_2!}
\frac{D^{N_3}\nu}{N_3!}\frac{D^{N_4}\lambda}{N_4!}\nonumber\\
&\times\frac{\prod_{m\ne m'}2\sinh\frac{\mu_m-\mu_{m'}}{2}
\prod_{k\ne k'}2\sinh\frac{\kappa_k-\kappa_{k'}}{2}
\prod_{n\ne n'}2\sinh\frac{\nu_n-\mu_{n'}}{2}
\prod_{l\ne l'}2\sinh\frac{\lambda_l-\lambda_{l'}}{2}}
{\prod_{m,k'}2\cosh\frac{\mu_m-\kappa_{k'}}{2}
\prod_{k,n'}2\cosh\frac{\kappa_k-\nu_{n'}}{2}
\prod_{n,l'}2\cosh\frac{\nu_n-\lambda_{l'}}{2}
\prod_{l,m'}2\cosh\frac{\lambda_l-\mu_{m'}}{2}},
\label{Z22N1N2N3N4}
\end{align}
with the integrations defined in \eqref{integrations}.
Surprisingly, we find that the BPS indices identified in the previous section actually works for one deformation.

Our strategy is the same as before.
For the open string formalism to work, we would like to rewrite the partition function $Z_k(N_1,N_2,N_3,N_4)$ using the combined determinant formula \eqref{vdmcauchy} of the Vandermonde determinant and the Cauchy determinant.
In general the extra exponential factors $e^{\pm\frac{M}{2}\sum_m\mu_m}$ and $e^{\pm\frac{M}{2}\sum_n\nu_n}$ cause difficulties in the analysis, hence we hope them to cancel among themselves.
Eventually we have found that the following two one-parameter deformations do not require extra cares: U$(N)_k\times$U$(N+M)_0\times$U$(N+2M)_{-k}\times$U$(N+M)_0$ and U$(N+M)_k\times$U$(N)_0\times$U$(N+M)_{-k}\times$U$(N)_0$.
For later convenience, among the general rank deformations, let us denote the two-parameter deformation
\begin{align}
\text{U}(N+M_2)_k\times\text{U}(N+M_1)_0\times
\text{U}(N+2M_1+M_2)_{-k}\times\text{U}(N+M_1)_0,
\label{gaugegroup}
\end{align}
as the deformation $(M_1,M_2)$.
We shall define the grand canonical partition function $\Xi_{k,(M_1,M_2)}(z)$ and the reduced grand potential $J_{k,(M_1,M_2)}(\mu)$ in the two-parameter rank deformation $(M_1,M_2)$ by
\begin{align}
\Xi_{k,(M_1,M_2)}(z)&=\sum_{N=0}^\infty z^{N}Z_{k,(M_1,M_2)}(N),
\nonumber\\
\sum_{n=-\infty}^\infty e^{J_{k,(M_1,M_2)}(\mu+2\pi in)}
&=\sum_{N=0}^\infty e^{(N+M_2)\mu}|Z_{k,(M_1,M_2)}(N)|,\quad
(J_{k,(M_1,M_2)}(\mu))^*=J_{k,(M_1,M_2)}(\mu^*),
\label{grand}
\end{align}
with
\begin{align}
Z_{k,(M_1,M_2)}(N)=Z_k(N+M_2,N+M_1,N+2M_1+M_2,N+M_1).
\label{ZkM1M2}
\end{align}
Note that we have correlated the power of $e^{\mu}$, $N+M_2$, to the first entry of the partition function \eqref{ZkM1M2}.
This is partially motivated by the discussions of reversing the rank sizes in the ABJM theory in section \ref{not_trivial1}.
The correlation of the power of $e^{\mu}$ with the entry of the partition function is very important.
Otherwise, we would encounter difficulties later in expressing the grand potential with the two-parameter rank deformation $J_{k,(M_1,M_2)}(\mu)$ in terms of the topological string theory.

\begin{figure}[!ht]
\vspace{0.5cm}
\begin{center}
\includegraphics[scale=0.45,angle=-90]{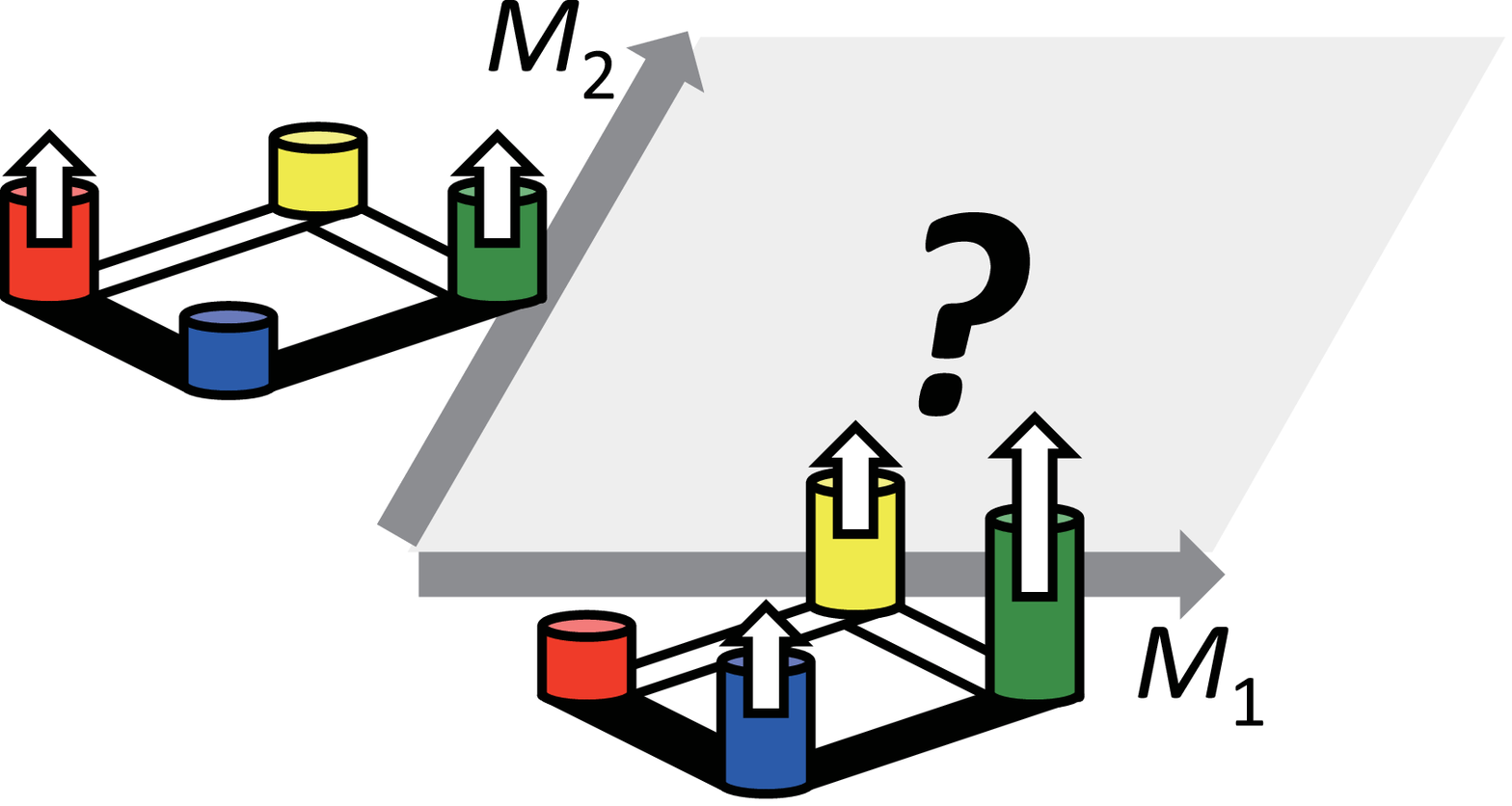}
\end{center}
\vspace{-0.5cm}
\caption{The schematic picture of the analysis studying the $(2,2)$ model.}
\label{221111_22only}
\end{figure}

In this notation $\Xi_k(z)$ and $J_k(\mu)$ without deformations appearing in section \ref{22_0000} should be denoted as $\Xi_{k,(0,0)}(z)$ and $J_{k,(0,0)}(\mu)$.
Also, the above two one-parameter deformations should be denoted respectively as $(M,0)$ and $(0,M)$ and the corresponding reduced grand potentials are $J_{k,(M,0)}(\mu)$ and $J_{k,(0,M)}(\mu)$, which are the main subjects in this section.
See figure \ref{221111_22only} for the schematic picture of the analysis in this section.

\subsection{Rank deformation $(M_1,M_2)=(M,0)$}
\label{22:0121}

\subsubsection{Open string formalism}
\label{0121fermi}

Let us start with the rank deformation U$(N)_k\times$U$(N+M)_0\times$U$(N+2M)_{-k}\times$U$(N+M)_0$.
Using the Cauchy-Vandermonde determinant \eqref{vdmcauchy}, we can express the partition function as
\begin{align}
&Z_{k,(M,0)}(N)=\int\frac{D^N\mu}{N!}\frac{D^{N+M}\kappa}{(N+M)!}
\frac{D^{N+2M}\nu}{(N+2M)!}\frac{D^{N+M}\lambda}{(N+M)!}\nonumber\\
&\qquad\times\det\begin{pmatrix}\bigl[Q(\mu,\kappa)\bigr]_{N\times(N+M)}\\
\bigl[E_l(\kappa)\bigr]_{M\times(N+M)}\end{pmatrix}
\det\begin{pmatrix}\bigl[Q'(\kappa,\nu)\bigr]_{(N+M)\times(N+2M)}\\
\bigl[E_{l'}(\nu)\bigr]_{M\times(N+2M)}\end{pmatrix}\nonumber\\
&\qquad\times\det\begin{pmatrix}\bigl[P'(\nu,\lambda)\bigr]
_{(N+2M)\times(N+M)}&\bigl[E_{a'}(\nu)\bigr]_{(N+2M)\times M}\end{pmatrix}
\nonumber\\
&\qquad\times
\det\begin{pmatrix}\bigl[P(\lambda,\mu)\bigr]
_{(N+M)\times N}&\bigl[E_a(\lambda)\bigr]_{(N+M)\times M}\end{pmatrix},
\label{ZkM1}
\end{align}
where
\begin{align}
&Q(\mu,\kappa)=\frac{1}{2\cosh\frac{\mu-\kappa}{2}},\quad
Q'(\kappa,\nu)=\frac{1}{2\cosh\frac{\kappa-\nu}{2}},\quad
E_l(\kappa)=e^{l\kappa},\nonumber\\
&P'(\nu,\lambda)=\frac{1}{2\cosh\frac{\nu-\lambda}{2}},\quad
P(\lambda,\mu)=\frac{1}{2\cosh\frac{\lambda-\mu}{2}},\quad
E_a(\kappa)=e^{a\kappa},
\label{QQEPPE}
\end{align}
with $l,l'=M-\frac{1}{2},M-\frac{3}{2},\cdots,\frac{3}{2},\frac{1}{2}$ and $a,a'=-M+\frac{1}{2},-M+\frac{3}{2},\cdots,-\frac{3}{2},-\frac{1}{2}$ (appearing in the matrices in this order).
Note that $Q$, $Q'$, $P'$ and $P$ are identical functions of two variables, except the origin of the two variables.
This redundant notation is helpful as it encodes the types of the two arguments and makes the matrix multiplication clearer.
Using the formula in appendix A of \cite{MM} (or a special case of our appendix \ref{detformula}), we can combine the four determinants into one
\begin{align}
Z_{k,(M,0)}(N)&=\int\frac{D^N\mu}{N!}
\det\begin{pmatrix}
\bigl[QQ'P'P(\mu,\mu')\bigr]_{N\times N}&\bigl[QQ'P'E_a(\mu)\bigr]_{N\times M}&\bigl[QQ'E_{a'}(\mu)\bigr]_{N\times M}\\
\bigl[E_lQ'P'P(\mu')\bigr]_{M\times N}&\bigl[E_lQ'P'E_a\bigr]_{M\times M}&\bigl[E_lQ'E_{a'}\bigr]_{M\times M}\\
\bigl[E_{l'}P'P(\mu')\bigr]_{M\times N}&\bigl[E_{l'}P'E_a\bigr]_{M\times M}&\bigl[E_{l'}E_{a'}\bigr]_{M\times M}
\end{pmatrix}.
\end{align}
With the use of the formula in appendix B of \cite{MM}, the grand canonical partition function becomes
\begin{align}
\Xi_{k,(M,0)}(z)&=\Det\begin{pmatrix}
1+zQQ'P'P&zQQ'P'E_a&zQQ'E_{a'}\\
E_lQ'P'P&E_lQ'P'E_a&E_lQ'E_{a'}\\
E_{l'}P'P&E_{l'}P'E_a&E_{l'}E_{a'}\end{pmatrix},
\end{align}
which can further be computed as
\begin{align}
\frac{\Xi_{k,(M,0)}(z)}{\Xi_k(z)}&=\det\begin{pmatrix}
E_lQ'(1+zP'PQQ')^{-1}P'E_a&E_lQ'(1+zP'PQQ')^{-1}E_{a'}\\
E_{l'}(1+zP'PQQ')^{-1}P'E_a&E_{l'}(1+zP'PQQ')^{-1}E_{a'}
\end{pmatrix},
\label{noreg}
\end{align}
with $\Xi_k(z)=\Det(1+zP'PQQ')$ being the original grand canonical partition function \eqref{22nodeform} without rank deformations.

The only remaining task is the computation of the matrix elements in \eqref{noreg}.
However, a naive evaluation leads us to the problems of infinity, $E_lQ'=E_l\bigl(2\cosh\pi il\bigr)^{-1}=\infty$ and $P'E_a=\bigl(2\cosh(-\pi ia)\bigr)^{-1}E_a=\infty$.
On the other hand, if we factor out the divergent factors there remain four identical blocks
\begin{align}
E_lQ'(1+zP'PQQ')^{-1}P'E_a
&=\bigl(2\cosh\pi il\bigr)^{-1}
E_{l}(1+zP'PQQ')^{-1}E_a\bigl(2\cosh(-\pi ia)\bigr)^{-1},\nonumber\\
E_lQ'(1+zP'PQQ')^{-1}E_a
&=\bigl(2\cosh\pi il\bigr)^{-1}E_{l}(1+zP'PQQ')^{-1}E_{a'},\nonumber\\
E_{l'}(1+zP'PQQ')^{-1}P'E_a
&=E_{l'}(1+zP'PQQ')^{-1}E_a\bigl(2\cosh(-\pi ia)\bigr)^{-1},
\end{align}
which makes the determinant vanishing.
This indicates that we can remove the divergence by appropriate elementary row/column operations.
Below we show this by an explicit regularization.

\subsubsection{Regularization}
\label{FGforM0_regularization}

To get rid of the indefinite property, let us regularize the expression \eqref{noreg} by changing $E_l$ and $E_a$ as
\begin{align}
E_l\to E_lR^{-1}\quad E_a\to RE_a\quad R=e^{i\epsilon q},
\end{align}
which implies
\begin{align} 
&E_lR^{-1}Q'
=\llangle 2\pi il|e^{-i\epsilon{\widehat q}}\frac{1}{2\cosh\frac{\widehat p}{2}}
=\llangle 2\pi il|\frac{1}{2\cosh\frac{{\widehat p}+\epsilon\hbar}{2}}
e^{-i\epsilon{\widehat q}}
\simeq\frac{e^{-\pi i l}}{\epsilon\hbar}\llangle 2\pi il|e^{-i\epsilon{\widehat q}}
=\frac{e^{-\pi i l}}{\epsilon\hbar}E_lR^{-1},\nonumber\\
&P'RE_a
=\frac{1}{2\cosh\frac{\widehat p}{2}}e^{i\epsilon{\widehat q}}|{-2\pi ia}\rrangle
=e^{i\epsilon{\widehat q}}
\frac{1}{2\cosh\frac{{\widehat p}+\epsilon\hbar}{2}}|{-2\pi ia}\rrangle
\simeq\frac{e^{\pi ia}}{\epsilon\hbar}e^{i\epsilon{\widehat q}}|{-2\pi ia}\rrangle
=\frac{e^{\pi ia}}{\epsilon\hbar}RE_a,
\end{align}
where we have used
\begin{align}
\frac{1}{2\cosh(x+\pi il)}=\frac{e^{-\pi il}}{2\sinh x},\quad
\frac{1}{2\cosh(x-\pi ia)}=\frac{e^{\pi ia}}{2\sinh x}.
\end{align}
Then, \eqref{noreg} can be regularized as
\begin{align}
&\frac{\Xi_{k,(M,0)}(z)}{\Xi_k(z)}=\frac{e^{\pi i(\sum a-\sum l)}}{\hbar^{2M}}
\nonumber\\
&\qquad\times\det\begin{pmatrix}
\epsilon^{-2}
E_lR^{-1}(1+zP'PQQ')^{-1}RE_a
&\epsilon^{-1}E_lR^{-1}(1+zP'PQQ')^{-1}E_{a'}\\
\epsilon^{-1}E_{l'}(1+zP'PQQ')^{-1}RE_a
&E_{l'}(1+zP'PQQ')^{-1}E_{a'}
\end{pmatrix}.
\end{align}
Now if we expand the regulator $R$ as
\begin{align}
R^{-1}\simeq 1-i\epsilon q,\quad
R\simeq 1+i\epsilon q,
\end{align}
we find that all the divergent terms in the matrix elements can be eliminated by the elementary row/column operations, and we obtain
\begin{align}
\frac{\Xi_{k,(M,0)}(z)}{\Xi_k(z)}
=\frac{e^{\pi i(\sum a-\sum l)}}{\hbar^{2M}}
\det\begin{pmatrix}
(E_lq)(1+zP'PQQ')^{-1}(qE_a)&(E_lq)(1+zP'PQQ')^{-1}E_{a'}\\
E_{l'}(1+zP'PQQ')^{-1}(qE_a)&E_{l'}(1+zP'PQQ')^{-1}E_{a'}
\end{pmatrix}.
\label{Xi/Xi0forM0} 
\end{align}

It is interesting to find that two functions $(E_{l'})$, $(E_lq)$ or $(E_{a'})$, $(qE_a)$ appear, which correspond to $\phi_m(q)$ and $\psi_m(q)$ introduced in \cite{MN3} to study the $(2,2)$ model without rank deformations.
Note that in \cite{MN3} we introduced these two functions technically from the Fourier transformation of the powers of the hyperbolic secant function \eqref{TWPYstructure} without knowing the deep meaning of its appearance.
After introducing the rank deformations, we see that there are in fact two functions $E_{l'}(\nu)$, $E_l(\kappa)$ or $E_{a'}(\nu)$, $E_a(\lambda)$ appearing in the determinant formula \eqref{ZkM1} which have distinct origins from the beginning:
one is associated to the vertices of non-vanishing levels in the quiver diagram, while the other is associated to those of vanishing levels.
Due to the vanishing levels we end up with an indefinite expression, which requires a regularization.
The function $\psi_m(q)$ turns out to be the regularized form of $E_l(\kappa)$ or $E_a(\lambda)$ associated to the vertices of vanishing levels.

\subsubsection{Instanton effects}

Using the open string formalism constructed in section \ref{0121fermi} and section \ref{FGforM0_regularization}, we can proceed to compute the exact values of the partition function one by one.
We have obtained the exact values $Z_{k,M}(N)=Z_k(N,N+M,N+2M,N+M)$ for $N=0,1,\cdots,N_{\text{max}}$, where $(k,M,N_{\text{max}})=(2,1,6), (3,1,6), (4,1,7), (4,2,7), (6,1,6), (6,2,6), (6,3,6)$.
We summarize the result of the first few exact values in appendix \ref{0121values}.

Using these exact values we can apply the numerical fitting to find the coefficients $C_k$, $B_{k,(M,0)}$, $A_{k,(M,0)}$ for the perturbative part and the instanton coefficients for the non-perturbative part.
For the perturbative part, we find that the coefficients
\begin{align}
C_{k}=\frac{1}{2\pi^2k},\quad
B_{k,(M,0)}=-\frac{1}{6k}-\frac{k}{3}+\frac{k}{2}\biggl(1-\frac{M}{k}\biggr)^2,\quad
A_{k,(M,0)}=4A^\text{ABJM}_{2k},
\label{0121CBA}
\end{align}
fit the exact values well.
For the non-perturbative part the instanton coefficients are listed in appendix \ref{0121Jnp}.

As we have reviewed in section \ref{22np}, the reduced grand potential has the structure of the free energy of topological strings only after we introduce the effective chemical potential $\mu_\text{eff}$.
Here $\mu_\text{eff}$ is determined such that the coefficient of $e^{-n\mu_\text{eff}}$ proportional to $\pi^{-2}$ is given by $\sim((n\mu_\text{eff})^2/2+n\mu_\text{eff}+1)/\pi^2$ as in \eqref{cancellation}.
Since the terms proportional to $\pi^{-2}$ in appendix \ref{0121Jnp} is simply a sign modification from the undeformed case, the expression of $\mu_\text{eff}$ turns out to be a sign modification from \eqref{mueff},
\begin{align}
\mu_\text{eff}=\mu+4(-1)^Me^{-\mu}
{}_4F_3\biggl(1,1,\frac{3}{2},\frac{3}{2};2,2,2;-16(-1)^Me^{-\mu}\biggr).
\end{align}
Using this effective chemical potential, we can express the reduced grand potential as in appendix \ref{0121Jnpeff}.

Now let us study the instanton effects in this reduced grand potential.
As before we expect that after the rewriting with the effective chemical potential, the bound states of worldsheet instantons and membrane instantons are incorporated simply in the pure worldsheet instanton terms.
Hence we can use all of the data without the pure membrane instanton effects to study the worldsheet instanton effects.
We find that the worldsheet instantons \eqref{pureWS} satisfy the same multi-covering structure as \eqref{22multicover}
\begin{align}
d_m(k,M)=\sum_{n|m}\frac{1}{n}\delta_{\frac{m}{n}}\biggl(\frac{k}{n},M\biggr),
\label{wsM}
\end{align}
where $\delta_d(k,M)$ are identified as
\begin{align}
&\delta_1(k,M)=\frac{16\cos\frac{M\pi}{k}}{(2\sin\frac{\pi}{k})^2},\quad
\delta_2(k,M)=\frac{-16-4\cos\frac{2M\pi}{k}}{(2\sin\frac{\pi}{k})^2},\quad
\delta_3(k,M)=\frac{48\cos\frac{M\pi}{k}}{(2\sin\frac{\pi}{k})^2},\nonumber\\
&\delta_4(k,M)=\frac{-128-64\cos\frac{2M\pi}{k}}{(2\sin\frac{\pi}{k})^2}+5,\quad
\delta_5(k,M)=\frac{880\cos\frac{M\pi}{k}+80\cos\frac{3M\pi}{k}}
{(2\sin\frac{\pi}{k})^2}-96\cos\frac{M\pi}{k}.
\label{M0ws}
\end{align}
Note that of course it is impossible to fix infinitely many coefficients of the worldsheet instanton from the finite data.
Our criterion for determining the coefficients is that the set of finite data fixes the next coefficient to vanish.

As a simplest attempt, we stick to the choice of two K\"ahler parameters.
The expression of the coefficient $B$ in \eqref{0121CBA} may indicate that the two K\"ahler parameters introduced in \eqref{parameters} are modified into
\begin{align}
T^\pm=\frac{\mu_\text{eff}}{k}\pm\pi i\bigg(1-\frac{M}{k}\biggr).
\label{Kahler220M2MM}
\end{align} 
From this expression we can classify the BPS indices by the degrees by comparing the worldsheet instantons \eqref{M0ws} with the general formula \eqref{top}.
Surprisingly, the result fits exactly with table \ref{BPS} obtained from the membrane instanton without rank deformations.
Besides, after determining the BPS indices, we can proceed to study the effects $e^{-n\mu_\text{eff}}$ which contain both worldsheet instantons and membrane instantons.
We continue to find the exact match.
This result indicates that, with the identification of the K\"{a}hler parameters \eqref{Kahler220M2MM}, our analysis in section \ref{22revisit} for the split of the BPS indices is completely meaningful, and that the $(2,2)$ model with the deformation $(M,0)$ is expressed in terms of the topological string theory with the set of the BPS indices in table \ref{BPS}.

\subsection{Rank deformation $(M_1,M_2)=(0,M)$}
\label{22:1010}

\subsubsection{Open string formalism}

Next let us consider the rank deformation of U$(N+M)_k\times$U$(N)_0\times$U$(N+M)_{-k}\times$U$(N)_0$.
As before we shall construct the open string formalism first to evaluate the exact values of the partition function.
Using \eqref{vdmcauchy} we can rewrite the partition function as
\begin{align}
&Z_{k,(0,M)}(N)=\int
\frac{D^{N+M}\mu}{(N+M)!}
\frac{D^N\kappa}{N!}
\frac{D^{N+M}\nu}{(N+M)!}
\frac{D^N\lambda}{N!}\nonumber\\
&\qquad\times\det
\begin{pmatrix}
\bigl[Q(\mu,\kappa)\bigr]_{(N+M)\times N}&\bigl[E_a(\mu)\bigr]_{(N+M)\times M}
\end{pmatrix}
\det
\begin{pmatrix}
\bigl[Q'(\kappa,\nu)\bigr]_{N\times (N+M)}\\\bigl[E'_{l'}(\nu)\bigr]_{M\times (N+M)}
\end{pmatrix}\nonumber\\
&\qquad\times\det
\begin{pmatrix}
\bigl[P'(\nu,\lambda)\bigr]_{(N+M)\times N}&\bigl[E'_{a'}(\nu)\bigr]_{(N+M)\times M}
\end{pmatrix}
\det
\begin{pmatrix}
\bigl[P(\lambda,\mu)\bigr]_{N\times (N+M)}\\\bigl[E_l(\mu)\bigr]_{M\times (N+M)}
\end{pmatrix},
\label{Z22_2020}
\end{align}
where $Q$, $Q'$, $P$, $P'$ and $E$ are given in \eqref{QQEPPE}.
$E'$ is the same as $E$.
We give them different notations by indicating that $E$ is contracted with $e^{\frac{ik}{4\pi}\mu^2}$ while $E'$ is contracted with $e^{-\frac{ik}{4\pi}\nu^2}$.
To combine the determinants the determinant formula in appendix A of \cite{MM} is not enough.
We have proved a generalized version of the determinant formula in appendix \ref{detformula}, so that we can combine all of the determinants into one
\begin{align}
&Z_{k,(0,M)}(N)=(-1)^{M^2}\int\frac{D^N\lambda}{N!}\nonumber\\
&\quad\times\det
\begin{pmatrix}
\bigl[PQQ'P'(\lambda,\lambda')\bigr]_{N\times N}
&\bigl[PQQ'E'_{a'}(\lambda)\bigr]_{N\times M}&\bigl[PE_a(\lambda)\bigr]_{N\times M}\\
\bigl[E_lQQ'P'(\lambda')\bigr]_{M\times N}
&\bigl[E_lQQ'E'_{a'}\bigr]_{M\times M}&\bigl[E_lE_a\bigr]_{M\times M}\\
\bigl[E_{l'}'P(\lambda')\bigr]_{M\times N}
&\bigl[E_{l'}'E'_{a'}\bigr]_{M\times M}&\bigl[0\bigr]_{M\times M}
\end{pmatrix}.
\end{align}
Applying the formula in \cite{MM} we finally obtain
\begin{align}
\frac{\Xi_{k,(0,M)}(z)}{\Xi_k(z)}
=\det
\begin{pmatrix}
E_l(1+zQQ'P'P)^{-1}E_a&E_lQ(1+zQ'P'PQ)^{-1}Q'E'_{a'}\\
-zE'_{l'}P'(1+zPQQ'P')^{-1}PE_a&E'_{l'}(1+zP'PQQ')^{-1}E'_{a'}
\end{pmatrix}.
\end{align}
Each element of the $2M\times 2M$ matrix can be computed in the small $z$ expansion as in \eqref{noreg}.
This time we do not need any regularizations.

\subsubsection{Instanton effects}

In this case we have obtained the exact values for $(k,M,N_\text{max})=(2,1,13)$, $(3,1,6)$, $(4,1,7)$, $(4,2,7)$, $(6,1,6)$, $(6,2,5)$ and $(6,3,5)$, which are given in appendix \ref{1010values}.
From these exact values we can start the numerical fitting,
\begin{align}
C_{k}=\frac{1}{2\pi^2k},\quad
B_{k,(0,M)}=-\frac{1}{6k}+\frac{k}{6}+\frac{M^2}{k}.
\label{1010CB}
\end{align}
We have not figured out the general expression of the constant map $A_{k,(0,M)}$ except
\begin{align}
A_{k,(0,0)}=4A^\text{ABJM}_{2k},\quad
A_{k,(0,\frac{k}{2})}=2A^\text{ABJM}_k.
\label{AM21}
\end{align}
For the other cases we obtain
\begin{align}
A_{3,(0,1)}&=4A^{\text{ABJM}}_6+\log 9,\quad
A_{4,(0,1)}=4A^{\text{ABJM}}_8+\log 16,\nonumber\\
A_{6,(0,1)}&=4A^{\text{ABJM}}_{12}+\log 36,\quad
A_{6,(0,2)}=4A^{\text{ABJM}}_{12}+\log 1296.
\label{AM22}
\end{align}
If we rewrite the coefficient $B_{k,(0,M)}$ as
\begin{align}
B_{k,(0,M)}=-\frac{1}{6k}-\frac{k}{3}
+\frac{k}{8}\biggl(1-\frac{2M}{k}\biggr)^2+\frac{k}{4}
+\frac{k}{8}\biggl(1+\frac{2M}{k}\biggr)^2,
\end{align}
it may suggest that the two K\"{a}hler parameters in the previous case further split into six
\begin{align}
T^\pm_1
=\frac{\mu_\text{eff}}{k}\pm\pi i\biggl(1-\frac{2M}{k}\biggr),\quad
T^\pm_2
=\frac{\mu_\text{eff}}{k}\pm\pi i,\quad
T^\pm_3
=\frac{\mu_\text{eff}}{k}\pm\pi i\biggl(1+\frac{2M}{k}\biggr).
\label{Kahler22M0M0}
\end{align}
As we see later in section \ref{220Mtop} these K\"ahler parameters are indeed consistent with the instanton coefficients.

As before, from the numerical analysis we can read off the reduced grand potential $J^\text{np}_{k,(0,M)}(\mu)$ and rewrite it in terms of the effective chemical potential $\widetilde J^\text{np}_{k,(0,M)}(\mu_\text{eff})$.
The results are given respectively in appendix \ref{1010Jnp} and appendix \ref{1010Jnpeff}.
This time from the signs of the terms proportional to $\pi^{-2}$ we define the effective chemical potential as
\begin{align}
\mu_\text{eff}=\mu+4e^{-\mu}
{}_4F_3\biggl(1,1,\frac{3}{2},\frac{3}{2};2,2,2;-16e^{-\mu}\biggr).
\end{align}
Then, the worldsheet instantons are given by \eqref{wsM} with
\begin{align}
&\delta_1(k,M)=\frac{8+8\cos\frac{2M\pi}{k}}{(2\sin\frac{\pi}{k})^2},\quad
\delta_2(k,M)=\frac{-12-8\cos\frac{2M\pi}{k}}{(2\sin\frac{\pi}{k})^2},\quad
\delta_3(k,M)=\frac{24+24\cos\frac{2M\pi}{k}}{(2\sin\frac{\pi}{k})^2},\nonumber\\
&\delta_4(k,M)=\frac{-88-96\cos\frac{2M\pi}{k}-8\cos\frac{4M\pi}{k}}
{(2\sin\frac{\pi}{k})^2}+5,\nonumber\\
&\delta_5(k,M)=\frac{400+480\cos\frac{2M\pi}{k}+80\cos\frac{4M\pi}{k}}
{(2\sin\frac{\pi}{k})^2}-48-48\cos\frac{2M\pi}{k}.
\label{delta0M}
\end{align}

\subsubsection{Topological strings}
\label{220Mtop}

So far we have studied the $(2,2)$ model with the rank deformation $(M_1,M_2)=(0,M)$.
We shall see whether the instanton effects match with the expression of the free energy of topological strings \eqref{top}.

In the analysis of the deformation $(M_1,M_2)=(M,0)$ we have found two K\"{a}hler parameters \eqref{Kahler220M2MM}.
Here in the analysis of the deformation $(M_1,M_2)=(0,M)$ the instanton effects are more complicated and it is natural to expect more K\"ahler parameters to appear.
However, after setting the deformation parameter $M$ to zero, these two deformations reduce to the same $(2,2)$ model.
Therefore we naturally guess that, in the deformation $(M_1,M_2)=(0,M)$, the degrees in the deformation $(M_1,M_2)=(M,0)$ are further split as $(d^+;d^-)\to(d^{+}_1,d^{+}_2,\cdots;d^{-}_1,d^{-}_2,\cdots)$ with the total Gopakumar-Vafa invariants preserved,
\begin{align}
N_{j_L,j_R}^{(d^+;d^-)}=\sum_{\sum_id^{+}_i=d^+,\sum_id^{-}_i=d^-}
N_{j_L,j_R}^{(d^{+}_1,d^{+}_2,\cdots;d^{-}_1,d^{-}_2,\cdots)},\quad
(-1)^{d-1}N_{j_L,j_R}^{(d^{+}_1,d^{+}_2,\cdots;d^{-}_1,d^{-}_2,\cdots)}\ge 0.
\end{align}

Let us first discuss the validity of the identification of the K\"ahler parameters \eqref{Kahler22M0M0} along with the BPS indices of $d=1$.
From the reality of the reduced grand potential, we expect that each K\"ahler parameter is accompanied with its complex conjugate
\begin{align}
T^{\pm}_i=\frac{\mu_\text{eff}}{k}\pm\pi i\biggl(1-b_i\frac{2M}{k}\biggr),
\end{align}
and that the BPS indices are symmetric under the exchange of the degrees for all the complex-conjugate pairs
\begin{align}
N_{j_L,j_R}^{(d^{+}_1,d^{+}_2,\cdots;d^{-}_1,d^{-}_2,\cdots)}
=N_{j_L,j_R}^{(d^{-}_1,d^{-}_2,\cdots;d^{+}_1,d^{+}_2,\cdots)}.
\end{align}
Under these assumptions, we shall match the topological string expression with the first worldsheet instanton.
Since we only have non-vanishing BPS indices for $(j_L,j_R)=(0,0)$ in $d=1$, we shall match $J^\text{WS}_{k}(\mu_\text{eff})$ in \eqref{top} with these substitutions against $\delta_1(k,M)$ in \eqref{delta0M}
\begin{align}
\sum_iN^{i}_{0,0}
\frac{2\cos(b_i\frac{2M\pi}{k})}{(2\sin\frac{\pi}{k})^2}
=\frac{8+8\cos\frac{2M\pi}{k}}{(2\sin\frac{\pi}{k})^2},
\end{align}
with the notation of the BPS indices simplified as
\begin{align}
N^{i}_{0,0}
=N^{(0,\cdots,0,\stackrel{i\text{-th}}{1},0,\cdots,0;0,0,\cdots,0)}_{0,0}
=N^{(0,0,\cdots,0;0,\cdots,0,\stackrel{i\text{-th}}{1},0,\cdots,0)}_{0,0}.
\end{align}
From the consistency $b_i$ has to be
\begin{align}
b_1=1,\quad b_2=0,\quad b_3=-1,
\end{align}
and the non-vanishing BPS indices satisfy
\begin{align}
N^2_{0,0}=4,\quad N^1_{0,0}+N^3_{0,0}=4.
\label{N2N13}
\end{align}
This also implies
\begin{align}
\sum_{i=1}^3N^i_{0,0}=8,\quad\sum_{i=1}^3N^i_{0,0}b_i^2=4,
\label{sumN}
\end{align}
which will be useful below.
Next, let us turn to the match of the quadratic polynomial in the instanton effects where we have both the first membrane instanton and other worldsheet instantons.
By matching the constant part without the proportionality $\pi^{-2}$ in \eqref{cancellation} obtained after cancelling the apparent singularities with $n=1$ and $(j_L,j_R)=(0,0)$ substituted, we obtain the equality of
\begin{align}
\sum_{i=1}^3
2N^i_{0,0}\biggl[\frac{1}{4\pi^2k}\frac{(\pi ik(1-b_i\frac{2M}{k}))^2}{2}
+\frac{1}{24}\biggl(\frac{2}{k}+k\biggr)\biggr]
+\sum_{m\ne k,m|k}\frac{1}{m}\delta_{\frac{k}{m}}\biggl(\frac{k}{m},M\biggr),
\label{aftercancel}
\end{align}
with $-5$, $-\frac{8}{9}$, $-\frac{83}{2}$ and $-\frac{13}{2}$ for $(k,M)=(2,1)$, $(3,1)$, $(4,1)$ and $(4,2)$ respectively (see appendix \ref{1010Jnpeff}).
Using \eqref{sumN} for the first term in \eqref{aftercancel} and \eqref{delta0M} for the second term, we end up with the same equation
\begin{align}
\sum_{i=1}^3N^i_{0,0}b_i=0,
\end{align}
for all of the four cases.
Finally along with \eqref{N2N13} we find that the only non-vanishing BPS indices with $d=1$ are $N_{0,0}^2=4$, $N_{0,0}^1=N_{0,0}^3=2$, namely,
\begin{align}
&N^{(0,1,0;0,0,0)}_{0,0}=N^{(0,0,0;0,1,0)}_{0,0}=4,\nonumber\\
&N^{(1,0,0;0,0,0)}_{0,0}=N^{(0,0,0;1,0,0)}_{0,0}
=N^{(0,0,1;0,0,0)}_{0,0}=N^{(0,0,0;0,0,1)}_{0,0}=2.
\end{align}

\begin{table}[ht!]
\begin{center}
\begin{tabular}{|c||c|c||c|}
\hline
$|{\bm d}|$&
\multicolumn{2}{c||}
{$\{{\bm d}=(d_1^{+},d_2^{+},d_3^{+};d_1^{-},d_2^{-},d_3^{-})\}$}
&$\pm N^{\bm d}_{j_L,j_R}(j_L,j_R)$\\
\hline\hline
&$(1,0,0;0,0,0)$&$(0,0,0;1,0,0)$&$2(0,0)$\\
\cline{2-4}
$1$&$(0,1,0;0,0,0)$&$(0,0,0;0,1,0)$&$4(0,0)$\\
\cline{2-4}
&$(0,0,1;0,0,0)$&$(0,0,0;0,0,1)$&$2(0,0)$\\
\hline
&$(0,2,0;0,0,0),(1,0,1;0,0,0)$&$(0,0,0;0,2,0),(0,0,0;1,0,1)$
&$(0,\frac{1}{2})$\\
\cline{2-4}
$2$&$(1,0,0;0,1,0),(0,1,0;0,0,1)$&$(0,1,0;1,0,0),(0,0,1;0,1,0)$
&$2(0,\frac{1}{2})$\\
\cline{2-4}
&\multicolumn{2}{c||}{$(1,0,0;1,0,0),(0,1,0;0,1,0),(0,0,1;0,0,1)$}
&$4(0,\frac{1}{2})$\\
\hline
&$\begin{array}{c}
(2,0,0;1,0,0),(0,2,0;0,0,1),\\[-4pt]
(1,1,0;0,1,0),(1,0,1;0,0,1)
\end{array}$&
$\begin{array}{c}
(1,0,0;2,0,0),(0,0,1;0,2,0),\\[-4pt]
(0,1,0;1,1,0),(0,0,1;1,0,1)
\end{array}$&
$2(0,1)$\\
\cline{2-4}
$3$&$\begin{array}{c}
(0,2,0;0,1,0),(1,0,1;0,1,0),\\[-4pt]
(1,1,0;1,0,0),(0,1,1;0,0,1)
\end{array}$&
$\begin{array}{c}
(0,1,0;0,2,0),(0,1,0;1,0,1),\\[-4pt]
(1,0,0;1,1,0),(0,0,1;0,1,1)
\end{array}$&
$4(0,1)$\\
\cline{2-4}
&$\begin{array}{c}
(0,0,2;0,0,1),(0,2,0;1,0,0),\\[-4pt]
(0,1,1;0,1,0),(1,0,1;1,0,0)
\end{array}$&
$\begin{array}{c}
(0,0,1;0,0,2),(1,0,0;0,2,0),\\[-4pt]
(0,1,0;0,1,1),(1,0,0;1,0,1)
\end{array}$&
$2(0,1)$\\
\hline
\end{tabular}
\caption{The BPS indices $N^{\bm d}_{j_L,j_R}$ identified for the $(2,2)$ model under the assumption of six K\"ahler parameters.
Here $\pm N^{\bm d}_{j_L,j_R}(j_L,j_R)$ on the top of the right column stands for the abbreviation of
$(-1)^{|{\bm d}|-1}\sum_{\{{\bm d}\}}\sum_{j_L,j_R}
N^{\bm d}_{j_L,j_R}(j_L,j_R)$.
}
\label{d123}
\end{center}
\end{table}

After fixing the K\"ahler parameters to be \eqref{Kahler22M0M0} we can proceed further to higher instantons.
Since we have generated six K\"ahler parameters \eqref{Kahler22M0M0} from a single parameter $M$, there are some essential ambiguities.
In fact, due to the relations
\begin{align}
2T^{\pm}_2=T^{\pm}_1+T^{\pm}_3,\quad
T^{+}_1+T^{-}_1=T^{+}_2+T^{-}_2=T^{+}_3+T^{-}_3,\quad
T^{\pm}_2+T^{\mp}_1=T^{\mp}_2+T^{\pm}_3,
\label{degeneracy}
\end{align}
we cannot distinguish which degrees the BPS indices belong to.
For example, for the degrees $(d^+;d^-)=(2,0)$, the independent degrees are
\begin{align}
2T_2^+=T_1^++T_3^+,\quad
T_2^++T_1^+,\quad
T_2^++T_3^+,\quad
2T_1^+,\quad
2T_3^+,
\end{align}
while for the degrees $(d^+;d^-)=(1,1)$, the independent degrees are
\begin{align}
T_2^-+T_1^+=T_2^++T_3^-,\quad
T_1^++T_3^-,\quad
T_2^++T_2^-=T_1^++T_1^-=T_3^++T_3^-.
\end{align}
If we match the free energy of topological strings with the unknown BPS indices against \eqref{delta0M} with the help of the $e^{-2\mu_\text{eff}}$ term in $(k,M)=(2,1)$, we can uniquely fix the BPS indices as in table \ref{d123} for $d=2$, aside from the essential ambiguity \eqref{degeneracy}.
For $d=3$ strictly speaking we cannot split the BPS indices in the first row and in the third row.
We fix the BPS indices by imposing an additional symmetry
\begin{align}
N^{(d^{+}_1,d^{+}_2,d^{+}_3;d^{-}_1,d^{-}_2,d^{-}_3)}_{j_L,j_R}
=N^{(d^{+}_3,d^{+}_2,d^{+}_1;d^{-}_3,d^{-}_2,d^{-}_1)}_{j_L,j_R},
\label{13sym}
\end{align}
of exchanging $T^\pm_1$ and $T^\pm_3$.
In section \ref{two1111}, we shall see that actually the separation is consistent by introducing new deformations.

\subsection{Attempt for general $(M_1,M_2)$}

In the above two subsections, we have studied two specific rank deformations with only one of the deformation parameters $(M_1,M_2)$ turned on.
This is basically because these two deformations are under good control in the open string formalism.

If we expect that there is a topological string expression for general $(M_1,M_2)$, we would expect that there is a natural expression for the perturbative coefficients $B_{k,(M_1,M_2)}$ and $A_{k,(M_1,M_2)}$ as well as the non-perturbative effects.
It may be natural to expect that the coefficient $B_{k,(M_1,M_2)}$ is generally given by a linear combination of six terms $1/k$, $M_1$, $M_2$, $M_1^2/k$, $M_1M_2/k$ and $M_2^2/k$.
Our expressions for $B_{k,(M_1,M_2)}$ in the above two deformations, however, is not enough to fix the coefficient of $M_1M_2/k$, since it vanishes for both $M_1=0$ and $M_2=0$.
For the constant part $A_{k,(M_1,M_2)}$ we might expect that it depends only on the deformation of $M_2$, though we cannot check it at this point.

The non-perturbative part may be more restrictive from the expression of the free energy of topological strings.
If we expect that the non-perturbative part is expressed by the topological string free energy \eqref{top}, the only possibility is to change the number of the K\"ahler parameters or the identification of these geometrical parameters with the gauge theory parameters.
The simplest choice to combine the above two subsections would be
\begin{align}
T^{\pm}_1&=\frac{\mu_\text{eff}}{k}
\pm\pi i\biggl(1-\frac{M_1}{k}-\frac{2M_2}{k}\biggr),\nonumber\\
T^{\pm}_2&=\frac{\mu_\text{eff}}{k}
\pm\pi i\biggl(1-\frac{M_1}{k}\biggr),\nonumber\\
T^{\pm}_3&=\frac{\mu_\text{eff}}{k}
\pm\pi i\biggl(1-\frac{M_1}{k}+\frac{2M_2}{k}\biggr).
\label{six}
\end{align}
Unfortunately, note that the K\"ahler parameters in \eqref{six} deformed by two parameters $(M_1,M_2)$ still cannot lift the degeneracies of the degrees in \eqref{degeneracy}.

So far we have proposed a natural interpolation of the result of the above two specific rank deformations.
To really trust this interpolation we need some checks.
In the following we shall point out through the brane construction that there are two other tractable one-parameter deformations which can be used to check our proposal.

\section{Two additional rank deformations}
\label{two1111}

In the previous section, after studying the two rank deformations $(M_1,0)$ and $(0,M_2)$, we have tried to guess the unified expression of the reduced grand potential $\widetilde J_{k,(M_1,M_2)}(\mu_\text{eff})$ for the rank deformation \eqref{gaugegroup}
\begin{align}
\text{U}(N+M_2)_k\times\text{U}(N+M_1)_0\times
\text{U}(N+2M_1+M_2)_{-k}\times\text{U}(N+M_1)_0,
\label{22group}
\end{align} 
with general two deformation parameters $(M_1,M_2)$.
We expect that the non-perturbative effects are given by the free energy of topological strings with the six K\"ahler parameters \eqref{six}.
There are, however, not enough data to check the validity of \eqref{six}.

\begin{figure}[!ht]
\vspace{0.5cm}
\begin{center}
\includegraphics[scale=0.45,angle=-90]{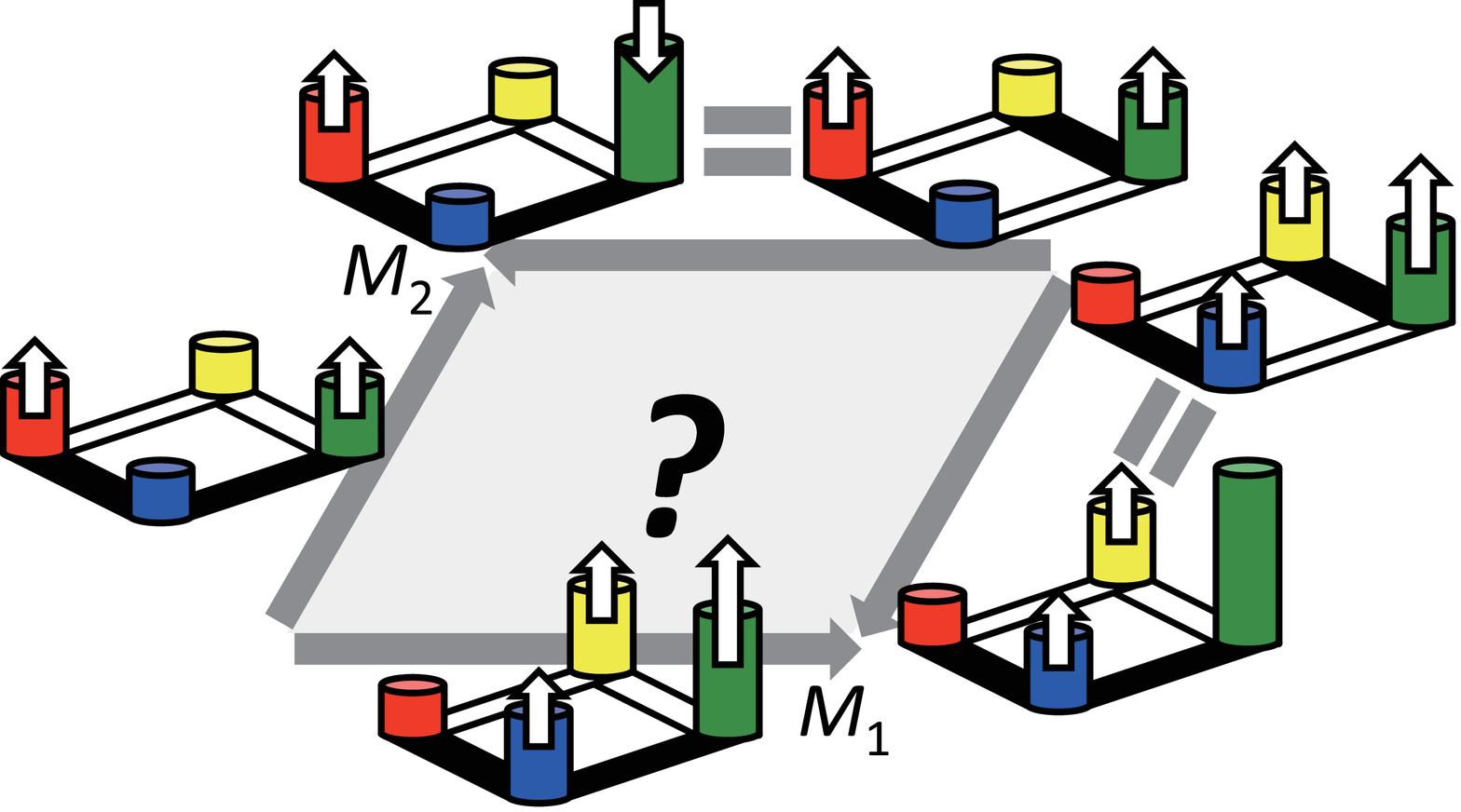}
\end{center}
\vspace{-0.5cm}
\caption{The schematic picture of the analysis studying the $(2,2)$ model using the $(1,1,1,1)$ model.}
\label{221111}
\end{figure}

In this section, we point out that the same theory can be investigated from its field-theoretical dual.
Namely, as reviewed in section \ref{transition}, by utilizing the Hanany-Witten transition in the brane picture, we can map the rank deformed $(2,2)$ model with gauge group \eqref{22group} to the rank deformed $(1,1,1,1)$ model with gauge group
\begin{align}
\text{U}(N+M_2)_k\times\text{U}(N+M_1)_{-k}\times
\text{U}(N+k-M_2)_k\times\text{U}(N+M_1)_{-k}.
\label{1111group}
\end{align}
In particular, with the duality between \eqref{22group} and \eqref{1111group}, we find that the two one-parameter rank deformations $(M_1,M_2)=(k/2-M,k/2)$ and $(M_1,M_2)=(k/2,k/2-M)$ of the $(2,2)$ model correspond respectively to the rank deformations of the $(1,1,1,1)$ model
\begin{align}
&\text{U}(N'+M)_k\times\text{U}(N')_{-k}\times\text{U}(N'+M)_k
\times\text{U}(N')_{-k},
\label{1111_0101}\\
&\text{U}(N')_k\times\text{U}(N'+M)_{-k}\times\text{U}(N'+2M)_k
\times\text{U}(N'+M)_{-k},
\label{1111_0121}
\end{align}
($N'=N+k/2-M$), which are tractable in the exact computation of the partition function.
Below we first explain briefly the duality between \eqref{22group} and \eqref{1111group} using the Hanany-Witten transition and continue to study the two additional rank deformations.
The schematic picture of the analysis in this section is given in figure \ref{221111}.

\subsection{Dualities from Hanany-Witten transitions}
\label{transition}

Here we shall explain the duality from the brane configurations.
It was known \cite{KOO,BHKK} that the quiver superconformal Chern-Simons theories are realized by NS5-branes, $(1,k)$5-branes as well as some D3-branes in the brane configurations.
Here the $(1,k)$5-brane and the NS5-brane correspond respectively to $s_a=+1$ and $s_a=-1$ which characterize the levels by 
\begin{align}
k_a=\frac{k}{2}(s_a-s_{a-1}),
\label{IIBlevels}
\end{align}
while the numbers of the D3-branes correspond to the rank of each gauge group.
Hence, from the viewpoint of the brane configurations, the $(2,2)$ model with $\{s_a\}=\{+1,+1,-1,-1\}$ studied in the previous section and the $(1,1,1,1)$ model with $\{s_a\}=\{+1,-1,+1,-1\}$ known as the orbifold ABJM theory are connected to each other under the exchange of the ordering of the 5-branes.

It is famous that the exchange of 5-branes will add/subtract the number of D3-branes in-between, called Hanany-Witten transition.
Though the original brane system studied in \cite{HW} consists of a D3-brane (in 0123 plane) stretched between a NS5-brane (in 012456 plane) and a D5-brane (in 012789 plane), we can generalize the argument to the system with $N$ D3-branes stretched between two types of general $(q,p)$5-branes \cite{KOO}.
In the case of our interest, we obtain the following equivalence
\begin{align}
\includegraphics[scale=0.3]{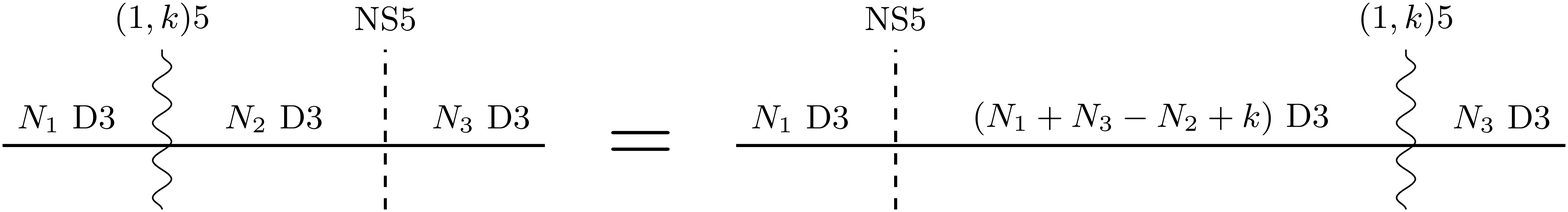}.
\label{HWrule}
\end{align}
In particular, if we apply the Hanany-Witten rule \eqref{HWrule} to map the $(2,2)$ model with $\{s_a\}=\{+1,+1,-1,-1\}$ to the $(1,1,1,1)$ model with $\{s_a\}=\{+1,-1,+1,-1\}$ by exchanging the $(1,k)$5-brane of $s_2=+1$ and the NS5-brane of $s_3=-1$, we obtain the dual theory \eqref{1111group}.

\subsection{Rank deformation $(M_1,M_2)=(k/2-M,k/2)$}

Let us start with the rank deformation $(M_1,M_2)=(k/2-M,k/2)$.
From the definition of the reduced grand potential \eqref{grand}, we find
\begin{align}
\sum_{n=-\infty}^\infty e^{J_{k,(\frac{k}{2}-M,\frac{k}{2})}(\mu+2\pi in)}
=\sum_{N=0}^\infty e^{(N+\frac{k}{2})\mu}
\biggl|Z_k\biggl(N+\frac{k}{2},N+\frac{k}{2}-M,N+\frac{3k}{2}-2M,N+\frac{k}{2}-M
\biggr)\biggr|,
\end{align}
by the direct substitution.
According to the Hanany-Witten transition \eqref{1111group}, we can rewrite the partition function into that of the $(1,1,1,1)$ model
\begin{align}
\sum_{n=-\infty}^\infty e^{J_{k,(\frac{k}{2}-M,\frac{k}{2})}(\mu+2\pi in)}
=\sum_{N=0}^\infty e^{(N+\frac{k}{2})\mu}
\biggl|Z_k^{(1,1,1,1)}\biggl(N+\frac{k}{2},N+\frac{k}{2}-M,
N+\frac{k}{2},N+\frac{k}{2}-M\biggr)\biggr|.
\label{HW1111_M0M0}
\end{align}
After shifting the summation variable $N$ by $k/2-M$, we can simplify the expression as
\begin{align}
\sum_{n=-\infty}^\infty e^{J_{k,(\frac{k}{2}-M,\frac{k}{2})}(\mu+2\pi in)}
=\sum_{N=\frac{k}{2}-M}^\infty e^{(N+M)\mu}
|Z^{(1,1,1,1)}_k(N+M,N,N+M,N)|,
\end{align}
which is (up to complex conjugation) nothing but the orbifold theory of the rank deformed ABJM theory U$(N)_k\times$U$(N+M)_{-k}$.
Hence we can skip the numerical computation by applying the technique established in \cite{HM} to generate the explicit expression of the grand potential directly from that of the U$(N)_k\times$U$(N+M)_{-k}$ theory $J^{\text{ABJM}}_{k,M}(\mu)$ obtained in \cite{MM,HO},\footnote{
Though in \cite{HM} the reduced grand potential is defined from the partition function without taking the absolute values, the derivation of \eqref{Jrepeat} works for our convention as well.
Indeed the derivation is based only on the relation between the density matrices in the closed string formalism of $Z^{\text{ABJM}}_k(N,N+M)$ and $Z_k^{(1,1,1,1)}(N,N+M,N,N+M)$.
Since the density matrices for $|Z^{\text{ABJM}}_k(N,N+M)|$ differs from the density matrix for $Z^{\text{ABJM}}_k(N,N+M)$ only by the overall phase $e^{\frac{\pi iM}{2}}$, the density matrices for $|Z^{\text{ABJM}}_k(N,N+M)|$ and $|Z_k^{(1,1,1,1)}(N,N+M,N,N+M)|$ also satisfy the required relation and we can repeat the argument in \cite{HM} in our convention.
}
\begin{align}
J_{k,(\frac{k}{2}-M,\frac{k}{2})}(\mu)
=\log\Biggl[\sum_{n=-\infty}^\infty
\exp\biggl[J^{\text{ABJM}}_{k,M}\biggl(\frac{\mu-\pi i}{2}-2\pi in\biggr)
+J^{\text{ABJM}}_{k,M}\biggl(\frac{\mu+\pi i}{2}+2\pi in\biggr)\biggr]\Biggr]+M\mu.
\label{Jrepeat}
\end{align}

For the perturbative part we obtain
\begin{align}
C_k=\frac{1}{2\pi^2 k},\quad
B_{k,(\frac{k}{2}-M,\frac{k}{2})}=-\frac{1}{6k}
+\frac{k}{24}+\frac{M}{2}+\frac{M^2}{2k},\quad
A_{k,(\frac{k}{2}-M,\frac{k}{2})}=2A^\text{ABJM}_k.
\end{align}
We can introduce the effective chemical potential $\mu_\text{eff}$ from the same criterion \eqref{cancellation} as in the $(M,0)$ case, which we find
\begin{align}
\mu_{\text{eff}}=\mu+4(-1)^{\frac{k}{2}-M}e^{-\mu}\,_4F_3
\biggl(1,1,\frac{3}{2},\frac{3}{2};2,2,2;-16(-1)^{\frac{k}{2}-M}e^{-\mu}\biggr),
\end{align}
for $k=2,4,6$ while
\begin{align}
\mu_{\text{eff}}=\mu
+2e^{-2\mu}\,_4F_3\biggl(1,1,\frac{3}{2},\frac{3}{2};2,2,2;-16e^{-2\mu}\biggr),
\end{align}
for $k=3$.
In appendix \ref{orbJnp} and appendix \ref{orbJnpeff} we list the non-perturbative part of the reduced grand potential in terms of $\mu$ and $\mu_{\text{eff}}$ respectively.

First we notice a non-trivial consistency check.
In section \ref{22:0121} we have studied the deformation $(M_1,M_2)=(0,M)$ of the $(2,2)$ model and in the current section we have studied the deformation $(M_1,M_2)=(k/2-M,k/2)$ through the duality.
Though in general these two deformations are complement with each other, both of the analyses are applicable at the point $M=k/2$.
We have found a consistency check of these two analyses.
In fact, we have compared the results of the exact values of the partition functions for $(k,M)=(2,1), (4,2), (6,3)$ and find them completely identical.
The same result is rederived for general even integers $k$ using a different method called closed string formalism in appendix \ref{22closed}.

If we express the worldsheet instanton effects by
\begin{align}
J^\text{WS}_{k,(\frac{k}{2}-M,\frac{k}{2})}(\mu)
=\sum_{m=1}^\infty d_m(k,M)e^{-m\frac{\mu}{k}},
\end{align}
we find that, surprisingly, all of the worldsheet instantons with odd instanton numbers $m$ are identically vanishing.
The first few even worldsheet instantons are given by 
\begin{align}
d_2(k,M)&=\frac{1}{\sin^2\frac{2\pi}{k}}
\Bigl(\cos\frac{2\pi(1+M)}{k}+\cos\frac{2\pi(1-M)}{k}\Bigr),\nonumber\\
d_4(k,M)&=\frac{1}{2\sin^2\frac{4\pi}{k}}
\Bigl(-3-8\cos\frac{4\pi}{k}-5\cos\frac{8\pi}{k}
-\cos\frac{4\pi(1+M)}{k}-\cos\frac{4\pi(1-M)}{k}\Bigr).
\end{align}

\subsection{Rank deformation $(M_1,M_2)=(k/2,k/2-M)$}

Now let us consider the rank deformation $(M_1,M_2)=(k/2,k/2-M)$.
Again after the direct substitution in the Hanany-Witten transition \eqref{1111group} and the shift of the summation variable $N$, we find
\begin{align}
\sum_{n=-\infty}^\infty e^{J_{k,(\frac{k}{2},\frac{k}{2}-M)}(\mu+2\pi in)}
=\sum_{N=\frac{k}{2}-M}^\infty e^{N\mu}
|Z^{(1,1,1,1)}_k(N,N+M,N+2M,N+M)|,
\end{align}
where we encounter the rank deformation of the $(1,1,1,1)$ model with gauge group U$(N)_k\times$U$(N+M)_{-k}\times$U$(N+2M)_k\times$U$(N+M)_{-k}$.
For this case we can repeat the exact computation of the partition function \eqref{noreg}, as in the $(2,2)$ model with gauge group U$(N)_k\times$U$(N+M)_0\times$U$(N+2M)_{-k}\times$U$(N+M)_0$.
In the current case, as all the Chern-Simons levels are non-zero, the matrix elements  are manifestly finite and we do not need any regularizations.
We have computed the exact values of the partition function $Z_k^{(1,1,1,1)}(N,N+M,N+2M,N+M)$ for $N=0,1,\cdots,N_\text{max}$ with $(k,M,N_\text{max})=(2,1,5),(3,1,3),(4,1,4),(4,2,4),(6,1,4),(6,2,4),(6,3,4)$.
The first few values are listed in appendix \ref{11110121values}.

This time we find the coefficients $C$ and $B$ are given by
\begin{align}
C_k=\frac{1}{2\pi^2k},\quad
B_{k,(\frac{k}{2},\frac{k}{2}-M)}
=-\frac{1}{6k}-\frac{k}{3}+\frac{k}{8}+\frac{k}{4}\biggl(1-\frac{2M}{k}\biggr)^2,
\end{align}
while $A$ is related to the $(2,2)$ model with the rank deformation $(M_1,M_2)=(0,M)$ given in \eqref{AM21} and \eqref{AM22}
\begin{align}
A_{k,(\frac{k}{2},\frac{k}{2}-M)}=A_{k,(0,\frac{k}{2}-M)},
\end{align}
for $k=2,4,6$ while $A_{3,(\frac{3}{2},\frac{1}{2})}=2A^\text{ABJM}_{3}-\log 4$ for $(k,M)=(3,1)$.
Furthermore, the reduced grand potentials are given in appendix \ref{1111Jnp} and appendix \ref{1111Jnpeff}, where $\mu_{\text{eff}}$ is determined as
\begin{align}
\mu_{\text{eff}}=\mu+4(-1)^{\frac{k}{2}}e^{-\mu}
\,_4F_3\biggl(1,1,\frac{3}{2},\frac{3}{2};2,2,2;-16(-1)^{\frac{k}{2}}e^{-\mu}\biggr),
\end{align}
for $k=2,4,6$ while
\begin{align}
\mu_\text{eff}=\mu+2e^{-2\mu}
\,_4F_3\biggl(1,1,\frac{3}{2},\frac{3}{2};2,2,2;-16e^{-2\mu}\biggr),
\end{align}
for $(k,M)=(3,1)$.

We find that the comments found in the last subsection also apply here.
In fact, at the point $M=k/2$, the analysis of the deformation $(M_1,M_2)=(k/2,k/2-M)$ in this subsection is consistent with that of the deformation $(M_1,M_2)=(M,0)$ in section \ref{22:0121}.
Also, the odd worldsheet instanton effects are identically vanishing again.

\subsection{Unifications}

In the previous section, we have considered two rank deformations $(M_1,0)$ and $(0,M_2)$.
From the rank deformation $(M_1,0)$, we have found the BPS indices are separated by the two K\"ahler parameters \eqref{Kahler220M2MM}.
Furthermore from the rank deformation $(0,M_2)$ we can further split the BPS indices by the six K\"ahler parameters \eqref{Kahler22M0M0}.
By combining these two deformations we have introduced \eqref{six} by interpolations.
Here with further rank deformations, we can confirm the consistency of these interpolations.

We have observed that the coefficient $C_k$ is independent of $(M_1,M_2)$.
This is rather trivial, however, as $C_k$ appears in the strict large $N$ limit of the free energy which depends only on the volume of the dual geometry.
The first non-trivial unification appears in the coefficient $B_{k,(M_1,M_2)}$.
From the four deformations, if we require $B_{k,(M_1,M_2)}$ to be given by a linear combination of six terms $1/k$, $M_1$, $M_2$, $M_1^2/k$, $M_1M_2/k$ and $M_2^2/k$, we uniquely find that
\begin{align}
B_{k,(M_1,M_2)}=-\frac{1}{6k}+\frac{k}{6}-M_1+\frac{M_1^2}{2k}+\frac{M_2^2}{k}.
\end{align}
The $(M_1,M_2)$-dependence of this $B$ can be expressed by quadratic terms in the imaginary part of the K\"{a}hler parameters \eqref{six}
\begin{align}
B_{k,(M_1,M_2)}=-\frac{1}{6k}-\frac{k}{3}
+\frac{k}{8}\biggl[
\Bigl(1-\frac{M_1}{k}-\frac{2M_2}{k}\Bigr)^2
+2\Bigl(1-\frac{M_1}{k}\Bigr)^2
+\Bigl(1-\frac{M_1}{k}+\frac{2M_2}{k}\Bigr)^2
\biggr].
\end{align}
This fact implies that the perturbative part of the grand potential $J_{k,(M_1,M_2)}^{\text{pert}}(\mu_\text{eff})$ can be expressed as a cubic polynomial in the K\"{a}hler parameters but without quadratic terms ($i,j,l=1,2,\cdots,6$)
\begin{align}
\frac{C_k}{3}\mu_{\text{eff}}^3+B_{k,(M_1,M_2)}\mu_{\text{eff}}
=\frac{k^2}{\pi^2}c^{ijl}T_iT_jT_l+(1+2k^2)c^iT_i,
\label{pertKahler}
\end{align}
where $(T_1,T_2,T_3,T_4,T_5,T_6)=(T_1^+,T_2^+,T_3^+,T_1^-,T_2^-,T_3^-)$ and $c^{ijl}$, $c^i$ are some constants (decomposition is not unique due to the identity \eqref{degeneracy}).
This observation is consistent with the Yukawa coupling \cite{LV,HO} and would be an additional positive evidence for the unification.
Also, it seems that the coefficient $A_{k,(M_1,M_2)}$ depends only on the argument $M_2$,
\begin{align}
A_{k,(M_1,M_2)}=A_{k,(0,M_2)}.
\end{align}

\begin{table}[ht!]
\begin{center}
\begin{tabular}{|c||c|c||c|}
\hline
$|{\bm d}|$&
\multicolumn{2}{c||}
{$\{{\bm d}=(d_1^{+},d_2^{+},d_3^{+};d_1^{-},d_2^{-},d_3^{-})\}$}
&$\pm N^{\bm d}_{j_L,j_R}(j_L,j_R)$\\
\hline\hline
&$\begin{array}{c}
(0,3,0;1,0,0),(1,1,1;1,0,0),\\[-4pt]
(0,2,1;0,1,0),(1,0,2;0,1,0),\\[-4pt]
(0,1,2;0,0,1)
\end{array}$&
$\begin{array}{c}
(1,0,0;0,3,0),(1,0,0;1,1,1),\\[-4pt]
(0,1,0;0,2,1),(0,1,0;1,0,2),\\[-4pt]
(0,0,1;0,1,2)
\end{array}$&
$2(0,\frac{3}{2})$\\
\cline{2-4}
&$\begin{array}{c}
(0,3,0;0,1,0),(1,1,1;0,1,0),\\[-4pt]
(1,2,0;1,0,0),(0,2,1;0,0,1)
\end{array}$&
$\begin{array}{c}
(0,1,0;0,3,0),(0,1,0;1,1,1),\\[-4pt]
(1,0,0;1,2,0),(0,0,1;0,2,1)
\end{array}$&
$4(0,\frac{3}{2})$\\
\cline{2-4}
&$\begin{array}{c}
(0,3,0;0,0,1),(1,1,1;0,0,1),\\[-4pt]
(1,2,0;0,1,0),(2,0,1;0,1,0),\\[-4pt]
(2,1,0;1,0,0)
\end{array}$&
$\begin{array}{c}
(0,0,1;0,3,0),(0,0,1;1,1,1),\\[-4pt]
(0,1,0;1,2,0),(0,1,0;2,0,1),\\[-4pt]
(1,0,0;2,1,0)
\end{array}$&
$2(0,\frac{3}{2})$\\
\cline{2-4}
$4$&\multicolumn{2}{c||}{$\begin{array}{c}
(2,0,0;2,0,0),(0,2,0;0,2,0),(0,0,2;0,0,2)\\[-4pt]
(1,1,0;1,1,0),(1,0,1;1,0,1),(0,1,1;0,1,1)\\[-4pt]
(0,2,0;1,1,1),(1,1,1;0,2,0)
\end{array}$}&
$\begin{array}{c}
(0,\frac{1}{2})\\
+11(0,\frac{3}{2})\\
+(\frac{1}{2},2)
\end{array}$\\
\cline{2-4}
&$\begin{array}{c}
(1,1,0;1,0,1),(1,0,1;0,1,1),\\[-4pt]
(0,2,0;0,1,1),(1,1,0;0,2,0),\\[-4pt]
(0,1,1;0,0,2),(2,0,0;1,1,0)
\end{array}$&
$\begin{array}{c}
(1,0,1;1,1,0),(0,1,1;1,0,1),\\[-4pt]
(0,1,1;0,2,0),(0,2,0;1,1,0),\\[-4pt]
(0,0,2;0,1,1),(1,1,0;2,0,0)
\end{array}$&
$8(0,\frac{3}{2})$\\
\cline{2-4}
&$\begin{array}{c}
(2,0,0;0,2,0),(2,0,0;1,0,1),\\[-4pt]
(0,2,0;0,0,2),(1,0,1;0,0,2),\\[-4pt]
(1,1,0;0,1,1)
\end{array}$&
$\begin{array}{c}
(0,2,0;2,0,0),(1,0,1;2,0,0),\\[-4pt]
(0,0,2;0,2,0),(0,0,2;1,0,1),\\[-4pt]
(0,1,1;1,1,0)
\end{array}$&
$(0,\frac{3}{2})$\\
\hline
\end{tabular}
\caption{The conjectured BPS indices $N^{\bm d}_{j_L,j_R}$ with $|{\bm d}|=4$ identified for the $(2,2)$ model under an additional assumption of exchanging $T^\pm_1$ and $T^\pm_3$.
Here $\pm N^{\bm d}_{j_L,j_R}(j_L,j_R)$ on the top of the right column stands for the abbreviation of
$(-1)^{|{\bm d}|-1}\sum_{\{{\bm d}\}}\sum_{j_L,j_R}
N^{\bm d}_{j_L,j_R}(j_L,j_R)$.}
\label{BPS6K}
\end{center}
\end{table}

We can also check the BPS indices identified in table \ref{d123}.
Interestingly, we find that our proposal reproduces correctly the vanishing odd worldsheet instantons.
For example, for the first worldsheet instanton, we find that it works for both the deformations $(M_1,M_2)=(k/2-M,k/2)$ and $(M_1,M_2)=(k/2,k/2-M)$
\begin{align}
2(e^{-T^+_1}+e^{-T^+_2})
+2(e^{-T^-_1}+e^{-T^-_2})
+2(e^{-T^+_2}+e^{-T^+_3})
+2(e^{-T^-_2}+e^{-T^-_3})
\Big|_{\begin{subarray}{l}M_1=k/2-M\\M_2=k/2\end{subarray}}&=0,\nonumber\\
2(e^{-T^+_1}+e^{-T^-_3})
+2(e^{-T^-_1}+e^{-T^+_3})
+4(e^{-T^{+}_2}+e^{-T^{-}_2})
\Big|_{\begin{subarray}{l}M_1=k/2\\M_2=k/2-M\end{subarray}}&=0,
\end{align}
with slightly different cancellations.
Also previously we have identified the BPS indices of $d=3$ by assuming the symmetry \eqref{13sym} of exchanging $T^\pm_1$ and $T^\pm_3$.
Here after obtaining additional data of the deformations $(M_1,M_2)=(k/2-M,k/2)$ and $(M_1,M_2)=(k/2,k/2-M)$, we can explicitly check that the identification of the BPS indices in table \ref{d123} is correct.

After having a non-trivial check for our proposal of describing the reduced grand potential in terms of the free energy of topological strings, we can now turn to the BPS indices with higher degrees $d=4$.
If we again assume the symmetry \eqref{13sym}, we can determine the BPS indices completely as in table \ref{BPS6K}.
Since there is no other way to check the result, we leave the BPS indices in table \ref{BPS6K} as a conjecture.

Finally, we stress that we have unified the four one-parameter deformations as a deformation with two parameters.
In the unification, it is important that in the definition of the reduced grand potential \eqref{grand} we have correlated the power of $e^\mu$ with the argument of the partition function \eqref{ZkM1M2} as partially motivated by the discussions in section \ref{not_trivial1}.
In fact, if we change the definition into, for example, that with the power of $e^\mu$ being simply $N$ or $\min(N+M_1,N+M_2)$, the unification never occurs. 

\section{Summary and conclusions}
\label{conclusion}

In this paper we have studied the $(2,2)$ model with the rank deformation of the two parameters $(M_1,M_2)$.
We start with the two one-parameter deformations by setting $M_1=0$ or $M_2=0$.
One of them splits the diagonal BPS indices with two K\"ahler parameters and the other further splits the BPS indices with six K\"ahler parameters.
Then we propose an expression with the linear interpolation of these two deformations.
Using the Hanany-Witten transition we can further check the validity of our proposal from the $(1,1,1,1)$ model with rank deformations.
To conclude, we have found that the reduced grand potential of a quiver superconformal Chern-Simons theory with the two-parameter rank deformation is expressed by the free energy of the topological string theory.
Compared with the previous works of rather trivial geometrical structures, local ${\mathbb P}^1\times{\mathbb P}^1$ or local ${\mathbb P}^2$, our model has a much richer structure.

There are apparently many questions related to our work.

Although we have claimed that we find a non-trivial check for our proposal of the topological string theory, we can only check the ``boundary'' of the moduli space (see figure \ref{221111}).
It is of course desirable to move into the ``bulk'' where both of $M_1$ and $M_2$ are neither zero nor $k/2$.
So far we are not aware of any good methods for this purpose.

From our analysis of the BPS indices using the $(2,2)$ model, we give a prediction of how the diagonal BPS indices on the local $D_5$ del Pezzo geometry in \cite{HKP} are split by various K\"ahler parameters.
We hope that these can be understood from the geometrical method.
It will be very interesting to compare our BPS indices with the results in \cite{MPTY} obtained by the refined topological vertex formalism \cite{AKMV,IKV} where the computation was done for the general six-parameter deformation.

It is surprising that all of the odd worldsheet instantons vanish when we study from the $(1,1,1,1)$ model side.
In other words, if we study the same rank deformation completely from the $(1,1,1,1)$ model side, we would never arrive at the correct K\"ahler parameters.
The vanishing is due to a fine cancellation happening with the specific K\"ahler parameters and the specific BPS indices.
Probably this reflects an unknown symmetry of the local $D_5$ del Pezzo geometry.

We have utilized the so-called open string formalism \cite{MM} for the numerical analysis.
It is also desirable to establish a closed string formalism for all of four boundary deformations.
This may encode the target Calabi-Yau threefold of the topological string theory in a direct way \cite{KaMa2}.
It may also enable us to study the membrane instantons from the WKB expansion \cite{OkuyamaOSp2}.

We have considered the rank deformations U$(N+M_2)_k\times$U$(N+M_1)_0\times$U$(N+2M_1+M_2)_{-k}\times$U$(N+M_1)_0$.
At present it is unclear whether the open string formalism is applicable to other rank deformations.
For example if we consider the deformation U$(N)_k\times$U$(N+M)_0\times$U$(N)_{-k}\times$U$(N+M)_0$ or U$(N+M)_k\times$U$(N)_0\times$U$(N+M)_{-k}\times$U$(N+2M)_0$ we apparently encounter a divergence where our regularization does not work.
We are not sure whether these divergences are solvable with technical improvements or essential to the system.

Probably it is not difficult to extend our work to the orthosymplectic groups O$(N+M_2)_k\times$Sp$(N+M_1)_0\times$O$(N+2M_1+M_2)_{-k}\times$Sp$(N+M_1)_0$ or Sp$(N+M_2)_k\times$O$(N+M_1)_0\times$Sp$(N+2M_1+M_2)_{-k}\times$O$(N+M_1)_0$ \cite{HLLLP2,ABJ,MePu,MS1,OkuyamaOSp2}.
It is interesting to see whether the results of these theories are given by the chiral projection \cite{Honda,MS2,MN5} or not.
The study would give us a deeper understanding of the instanton effects.

It is also not difficult to study the instanton effects of the vacuum expectation values of the BPS Wilson loop in the $(2,2)$ model \cite{OWZ,CDT}.
The Giambelli compatibility proved for the ABJM theory \cite{HHMO,MaMo} seems to work parallelly.

Of course it is interesting to ask whether general $(p,q)$ models are all described by the free energy of the topological string theory.
From our experience of the vanishing of the odd worldsheet instantons in the $(1,1,1,1)$ model, the possibility has even enhanced.

We hope to answer these questions in the future.

\appendix

\section{Notation for bras and kets}
\label{notations}

In this paper we use $\langle q|$ and $|q\rangle$ for position eigenstates, while $\llangle p|$ and $|p\rrangle$ denote the momentum eigenstates, which are normalized as
\begin{align}
\langle q_1|q_2\rangle&=2\pi\delta(q_1-q_2),\quad
\llangle p_1|p_2\rrangle=2\pi\delta(p_1-p_2),\quad
\langle q|p\rrangle=\sqrt{\frac{2\pi}{\hbar}}e^{\frac{i}{\hbar}qp},\quad
\llangle p|q\rangle=\sqrt{\frac{2\pi}{\hbar}}e^{-\frac{i}{\hbar}qp}.
\end{align}
We also use the following formula
\begin{align}
&
e^{-\frac{i}{2\hbar}{\widehat p}^2}
e^{-\frac{i}{2\hbar}{\widehat q}^2}
f({\widehat p})
e^{\frac{i}{2\hbar}{\widehat q}^2}
e^{\frac{i}{2\hbar}{\widehat p}^2}
=
f({\widehat q}),
\nonumber\\
&\llangle p|
e^{\frac{i}{2\hbar}{\widehat q}^2}
e^{\frac{i}{2\hbar}{\widehat p}^2}
=\frac{1}{\sqrt{-i}}e^{-\frac{i}{2\hbar}p^2}\langle{p}|,\quad
e^{-\frac{i}{2\hbar}{\widehat p}^2}
e^{-\frac{i}{2\hbar}{\widehat q}^2}
|p\rrangle
=\frac{1}{\sqrt{i}}e^{\frac{i}{2\hbar}p^2}|{p}\rangle,\nonumber\\
&\langle q_1|\frac{1}{2\cosh\frac{\widehat p}{2}}|q_2\rangle
=\frac{1}{2k\cosh\frac{q_1-q_2}{2k}},\quad
\langle q_1|\frac{1}{(2\cosh\frac{\widehat p}{2})^2}|q_2\rangle
=\frac{q_1-q_2}{4\pi k^2\sinh\frac{q_1-q_2}{2k}}.
\label{pket_qket_rotate}
\end{align}

\section{Determinantal formula}
\label{detformula}

In this appendix we shall prove a determinantal formula generalizing that proved in appendix A of \cite{MM}.

\noindent
{\bf Formula.}
Let $(\phi_i)_{1\le i\le N+M_1}$ and $(\psi_j)_{1\le j\le N+M_2}$ be arrays of functions of $x$ and let $(\xi_{ik})_{\begin{subarray}{c}1\le i\le N+M_1\\N+1\le k\le N+M_1\end{subarray}}$ and $(\eta_{lj})_{\begin{subarray}{c}N+1\le l\le N+M_2\\1\le j\le N+M_2\end{subarray}}$ be arrays of constants.
Then we have
\begin{align}
&\int\frac{d^Nx}{N!}
\det\begin{pmatrix}\bigl[\phi_i(x_k)\bigr]
_{\begin{subarray}{c}1\le i\le N+M_1\\1\le k\le N\end{subarray}}&
\bigl[\xi_{ik}\bigr]
_{\begin{subarray}{c}1\le i\le N+M_1\\N+1\le k\le N+M_1\end{subarray}}
\end{pmatrix}
\det\begin{pmatrix}[\psi_j(x_l)]
_{\begin{subarray}{c}1\le l\le N\\1\le j\le N+M_2\end{subarray}}\\
\bigl[\eta_{lj}\bigr]
_{\begin{subarray}{c}N+1\le l\le N+M_2\\1\le j\le N+M_2\end{subarray}}
\end{pmatrix}\nonumber\\
&=(-1)^{M_1M_2}\det\begin{pmatrix}\bigl[\phi_i\circ\psi_j\bigr]
_{\begin{subarray}{c}1\le i\le N+M_1\\1\le j\le N+M_2\end{subarray}}&
\bigl[\xi_{ik}\bigr]
_{\begin{subarray}{c}1\le i\le N+M_1\\N+1\le k\le N+M_1\end{subarray}}\\
\bigl[\eta_{lj}\bigr]
_{\begin{subarray}{c}N+1\le l\le N+M_2\\1\le j\le N+M_2\end{subarray}}&
\bigl[0\bigr]
_{\begin{subarray}{c}N+1\le l\le N+M_2\\N+1\le k\le N+M_1\end{subarray}}
\end{pmatrix},
\label{MMgeneralized}
\end{align}
with
\begin{align}
\phi_i\circ\psi_j=\int dx\phi_i(x)\psi_j(x).
\end{align}

\noindent
{\it Proof.}
The case of $M_2=0$ was already proved in \cite{MM}.
We can rederive it by assuming the case of $M_1=M_2=0$ using the Laplace expansion.
\begin{align}
&\int\frac{d^Nx}{N!}
\det\begin{pmatrix}\bigl[\phi_i(x_k)\bigr]
_{\begin{subarray}{c}1\le i\le N+M_1\\1\le k\le N\end{subarray}}&
\bigl[\xi_{ik}\bigr]
_{\begin{subarray}{c}1\le i\le N+M_1\\N+1\le k\le N+M_1\end{subarray}}
\end{pmatrix}
\det\begin{pmatrix}[\psi_j(x_l)]
_{\begin{subarray}{c}1\le l\le N\\1\le j\le N\end{subarray}}
\end{pmatrix}\nonumber\\
&=\int\frac{d^Nx}{N!}\sum_{(N+M_1)=(N)+(M_1)}\varepsilon^{(N),(M_1)}
\det_{N}\begin{pmatrix}\bigl[\phi_{i}(x_k)\bigr]_{i\in(N)}\end{pmatrix}
\det_{M_1}\begin{pmatrix}\bigl[\xi_{i'k}\bigr]_{i'\in(M_1)}\end{pmatrix}
\det_{N}\begin{pmatrix}[\psi_j(x_l)]
\end{pmatrix}\nonumber\\
&=\sum_{(N+M_1)=(N)+(M_1)}\varepsilon^{(N),(M_1)}
\det_{N}\begin{pmatrix}\bigl[\phi_i\circ\psi_j\bigr]_{i\in(N)}\end{pmatrix}
\det_{M_1}\begin{pmatrix}\bigl[\xi_{i'k}\bigr]_{i'\in(M_1)}\end{pmatrix}
\nonumber\\
&=\det\begin{pmatrix}\bigl[\phi_i\circ\psi_j\bigr]
_{\begin{subarray}{c}1\le i\le N+M_1\\1\le j\le N\end{subarray}}
&\bigl[\xi_{ik}\bigr]
_{\begin{subarray}{c}1\le i\le N+M_1\\N+1\le k\le N+M_1\end{subarray}}
\end{pmatrix}.
\end{align}
Here in the Laplace expansion on the second line we have partitioned the set $(N+M_1)=\{1,\cdots,N,N+1,\cdots,N+M_1\}$ by two disjoint sets $(N)$ and $(M_1)$ which contain $N$ and $M_1$ elements respectively and $\varepsilon^{(N),(M_1)}$ denotes the sign of the permutation determined by these two disjoint sets.

Similarly we can derive the general case using the Laplace expansion if we assume the case of $M_2=0$.
\begin{align}
&\int\frac{d^Nx}{N!}
\det\begin{pmatrix}\bigl[\phi_i(x_k)\bigr]
_{\begin{subarray}{c}1\le i\le N+M_1\\1\le k\le N\end{subarray}}&
\bigl[\xi_{ik}\bigr]
_{\begin{subarray}{c}1\le i\le N+M_1\\N+1\le k\le N+M_1\end{subarray}}
\end{pmatrix}
\det\begin{pmatrix}[\psi_j(x_l)]
_{\begin{subarray}{c}1\le l\le N\\1\le j\le N+M_2\end{subarray}}\\
\bigl[\eta_{lj}\bigr]
_{\begin{subarray}{c}N+1\le l\le N+M_2\\1\le j\le N+M_2\end{subarray}}
\end{pmatrix}\nonumber\\
&=\int\frac{d^Nx}{N!}
\det_{N+M_1}\begin{pmatrix}\bigl[\phi_i(x_k)\bigr]
_{(N+M_1)\times N}&
\bigl[\xi_{ik}\bigr]
_{(N+M_1)\times M_1}
\end{pmatrix}\nonumber\\
&\qquad\times\sum_{(N+M_2)=(N)+(M_2)}\varepsilon^{(N),(M_2)}
\det_{N}\begin{pmatrix}\bigl[\psi_{j}(x_l)\bigr]_{j\in(N)}\end{pmatrix}
\det_{M_2}\begin{pmatrix}\bigl[\eta_{lj'}\bigr]_{j'\in(M_2)}\end{pmatrix}
\nonumber\\
&=\sum_{(N+M_2)=(N)+(M_2)}\varepsilon^{(N),(M_2)}
\det_{N+M_1}\begin{pmatrix}
\bigl[\phi_i\circ\psi_{j}\bigr]_{j\in(N)}&\bigl[\xi_{ik}\bigr]\end{pmatrix}
\det_{M_2}\begin{pmatrix}\bigl[\eta_{lj'}\bigr]_{j'\in(M_2)}\end{pmatrix}.
\end{align}
The final expression is nothing but the Laplace expansion of the desired result \eqref{MMgeneralized},
\begin{align}
&\det\begin{pmatrix}\bigl[\phi_i\circ\psi_j\bigr]
&\bigl[\xi_{ik}\bigr]\\\bigl[\eta_{lj}\bigr]&\bigl[0\bigr]\end{pmatrix}
\nonumber\\
&=\sum_{(N+M_1+M_2)=(N+M_1)+(M_2)}\varepsilon^{(N+M_1),(M_2)}
\det_{N+M_1}\begin{pmatrix}
\bigl[\phi_i\circ\psi_{j}\bigr]_{j\in(N)}&\bigl[\xi_{ik}\bigr]\end{pmatrix}
\det_{M_2}\begin{pmatrix}\bigl[\eta_{lj'}\bigr]_{j'\in(M_2)}\end{pmatrix},
\label{phipsixieta0}
\end{align}
except the difference in the summations and the signs.
Note that in the second determinant on the right-hand side of \eqref{phipsixieta0} we need to choose $M_2$ columns out of the blocks of $[\eta_{lj}]$ and $[0]$ from the left-hand side.
Since the determinant with the columns from $[0]$ is apparently vanishing, the second determinant consists only of the columns from $[\eta_{lj}]$.
This also implies that the first determinant contains all of the columns from $[\xi_{ik}]$, which means these $M_1$ columns out of $N+M_1$ are chosen inevitably.
This associates the two partitions $(N+M_2)=(N)+(M_2)$ and $(N+M_1+M_2)=(N+M_1)+(M_2)$.
Then, the sign difference between $\varepsilon^{(N),(M_2)}$ and $\varepsilon^{(N+M_1),(M_2)}$ reduces to $(-1)^{M_1M_2}$.

\section{Exact values and reduced grand potentials}
\label{valuesofZandJandJtilde}

In this appendix we list the first few absolute values of the exact values, the non-perturbative part of the reduced grand potential $J^\text{np}_{k,(M_1,M_2)}(\mu)$ obtained from the numerical fitting, and the expression of the reduced grand potential as a function of the effective chemical potential $\widetilde J^\text{np}_{k,(M_1,M_2)}(\mu_\text{eff})$.

\subsection{Rank deformation $(M_1,M_2)=(M,0)$}

\subsubsection{Exact values}
\label{0121values}

We list the absolute values of the exact values of the partition function $Z_{k,(M,0)}(N)=Z_k(N,N+M,N+2M,N+M)$ of the $(2,2)$ model.
\begin{align*}
&
|Z_{2,(1,0)}(0)|=\frac{1}{8 \pi },
\quad
|Z_{2,(1,0)}(1)|=\frac{-60+7 \pi ^2}{4608 \pi ^3},
\quad
|Z_{2,(1,0)}(2)|=\frac{3120-2200 \pi ^2+191 \pi ^4}{4915200 \pi ^5},
\nonumber\\&\quad
|Z_{2,(1,0)}(3)|=\frac{-8205120+14876400 \pi
^2-6626956 \pi ^4+527265 \pi ^6}{416179814400 \pi ^7},
\nonumber\\&
|Z_{3,(1,0)}(0)|=\frac{1}{18 \pi },
\quad
|Z_{3,(1,0)}(1)|=\frac{-1944+297 \pi ^2-16 \sqrt{3} \pi ^3}{472392 \pi ^3},
\nonumber\\&\quad
|Z_{3,(1,0)}(2)|=\frac{6659415-5546475 \pi ^2+86400 \sqrt{3} \pi ^3+933552
\pi ^4-89600 \sqrt{3} \pi ^5}{45916502400 \pi ^5},
\nonumber\\&\quad
|Z_{3,(1,0)}(3)|=\bigl(-7150833900+14082825645 \pi ^2-109544400 \sqrt{3} \pi ^3-9782255385 \pi ^4
\nonumber\\&\qquad
+422830800 \sqrt{3}
\pi ^5+1913656656 \pi ^6-236454400 \sqrt{3} \pi ^7\bigr)\big/\bigl(2186911176307200 \pi ^7\bigr),
\nonumber\\&
|Z_{4,(1,0)}(0)|=\frac{1}{32 \pi },
\quad
|Z_{4,(1,0)}(1)|=\frac{-264+56 \pi ^2-9 \pi ^3}{147456 \pi ^3},
\nonumber\\&\quad
|Z_{4,(1,0)}(2)|=\frac{23280-23800 \pi ^2+900 \pi ^3+6834 \pi ^4-1575 \pi ^5}{471859200
\pi ^5},
\nonumber\\&\quad
|Z_{4,(1,0)}(3)|=\bigl(-46650240+108074400 \pi ^2-1587600 \pi ^3-104993672 \pi ^4+7673400 \pi ^5
\nonumber\\&\qquad
+37423080 \pi ^6-9624825 \pi
^7\bigr)\big/\bigl({53271016243200 \pi ^7}\bigr),
\nonumber\\&
|Z_{4,(2,0)}(0)|=\frac{-8+\pi ^2}{4096 \pi ^2},
\quad
|Z_{4,(2,0)}(1)|=\frac{-120+101 \pi ^2-9 \pi ^4}{1179648 \pi ^4},
\nonumber\\&\quad
|Z_{4,(2,0)}(2)|=\frac{-348480+875400 \pi ^2-504824 \pi ^4+42525 \pi
^6}{135895449600 \pi ^6},
\nonumber\\&\quad
|Z_{4,(2,0)}(3)|=\frac{-8930880+46293240 \pi ^2-71819104 \pi ^4+37169497 \pi ^6-3075975 \pi ^8}{213084064972800 \pi ^8},
\nonumber\\&
|Z_{6,(1,0)}(0)|=\frac{1}{72 \pi },
\quad
|Z_{6,(1,0)}(1)|=\frac{-16524+5967 \pi ^2-784 \sqrt{3} \pi ^3}{30233088 \pi ^3},
\nonumber\\&\quad
|Z_{6,(1,0)}(2)|=\frac{246168720-360757800 \pi ^2+7430400 \sqrt{3}
\pi ^3+199970181 \pi ^4-31249600 \sqrt{3} \pi ^5}{23509249228800 \pi ^5},
\nonumber\\&\quad
|Z_{6,(1,0)}(3)|=\bigl(-2337074389440+7402297084560 \pi ^2+42446073600 \sqrt{3} \pi ^3
\nonumber\\&\qquad
-11270868001092
\pi ^4+65870582400 \sqrt{3} \pi ^5+7818461926143 \pi ^6-1247608149200 \sqrt{3} \pi ^7\bigr)
\nonumber\\&\qquad
\big/\bigl(17915176356308582400 \pi ^7\bigr),
\nonumber\\&
|Z_{6,(2,0)}(0)|=\frac{-9+\pi ^2}{46656 \pi ^2},
\quad
|Z_{6,(2,0)}(1)|=\frac{-7776+8559 \pi ^2-280 \sqrt{3} \pi ^3-633 \pi ^4}{1088391168 \pi ^4},
\nonumber\\&\quad
|Z_{6,(2,0)}(2)|=\bigl(-1966725360+6128411400
\pi ^2-92016000 \sqrt{3} \pi ^3-5379966783 \pi ^4
\nonumber\\&\qquad
+312372000 \sqrt{3} \pi ^5+317156575 \pi ^6\bigr)\big/\bigl(15233993500262400 \pi ^6\bigr),
\nonumber\\&\quad
|Z_{6,(2,0)}(3)|=\bigl(-8825832005760+53865327635280
\pi ^2-395388604800 \sqrt{3} \pi ^3
\nonumber\\&\qquad
-119607780893328 \pi ^4+3408494644800 \sqrt{3} \pi ^5+101128503159189 \pi ^6
\nonumber\\&\qquad
-7965819885240 \sqrt{3} \pi ^7-4870045142375
\pi ^8\bigr)\big/\bigl(5804517139443980697600 \pi ^8\bigr),
\nonumber\\&
|Z_{6,(3,0)}(0)|=\frac{-81+30 \pi ^2-4 \sqrt{3} \pi ^3}{10077696 \pi ^3},
\nonumber\\&\quad
|Z_{6,(3,0)}(1)|=\frac{393660-870183 \pi ^2+19440 \sqrt{3} \pi ^3+243594 \pi ^4-31276 \sqrt{3}
\pi ^5}{1410554953728 \pi ^5},
\nonumber\\&\quad
|Z_{6,(3,0)}(2)|=\bigl(-5201035920+28455057000 \pi ^2-234446400 \sqrt{3} \pi ^3-40728240261 \pi ^4
\nonumber\\&\qquad
+1180375200 \sqrt{3} \pi ^5+10354413510
\pi ^6-1314398900 \sqrt{3} \pi ^7\bigr)\big/\bigl(1096847532018892800 \pi ^7\bigr),
\nonumber\\&\quad
|Z_{6,(3,0)}(3)|=\bigl(396230335260480-4042937279954160 \pi ^2+15588255755520 \sqrt{3}
\pi ^3
\nonumber\\&\qquad
+13336677888588684 \pi ^4-160487480260800 \sqrt{3} \pi ^5-16079242635785673 \pi ^6
\nonumber\\&\qquad
+508270842290736 \sqrt{3} \pi ^7+3910596820235622 \pi ^8-493536223635700
\sqrt{3} \pi ^9\bigr)
\nonumber\\&\qquad
\big/\bigl(7522654212719398984089600 \pi ^9\bigr).
\end{align*}

\subsubsection{Reduced grand potential}
\label{0121Jnp}

We list the non-perturbative part of the reduced grand potential $J^\text{np}_{k,(M,0)}(\mu)$.
\begin{align*}
J_{2,(1,0)}^\text{np}&=\biggl[-\frac{2\mu^2+2\mu+2}{\pi^2}-2\biggr]e^{-\mu}
+\biggl[-\frac{26\mu^2+\mu+9/2}{2\pi^2}-14\biggr]e^{-2\mu}\\
&\quad+\biggl[-\frac{736\mu^2-608\mu/3+616/9}{6\pi^2}-\frac{416}{3}\biggr]e^{-3\mu}\\
&\quad+\biggl[-\frac{2701\mu^2-13949\mu/12+11291/48}{2\pi^2}-1582\biggr]e^{-4\mu}
+{\cal O}(e^{-5\mu}),\\
J_{3,(1,0)}^\text{np}&=\frac{8}{3}e^{-\frac{1}{3}\mu}-6e^{-\frac{2}{3}\mu}
+\biggl[-\frac{4\mu^2+4\mu+4}{3\pi^2}+\frac{88}{9}\biggr]e^{-\mu}
-\frac{238}{9}e^{-\frac{4}{3}\mu}+\frac{848}{15}e^{-\frac{5}{3}\mu}\\
&\quad+\biggl[-\frac{26\mu^2+\mu+9/2}{3\pi^2}-\frac{1540}{9}\biggr]e^{-2\mu}
+\frac{82672}{189}e^{-\frac{7}{3}\mu}-\frac{11866}{9}e^{-\frac{8}{3}\mu}
+{\cal O}(e^{-3\mu}),\\
J_{4,(1,0)}^\text{np}&=4\sqrt{2}e^{-\frac{1}{4}\mu}-8e^{-\frac{1}{2}\mu}
+\frac{32\sqrt{2}}{3}e^{-\frac{3}{4}\mu}
+\biggl[-\frac{\mu^2+\mu+1}{\pi^2}-58\biggr]e^{-\mu}\\
&\quad+\frac{776\sqrt{2}}{5}e^{-\frac{5}{4}\mu}-\frac{2432}{3}e^{-\frac{3}{2}\mu}
+\frac{14144\sqrt{2}}{7}e^{-\frac{7}{4}\mu}\\
&\quad+\biggl[-\frac{13\mu^2+\mu/2+9/4}{2\pi^2}-10190\biggr]e^{-2\mu}
+{\cal O}(e^{-\frac{9}{4}\mu}),\\
J_{4,(2,0)}^\text{np}&=-8e^{-\frac{1}{2}\mu}
+\biggl[\frac{\mu^2+\mu+1}{\pi^2}-32\biggr]e^{-\mu}
-\frac{512}{3}e^{-\frac{3}{2}\mu}\\
&\quad+\biggl[-\frac{13\mu^2+\mu/2+9/4}{2\pi^2}-1022\biggr]e^{-2\mu}
+{\cal O}(e^{-\frac{5}{2}\mu}),\\
J_{6,(1,0)}^\text{np}&=8\sqrt{3}e^{-\frac{1}{6}\mu}-\frac{50}{3}e^{-\frac{1}{3}\mu}
+24\sqrt{3}e^{-\frac{1}{2}\mu}-158e^{-\frac{2}{3}\mu}
+\frac{1952\sqrt{3}}{5}e^{-\frac{5}{6}\mu}\\
&\quad+\biggl[-\frac{2\mu^2+2\mu+2}{3\pi^2}-\frac{28432}{9}\biggr]e^{-\mu}
+\frac{183040\sqrt{3}}{21}e^{-\frac{7}{6}\mu}
-\frac{650522}{9}e^{-\frac{4}{3}\mu}
+{\cal O}(e^{-\frac{3}{2}\mu}),\\
J_{6,(2,0)}^\text{np}&=8e^{-\frac{1}{6}\mu}-\frac{46}{3}e^{-\frac{1}{3}\mu}
+\frac{68}{3}e^{-\frac{1}{2}\mu}-94e^{-\frac{2}{3}\mu}
+\frac{1568}{5}e^{-\frac{5}{6}\mu}\\
&\quad+\biggl[\frac{2\mu^2+2\mu+2}{3\pi^2}-\frac{11912}{9}\biggr]e^{-\mu}
+\frac{109616}{21}e^{-\frac{7}{6}\mu}
-\frac{195098}{9}e^{-\frac{4}{3}\mu}+{\cal O}(e^{-\frac{3}{2}\mu}),\\
J_{6,(3,0)}^\text{np}&=-\frac{44}{3}e^{-\frac{1}{3}\mu}-61e^{-\frac{2}{3}\mu}
+\biggl[-\frac{2\mu^2+2\mu+2}{3\pi^2}-\frac{4834}{9}\biggr]e^{-\mu}
-\frac{94645}{18}e^{-\frac{4}{3}\mu}+{\cal O}(e^{-\frac{5}{3}\mu}).
\end{align*}

\subsubsection{Reduced grand potential in effective chemical potential}
\label{0121Jnpeff}

We list the non-perturbative part of the reduced grand potential $\widetilde J^\text{np}_{k,(M,0)}(\mu_\text{eff})$ as a function of the effective chemical potential.
For completeness we also quote the results for the undeformed case $(M_1,M_2)=(0,0)$ \cite{MN3}. 
\begin{align*}
\widetilde J_{1,(0,0)}^\text{np}
&=\frac{2(\mu_{\text{eff}}^2+2\mu_{\text{eff}}+2)}{\pi^2}e^{-\mu_{\text{eff}}}
+\biggl[-\frac{9(2\mu_{\text{eff}}^2+2\mu_{\text{eff}}+1)}{2\pi^2}
+2\biggr]e^{-2\mu_{\text{eff}}}\nonumber\\
&\quad+\biggl[\frac{164(9\mu_{\text{eff}}^2+6\mu_{\text{eff}}+2)}{27\pi^2}
-16\biggr]e^{-3\mu_{\text{eff}}}
+\biggl[-\frac{777(8\mu_{\text{eff}}^2+4\mu_{\text{eff}}+1)}{16\pi^2}
+138\biggr]e^{-4\mu_{\text{eff}}}\nonumber\\
&\quad+{\cal O}(e^{-5\mu_{\text{eff}}}),\nonumber\\
\widetilde J_{2,(0,0)}^\text{np}
&=4e^{-\frac{1}{2}\mu_{\text{eff}}}
+\biggl[\frac{\mu_{\text{eff}}^2+2\mu_{\text{eff}}+2}{\pi^2}
-7\biggr]e^{-\mu_{\text{eff}}}
+\frac{40}{3}e^{-\frac{3}{2}\mu_{\text{eff}}}\nonumber\\
&\quad+\biggl[-\frac{9(2\mu_{\text{eff}}^2+2\mu_{\text{eff}}+1)}{4\pi^2}
-\frac{75}{2}\biggr]e^{-2\mu_{\text{eff}}}
+\frac{724}{5}e^{-\frac{5}{2}\mu_{\text{eff}}}\nonumber\\
&\quad+\biggl[\frac{82(9\mu_{\text{eff}}^2+6\mu_{\text{eff}}+2)}{27\pi^2}
-\frac{1318}{3}\biggr]e^{-3\mu_{\text{eff}}}
+\frac{7704}{7}e^{-\frac{7}{2}\mu_{\text{eff}}}\nonumber\\
&\quad+\biggl[-\frac{777(8\mu_{\text{eff}}^2+4\mu_{\text{eff}}+1)}{32\pi^2}
-\frac{13847}{4}\biggr]e^{-4\mu_{\text{eff}}}
+{\cal O}(e^{-\frac{9}{2}\mu_{\text{eff}}}),\nonumber\\
\widetilde J_{2,(1,0)}^\text{np}
&=\biggl[-\frac{\mu_\text{eff}^2+2\mu_\text{eff}+2}{\pi^2}
-4\biggr]e^{-\mu_\text{eff}}
+\biggl[-\frac{9(2\mu_\text{eff}^2+2\mu_\text{eff}+1)}{4\pi^2}
-7\biggr]e^{-2\mu_\text{eff}}\\
&\quad+\biggl[-\frac{82(9\mu_\text{eff}^2+6\mu_\text{eff}+2)}{27\pi^2}
-\frac{136}{3}\biggr]e^{-3\mu_\text{eff}}\\
&\quad+\biggl[-\frac{777(8\mu_\text{eff}^2+4\mu_\text{eff}+1)}{32\pi^2}
-\frac{621}{2}\biggr]e^{-4\mu_\text{eff}}
+{\cal O}(e^{-5\mu_\text{eff}}),\nonumber\\
\widetilde J_{3,(0,0)}^\text{np}
&=\frac{16}{3}e^{-\frac{1}{3}\mu_{\text{eff}}}
-4e^{-\frac{2}{3}\mu_{\text{eff}}}
+\biggl[\frac{2(\mu_{\text{eff}}^2+2\mu_{\text{eff}}+2)}{3\pi^2}
+\frac{112}{9}\biggr]e^{-\mu_{\text{eff}}}
-61e^{-\frac{4}{3}\mu_{\text{eff}}}
+\frac{3376}{15}e^{-\frac{5}{3}\mu_{\text{eff}}}\nonumber\\
&\quad+\biggl[-\frac{3(2\mu_{\text{eff}}^2+2\mu_{\text{eff}}+1)}{2\pi^2}
-\frac{2266}{3}\Bigr]e^{-2\mu_{\text{eff}}}
+\frac{52880}{21}e^{-\frac{7}{3}\mu_{\text{eff}}}
-\frac{51655}{6}e^{-\frac{8}{3}\mu_{\text{eff}}}
+{\cal O}(e^{-3\mu_{\text{eff}}}),\nonumber\\
\widetilde J_{3,(1,0)}^\text{np}
&=\frac{8}{3}e^{-\frac{1}{3}\mu_\text{eff}}-6e^{-\frac{2}{3}\mu_\text{eff}}
+\biggl[-\frac{2(\mu_\text{eff}^2+2\mu_\text{eff}+2)}{3\pi^2}
+\frac{74}{9}\biggr]e^{-\mu_\text{eff}}
-30e^{-\frac{4}{3}\mu_\text{eff}}+\frac{1088}{15}e^{-\frac{5}{3}\mu_\text{eff}}\\
&\quad+\biggl[-\frac{3(2\mu_\text{eff}^2+2\mu_\text{eff}+1)}{2\pi^2}
-211\biggr]e^{-2\mu_\text{eff}}
+\frac{12160}{21}e^{-\frac{7}{3}\mu_\text{eff}}
-\frac{5126}{3}e^{-\frac{8}{3}\mu_\text{eff}}
+{\cal O}(e^{-3\mu_\text{eff}}),\nonumber \\
\widetilde J_{4,(0,0)}^\text{np}
&=8e^{-\frac{1}{4}\mu_\text{eff}}
-8e^{-\frac{1}{2}\mu_\text{eff}}
+\frac{80}{3}e^{-\frac{3}{4}\mu_\text{eff}}
+\biggl[\frac{\mu_\text{eff}^2+2\mu_\text{eff}+2}{2\pi^2}
-\frac{197}{2}\biggr]e^{-\mu_\text{eff}}
+\frac{1928}{5}e^{-\frac{5}{4}\mu_\text{eff}}
\nonumber\\&\quad
-\frac{4784}{3}e^{-\frac{3}{2}\mu_\text{eff}}
+\frac{44976}{7}e^{-\frac{7}{4}\mu_\text{eff}}
+\biggl[-\frac{9(2\mu_\text{eff}^2+2\mu_\text{eff}+1)}{8\pi^2}
-\frac{101875}{4}\biggr]e^{-2\mu_\text{eff}}
+{\mathcal O}(e^{-\frac{9}{4}\mu_\text{eff}}),\nonumber\\
\widetilde J_{4,(1,0)}^\text{np}
&=4\sqrt{2}e^{-\frac{1}{4}\mu_\text{eff}}
-8e^{-\frac{1}{2}\mu_\text{eff}}
+\frac{32\sqrt{2}}{3}e^{-\frac{3}{4}\mu_\text{eff}}
+\biggl[-\frac{\mu_\text{eff}^2+2\mu_\text{eff}+2}{2\pi^2}
-59\biggr]e^{-\mu_\text{eff}}\\
&\quad+\frac{756\sqrt{2}}{5}e^{-\frac{5}{4}\mu_\text{eff}}
-\frac{2384}{3}e^{-\frac{3}{2}\mu_\text{eff}}
+\frac{13920\sqrt{2}}{7}e^{-\frac{7}{4}\mu_\text{eff}}\\
&\quad+\biggl[-\frac{9(2\mu_\text{eff}^2+2\mu_\text{eff}+1)}{8\pi^2}
-\frac{19917}{2}\biggr]e^{-2\mu_\text{eff}}
+{\cal O}(e^{-\frac{9}{4}\mu_\text{eff}}),\\
\widetilde J_{4,(2,0)}^\text{np}
&=-8e^{-\frac{1}{2}\mu_\text{eff}}
+\biggl[\frac{\mu_\text{eff}^2+2\mu_\text{eff}+2}{2\pi^2}-\frac{57}{2}\biggr]
e^{-\mu_\text{eff}}
-\frac{560}{3}e^{-\frac{3}{2}\mu_\text{eff}}\\
&\quad+\biggl[-\frac{9(2\mu_\text{eff}^2+2\mu_\text{eff}+1)}{8\pi^2}
-\frac{4607}{4}\biggr]e^{-2\mu_\text{eff}}
+{\cal O}(e^{-\frac{5}{2}\mu_\text{eff}})\\
\widetilde J_{6,(0,0)}^\text{np}&=16e^{-\frac{1}{6}\mu_\text{eff}}
-\frac{52}{3}e^{-\frac{1}{3}\mu_\text{eff}}
+\frac{148}{3}e^{-\frac{1}{2}\mu_\text{eff}}
-189e^{-\frac{2}{3}\mu_\text{eff}}
+\frac{4336}{5}e^{-\frac{5}{6}\mu_\text{eff}}
\nonumber\\&\quad
+\biggl[\frac{\mu_\text{eff}^2+2\mu_\text{eff}+2}{3\pi^2}
-\frac{38137}{9}\biggr]e^{-\mu_\text{eff}}
+\frac{148752}{7}e^{-\frac{7}{6}\mu_\text{eff}}
-\frac{214253}{2}e^{-\frac{4}{3}\mu_\text{eff}}
+{\mathcal O}(e^{-\frac{3}{2}\mu_\text{eff}}),\nonumber \\
\widetilde J_{6,(1,0)}^\text{np}
&=8\sqrt{3}e^{-\frac{1}{6}\mu_\text{eff}}
-\frac{50}{3}e^{-\frac{1}{3}\mu_\text{eff}}
+24\sqrt{3}e^{-\frac{1}{2}\mu_\text{eff}}
-158e^{-\frac{2}{3}\mu_\text{eff}}
+\frac{1952\sqrt{3}}{5}e^{-\frac{5}{6}\mu_\text{eff}}\\
&\quad+\biggl[-\frac{\mu_\text{eff}^2+2\mu_\text{eff}+2}{3\pi^2}
-\frac{28430}{9}\biggr]e^{-\mu_\text{eff}}
+\frac{60976\sqrt{3}}{7}e^{-\frac{7}{6}\mu_\text{eff}}
-72258e^{-\frac{4}{3}\mu_\text{eff}}
+{\cal O}(e^{-\frac{3}{2}\mu_\text{eff}}),\\
\widetilde J_{6,(2,0)}^\text{np}
&=8e^{-\frac{1}{6}\mu_\text{eff}}
-\frac{46}{3}e^{-\frac{1}{3}\mu_\text{eff}}
+\frac{68}{3}e^{-\frac{1}{2}\mu_\text{eff}}
-94e^{-\frac{2}{3}\mu_\text{eff}}
+\frac{1568}{5}e^{-\frac{5}{6}\mu_\text{eff}}\\
&\quad+\biggl[\frac{\mu_\text{eff}^2+2\mu_\text{eff}+2}{3\pi^2}
-\frac{11887}{9}\biggr]e^{-\mu_\text{eff}}
+\frac{36576}{7}e^{-\frac{7}{6}\mu_\text{eff}}
-21698e^{-\frac{4}{3}\mu_\text{eff}}
+{\cal O}(e^{-\frac{3}{2}\mu_\text{eff}}),\\
\widetilde J_{6,(3,0)}^\text{np}
&=-\frac{44}{3}e^{-\frac{1}{3}\mu_\text{eff}}
-61e^{-\frac{2}{3}\mu_\text{eff}}
+\biggl[-\frac{\mu_\text{eff}^2+2\mu_\text{eff}+2}{3\pi^2}
-\frac{4880}{9}\biggr]e^{-\mu_\text{eff}}
-\frac{10477}{2}e^{-\frac{4}{3}\mu_\text{eff}}\\
&\quad+{\cal O}(e^{-\frac{5}{3}\mu_\text{eff}}).
\end{align*}

\subsection{Rank deformation $(M_1,M_2)=(0,M)$}

\subsubsection{Exact values}
\label{1010values}

We list the absolute values of the exact values of the partition function $Z_{k,(0,M)}(N)=Z_k(N+M,N,N+M,N)$ of the $(2,2)$ model.
\begin{align*}
&
|Z_{2,(0,1)}(0)|=\frac{1}{2},
\quad
|Z_{2,(0,1)}(1)|=-\frac{1}{128}+\frac{3}{32 \pi ^2},
\quad
|Z_{2,(0,1)}(2)|=\frac{912-536 \pi ^2+45 \pi ^4}{147456 \pi ^4},
\nonumber\\&\quad
|Z_{2,(0,1)}(3)|=\frac{498240-800400 \pi ^2+375772
\pi ^4-30375 \pi ^6}{2123366400 \pi ^6},
\nonumber\\&
|Z_{3,(0,1)}(0)|=\frac{1}{3},
\quad
|Z_{3,(0,1)}(1)|=\frac{135+6 \sqrt{3} \pi -16 \pi ^2}{2916 \pi ^2},
\nonumber\\&\quad
|Z_{3,(0,1)}(2)|=\frac{78003+3240 \sqrt{3} \pi -57753 \pi ^2-4632 \sqrt{3} \pi ^3+7424
\pi ^4}{34012224 \pi ^4},
\nonumber\\&\quad
|Z_{3,(0,1)}(3)|=\bigl(160711695+5445630 \sqrt{3} \pi -320778225 \pi ^2-31707450 \sqrt{3} \pi ^3+190523016
\pi ^4
\nonumber\\&\qquad
+20385504 \sqrt{3} \pi ^5-25676800 \pi ^6\bigr)\big/\bigl(2479491129600 \pi ^6\bigr),
\nonumber\\&
|Z_{4,(0,1)}(0)|=\frac{1}{4},
\quad
|Z_{4,(0,1)}(1)|=\frac{7+\pi -\pi ^2}{256 \pi ^2},
\quad
|Z_{4,(0,1)}(2)|=\frac{318+48 \pi -278 \pi ^2-68 \pi ^3+45 \pi ^4}{294912 \pi ^4},
\nonumber\\&\quad
|Z_{4,(0,1)}(3)|=\frac{207000+28080
\pi -476700 \pi ^2-144000 \pi ^3+338075 \pi ^4+112239 \pi ^5-60750 \pi ^6}{8493465600 \pi ^6},
\nonumber\\&
|Z_{4,(0,2)}(0)|=\frac{1}{8},
\quad
|Z_{4,(0,2)}(1)|=\frac{1}{512} \bigl(-1+\frac{10}{\pi ^2}\bigr),
\quad
|Z_{4,(0,2)}(2)|=\frac{1032-904 \pi ^2+81 \pi ^4}{1179648 \pi ^4},
\nonumber\\&\quad
|Z_{4,(0,2)}(3)|=\frac{359280-918600
\pi ^2+589034 \pi ^4-50625 \pi ^6}{16986931200 \pi ^6},
\nonumber\\&
|Z_{6,(0,1)}(0)|=\frac{1}{6},
\quad
|Z_{6,(0,1)}(1)|=\frac{1188+168 \sqrt{3} \pi -211 \pi ^2}{93312 \pi ^2},
\nonumber\\&\quad
|Z_{6,(0,1)}(2)|=\frac{3090960+508032 \sqrt{3} \pi -3506328 \pi ^2-802560 \sqrt{3}
\pi ^3+737629 \pi ^4}{8707129344 \pi ^4},
\nonumber\\&\quad
|Z_{6,(0,1)}(3)|=\bigl(58116813120+9376456320 \sqrt{3} \pi -169521368400 \pi ^2-47614651200 \sqrt{3} \pi ^3
\nonumber\\&\qquad
+158711483916
\pi ^4+49841986536 \sqrt{3} \pi ^5-39273564275 \pi ^6\bigr)\big/\bigl(10155995666841600 \pi ^6\bigr),
\nonumber\\&
|Z_{6,(0,2)}(0)|=
\frac{1}{36},
\quad
|Z_{6,(0,2)}(1)|=
\frac{1512
-156 \sqrt{3} \pi
-67 \pi^2}{279936 \pi^2},
\nonumber\\&\quad
|Z_{6,(0,2)}(2)|=
\frac{10345968
-648000 \sqrt{3} \pi
-13099320 \pi^2
+2017776 \sqrt{3} \pi^3
+144767 \pi^4}{52242776064 \pi^4},
\nonumber\\&\quad
|Z_{6,(0,2)}(3)|=
\bigl(113122137600
-3919803840 \sqrt{3} \pi
-420613462800 \pi^2
+33112087200 \sqrt{3} \pi^3
\nonumber\\&\qquad
+397554956280 \pi^4
-72852108012 \sqrt{3} \pi^5
+2257580275 \pi^6\bigr)\big/\bigl(30467987000524800 \pi^6\bigr),
\nonumber\\&
|Z_{6,(0,3)}(0)|=\frac{1}{72},
\quad
|Z_{6,(0,3)}(1)|=\frac{4428-768 \sqrt{3} \pi -25 \pi ^2}{1119744 \pi ^2},
\nonumber\\&\quad
|Z_{6,(0,3)}(2)|=\frac{16737840-1824768 \sqrt{3} \pi -21426120 \pi ^2+4803072
\sqrt{3} \pi ^3-547049 \pi ^4}{104485552128 \pi ^4},
\nonumber\\&\quad
|Z_{6,(0,3)}(3)|=\bigl(387279558720-26739486720 \sqrt{3} \pi -1500837498000 \pi ^2+188336793600 \sqrt{3} \pi ^3
\nonumber\\&\qquad
+1377347910876
\pi ^4-341246267136 \sqrt{3} \pi ^5+53219751025 \pi ^6\bigr)\big/\bigl(121871948002099200 \pi ^6\bigr).
\end{align*}

\subsubsection{Reduced grand potential}
\label{1010Jnp}

We list the non-perturbative part of the reduced grand potential $J^\text{np}_{k,(0,M)}(\mu)$.
\begin{align*}
J_{2,(0,1)}^\text{np}&=\biggl[\frac{2\mu^2+2\mu+2}{\pi^2}-2\biggr]e^{-\mu}
+\biggl[-\frac{26\mu^2+\mu+9/2}{2\pi^2}+18\biggr]e^{-2\mu}
\nonumber\\&\quad
+\biggl[\frac{736\mu^2-608\mu/3+616/9}{6\pi^2}-\frac{608}{3}\biggr]e^{-3\mu}
\nonumber\\&\quad
+\biggl[-\frac{2701\mu^2-13949\mu/12+11291/48}{2\pi^2}+2514\biggr]e^{-4\mu}
+{\cal O}(e^{-5\mu}),\\
J_{3,(0,1)}^\text{np}&=\frac{4}{3}e^{-\frac{1}{3}\mu}-2e^{-\frac{2}{3}\mu}
+\biggl[\frac{4\mu^2+4\mu+4}{3\pi^2}+\frac{20}{9}\biggr]e^{-\mu}
-\frac{88}{9}e^{-\frac{4}{3}\mu}+\frac{108}{5}e^{-\frac{5}{3}\mu}
\nonumber\\&\quad
+\biggl[-\frac{26\mu^2+\mu+9/2}{3\pi^2}-\frac{298}{9}\biggr]e^{-2\mu}
+\frac{25208}{189}e^{-\frac{7}{3}\mu}
-\frac{2873}{9}e^{-\frac{8}{3}\mu}
+{\cal O}(e^{-3\mu}),\\
J_{4,(0,1)}^\text{np}&=4e^{-\frac{1}{4}\mu}-6e^{-\frac{1}{2}\mu}
+\frac{40}{3}e^{-\frac{3}{4}\mu}
+\biggl[\frac{\mu^2+\mu+1}{\pi^2}-38\biggr]e^{-\mu}
+\frac{544}{5}e^{-\frac{5}{4}\mu}
\nonumber\\&\quad
-336e^{-\frac{3}{2}\mu}
+\frac{7200}{7}e^{-\frac{7}{4}\mu}
+\biggl[-\frac{26\mu^2+\mu+9/2}{4\pi^2}-3150\biggr]e^{-2\mu}
+{\cal O}(e^{-\frac{9}{4}\mu}),\\
J_{4,(0,2)}^\text{np}&=\biggl[\frac{\mu^2+\mu+1}{\pi^2}\biggr]e^{-\mu}
+\biggl[-\frac{26\mu^2+\mu+9/2}{4\pi^2}+2\biggr]e^{-2\mu}
+{\cal O}(e^{-3\mu}),\\
J_{6,(0,1)}^\text{np}&=12e^{-\frac{1}{6}\mu}-\frac{46}{3}e^{-\frac{1}{3}\mu}
+36e^{-\frac{1}{2}\mu}-128e^{-\frac{2}{3}\mu}+\frac{2652}{5}e^{-\frac{5}{6}\mu}
\nonumber\\&\quad
+\biggl[\frac{2\mu^2+2\mu+2}{3\pi^2}-\frac{20990}{9}\biggr]e^{-\mu}
+\frac{73624}{7}e^{-\frac{7}{6}\mu}
-\frac{429089}{9}e^{-\frac{4}{3}\mu}
+{\cal O}(e^{-\frac{3}{2}\mu}),\\
J_{6,(0,2)}^\text{np}&=4e^{-\frac{1}{6}\mu}-\frac{22}{3}e^{-\frac{1}{3}\mu}
+\frac{40}{3}e^{-\frac{1}{2}\mu}-32e^{-\frac{2}{3}\mu}
+\frac{484}{5}e^{-\frac{5}{6}\mu}
\nonumber\\&\quad
+\biggl[\frac{2\mu^2+2\mu+2}{3\pi^2}-\frac{2570}{9}\biggr]e^{-\mu}
+\frac{18520}{21}e^{-\frac{7}{6}\mu}
-\frac{24401}{9}e^{-\frac{4}{3}\mu}
+{\cal O}(e^{-\frac{3}{2}\mu}),\\
J_{6,(0,3)}^\text{np}&=-\frac{4}{3}e^{-\frac{1}{3}\mu}+3e^{-\frac{2}{3}\mu}
+\biggl[\frac{2\mu^2+2\mu+2}{3\pi^2}+\frac{70}{9}\biggr]e^{-\mu}
+\frac{203}{18}e^{-\frac{4}{3}\mu}
+{\cal O}(e^{-\frac{3}{2}\mu}).
\end{align*}

\subsubsection{Reduced grand potential in effective chemical potential}
\label{1010Jnpeff}

We list the non-perturbative part of the reduced grand potential $\widetilde J^\text{np}_{k,(0,M)}(\mu_\text{eff})$ as a function of the effective chemical potential.
\begin{align*}
\widetilde J_{2,(0,1)}^\text{np}
&=\biggl[\frac{\mu_\text{eff}^2+2\mu_\text{eff}+2}{\pi^2}
-5\biggr]e^{-\mu_\text{eff}}
+\biggl[-\frac{9(2\mu_\text{eff}^2+2\mu_\text{eff}+1)}{4\pi^2}
+\frac{23}{2}\biggr]e^{-2\mu_\text{eff}}
\nonumber\\&\quad
+\biggl[\frac{82(9\mu_\text{eff}^2+6\mu_\text{eff}+2)}{27\pi^2}
-\frac{242}{3}\biggr]e^{-3\mu_\text{eff}}
+\biggl[-\frac{777(8\mu_\text{eff}^2+4\mu_\text{eff}+1)}{32\pi^2}
+\frac{2411}{4}\biggr]e^{-4\mu_\text{eff}}
\nonumber\\&\quad
+{\cal O}(e^{-5\mu_\text{eff}}),\\
\widetilde J_{3,(0,1)}^\text{np}
&=\frac{4}{3}e^{-\frac{1}{3}\mu_\text{eff}}-2e^{-\frac{2}{3}\mu_\text{eff}}
+\biggl[\frac{2(\mu_\text{eff}^2+2\mu_\text{eff}+2)}{3\pi^2}
-\frac{8}{9}\biggr]e^{-\mu_\text{eff}}
-8e^{-\frac{4}{3}\mu_\text{eff}}+\frac{244}{15}e^{-\frac{5}{3}\mu_\text{eff}}
\nonumber\\&\quad
+\biggl[-\frac{3(2\mu_\text{eff}^2+2\mu_\text{eff}+1)}{2\pi^2}
-\frac{68}{3}\biggr]e^{-2\mu_\text{eff}}
+\frac{1712}{21}e^{-\frac{7}{3}\mu_\text{eff}}
-\frac{539}{3}e^{-\frac{8}{3}\mu_\text{eff}}
+{\cal O}(e^{-3\mu_\text{eff}}),\\
\widetilde J_{4,(0,1)}^\text{np}
&=4e^{-\frac{1}{4}\mu_\text{eff}}
-6e^{-\frac{1}{2}\mu_\text{eff}}
+\frac{40}{3}e^{-\frac{3}{4}\mu_\text{eff}}
+\biggl[\frac{\mu_\text{eff}^2+2\mu_\text{eff}+2}{2\pi^2}
-\frac{83}{2}\biggr]e^{-\mu_\text{eff}}
+\frac{564}{5}e^{-\frac{5}{4}\mu_\text{eff}}
\nonumber\\&\quad
-348e^{-\frac{3}{2}\mu_\text{eff}}
+\frac{7480}{7}e^{-\frac{7}{4}\mu_\text{eff}}
+\biggl[-\frac{9(2\mu_\text{eff}^2+2\mu_\text{eff}+1)}{8\pi^2}
-\frac{13201}{4}\biggr]e^{-2\mu_\text{eff}}
+{\cal O}(e^{-\frac{9}{4}\mu_\text{eff}}),\\
\widetilde J_{4,(0,2)}^\text{np}
&=\biggl[\frac{\mu_\text{eff}^2+2\mu_\text{eff}+2}{2\pi^2}
-\frac{13}{2}\biggr]e^{-\mu_\text{eff}}
+\biggl[-\frac{9(2\mu_\text{eff}^2+2\mu_\text{eff}+1)}{8\pi^2}
+\frac{21}{4}\biggr]e^{-2\mu_\text{eff}}
+{\cal O}(e^{-3\mu_\text{eff}}),\\
\widetilde J_{6,(0,1)}^\text{np}
&=12e^{-\frac{1}{6}\mu_\text{eff}}
-\frac{46}{3}e^{-\frac{1}{3}\mu_\text{eff}}
+36e^{-\frac{1}{2}\mu_\text{eff}}
-128e^{-\frac{2}{3}\mu_\text{eff}}
+\frac{2652}{5}e^{-\frac{5}{6}\mu_\text{eff}}
\nonumber\\&\quad
+\biggl[\frac{\mu_\text{eff}^2+2\mu_\text{eff}+2}{3\pi^2}
-\frac{21031}{9}\biggr]e^{-\mu_\text{eff}}
+\frac{73680}{7}e^{-\frac{7}{6}\mu_\text{eff}}
-47697e^{-\frac{4}{3}\mu_\text{eff}}
+{\cal O}(e^{-\frac{3}{2}\mu_\text{eff}}),\\
\widetilde J_{6,(0,2)}^\text{np}
&=4e^{-\frac{1}{6}\mu_\text{eff}}
-\frac{22}{3}e^{-\frac{1}{3}\mu_\text{eff}}
+\frac{40}{3}e^{-\frac{1}{2}\mu_\text{eff}}
-32e^{-\frac{2}{3}\mu_\text{eff}}
+\frac{484}{5}e^{-\frac{5}{6}\mu_\text{eff}}
\nonumber\\&\quad
+\biggl[\frac{\mu_\text{eff}^2+2\mu_\text{eff}+2}{3\pi^2}
-\frac{2629}{9}\biggr]e^{-\mu_\text{eff}}
+\frac{6192}{7}e^{-\frac{7}{6}\mu_\text{eff}}
-2721e^{-\frac{4}{3}\mu_\text{eff}}
+{\cal O}(e^{-\frac{3}{2}\mu_\text{eff}}),\\
\widetilde J_{6,(0,3)}^\text{np}
&=-\frac{4}{3}e^{-\frac{1}{3}\mu_\text{eff}}
+3e^{-\frac{2}{3}\mu_\text{eff}}
+\biggl[\frac{\mu_\text{eff}^2+2\mu_\text{eff}+2}{3\pi^2}
-\frac{19}{9}\biggr]e^{-\mu_\text{eff}}
+\frac{19}{2}e^{-\frac{4}{3}\mu_\text{eff}}
+{\cal O}(e^{-\frac{5}{3}\mu_\text{eff}}).
\end{align*}

\subsection{Rank deformation $(M_1,M_2)=(k/2-M,k/2)$}

Here we omit the results for $M=k/2$, as we observe that they coincide with those given in appendix \ref{1010Jnp} and appendix \ref{1010Jnpeff}.

\subsubsection{Reduced grand potential}
\label{orbJnp}

We list the non-perturbative part of the reduced grand potential $J^\text{np}_{k,(k/2-M,k/2)}(\mu)$.
\begin{align*}
J_{2,(1,1)}^\text{np}&=\biggl[-\frac{2\mu^2+2\mu+2}{\pi^2}+2\biggr]e^{-\mu}
+\biggl[-\frac{26\mu^2+\mu+9/2}{2\pi^2}+18\biggr]e^{-2\mu}\\
&\quad+\biggl[-\frac{736\mu^2-608\mu/3+616/9}{6\pi^2}+\frac{608}{3}\biggr]e^{-3\mu}\\
&\quad+\biggl[-\frac{2701\mu^2-13949\mu/12+11291/48}{2\pi^2}+2514\biggr]e^{-4\mu}
+{\cal O}(e^{-5\mu}),\\
J_{3,(\frac{3}{2},\frac{3}{2})}^\text{np}
&=-\frac{4}{3}e^{-\frac{2}{3}\mu}+3e^{-\frac{4}{3}\mu}
+\biggl[\frac{4\mu^2+2\mu+1}{6\pi^2}+\frac{70}{9}\biggr]e^{-2\mu}
+\frac{203}{18}e^{-\frac{8}{3}\mu}
+{\cal O}(e^{-\frac{10}{3}\mu}),\\
J_{3,(\frac{1}{2},\frac{3}{2})}^\text{np}
&=\frac{2}{3}e^{-\frac{2}{3}\mu}+2e^{-\frac{4}{3}\mu}
+\biggl[\frac{4\mu^2+2\mu+1}{6\pi^2}-\frac{92}{9}\biggr]e^{-2\mu}
-\frac{26}{9}e^{-\frac{8}{3}\mu}
+{\cal O}(e^{-\frac{10}{3}\mu}),\\
J_{4,(2,2)}^\text{np}&=
\biggl[\frac{\mu^2+\mu+1}{\pi^2}\biggr]e^{-\mu}
+\biggl[-\frac{26\mu^2+\mu+9/2}{4\pi^2}+2\biggr]e^{-2\mu}
+{\cal O}(e^{-3\mu}),\\
J_{4,(1,2)}^\text{np}&=
\biggl[-\frac{\mu^2+\mu+1}{\pi^2}+6\biggr]e^{-\mu}
+\biggl[-\frac{26\mu^2+\mu+9/2}{4\pi^2}+50\biggr]e^{-2\mu}
+{\cal O}(e^{-3\mu}),\\
J_{6,(3,3)}^\text{np}&=\frac{4}{3}e^{-\frac{1}{3}\mu}+3e^{-\frac{2}{3}\mu}
+\biggl[-\frac{2\mu^2+2\mu+2}{3\pi^2}-\frac{70}{9}\biggr]e^{-\mu}
+\frac{203}{18}e^{-\frac{4}{3}\mu}
+{\cal O}(e^{-\frac{5}{3}\mu}),\\
J_{6,(2,3)}^\text{np}&=\frac{2}{3}e^{-\frac{1}{3}\mu}+2e^{-\frac{2}{3}\mu}
+\biggl[\frac{2\mu^2+2\mu+2}{3\pi^2}-\frac{92}{9}\biggr]e^{-\mu}
-\frac{26}{9}e^{-\frac{4}{3}\mu}
+{\cal O}(e^{-\frac{5}{3}\mu}),\\
J_{6,(1,3)}^\text{np}&=-\frac{2}{3}e^{-\frac{1}{3}\mu}+2e^{-\frac{2}{3}\mu}
+\biggl[-\frac{2\mu^2+2\mu+2}{3\pi^2}+\frac{92}{9}\biggr]e^{-\mu}
-\frac{26}{9}e^{-\frac{4}{3}\mu}
+{\cal O}(e^{-\frac{5}{3}\mu}).
\end{align*}

\subsubsection{Reduced grand potential in effective chemical potential}
\label{orbJnpeff}

We list the non-perturbative part of the reduced grand potential $\widetilde J^\text{np}_{k,(k/2-M,k/2)}(\mu_\text{eff})$ as a function of the effective chemical potential.
\begin{align*}
\widetilde J_{2,(1,1)}^\text{np}
&=\biggl[-\frac{\mu_\text{eff}^2+2\mu_\text{eff}+2}{\pi^2}
+2\biggr]e^{-\mu_\text{eff}}
+\biggl[-\frac{9(2\mu_\text{eff}^2+2\mu_\text{eff}+1)}{4\pi^2}
+10\biggr]e^{-2\mu_\text{eff}}\\
&\quad+\biggl[-\frac{82(9\mu_\text{eff}^2+6\mu_\text{eff}+2)}{27\pi^2}
+\frac{212}{3}\biggr]e^{-3\mu_\text{eff}}
+\biggl[-\frac{777(8\mu_\text{eff}^2+4\mu_\text{eff}+1)}{32\pi^2}
+554\biggr]e^{-4\mu_\text{eff}}\\
&\quad+{\cal O}(e^{-5\mu_\text{eff}}),\\
\widetilde J_{3,(\frac{3}{2},\frac{3}{2})}^\text{np}
&=-\frac{4}{3}e^{-\frac{2}{3}\mu_\text{eff}}+3e^{-\frac{4}{3}\mu_\text{eff}}
+\biggl[\frac{2\mu_\text{eff}^2+2\mu_\text{eff}+1}{6\pi^2}
+\frac{275}{36}\biggr]e^{-2\mu_\text{eff}}
+\frac{19}{2}e^{-\frac{8}{3}\mu_\text{eff}}
+{\cal O}(e^{-\frac{10}{3}\mu_\text{eff}}),\\
\widetilde J_{3,(\frac{1}{2},\frac{3}{2})}^\text{np}
&=\frac{2}{3}e^{-\frac{2}{3}\mu_\text{eff}}+2e^{-\frac{4}{3}\mu_\text{eff}}
+\biggl[\frac{2\mu_\text{eff}^2+2\mu_\text{eff}+1}{6\pi^2}
-\frac{421}{36}\biggr]e^{-2\mu_\text{eff}}
-2e^{-\frac{8}{3}\mu_\text{eff}}
+{\cal O}(e^{-\frac{10}{3}\mu_\text{eff}}),\\
\widetilde J_{4,(2,2)}^\text{np}
&=\biggl[\frac{\mu_\text{eff}^2+2\mu_\text{eff}+2}{2\pi^2}
-\frac{1}{2}\biggr]e^{-\mu_\text{eff}}
+\biggl[-\frac{9(2\mu_\text{eff}^2+2\mu_\text{eff}+1)}{8\pi^2}
+\frac{9}{4}\biggr]e^{-2\mu_\text{eff}}
+{\cal O}(e^{-3\mu_\text{eff}}),\\
\widetilde J_{4,(1,2)}^\text{np}
&=\biggl[-\frac{\mu_\text{eff}^2+2\mu_\text{eff}+2}{2\pi^2}
+9\biggr]e^{-\mu_\text{eff}}
+\biggl[-\frac{9(2\mu_\text{eff}^2+2\mu_\text{eff}+1)}{8\pi^2}
+\frac{55}{2}\biggr]e^{-2\mu_\text{eff}}
+{\cal O}(e^{-3\mu_\text{eff}}),\\
\widetilde J_{6,(3,3)}^\text{np}
&=\frac{4}{3}e^{-\frac{1}{3}\mu_\text{eff}}
+3e^{-\frac{2}{3}\mu_\text{eff}}
+\biggl[-\frac{\mu_\text{eff}^2+2\mu_\text{eff}+2}{3\pi^2}
-\frac{62}{9}\biggr]e^{-\mu_\text{eff}}
+\frac{19}{2}e^{-\frac{4}{3}\mu_\text{eff}}
+{\cal O}(e^{-\frac{5}{3}\mu_\text{eff}}),\\
\widetilde J_{6,(2,3)}^\text{np}
&=\frac{2}{3}e^{-\frac{1}{3}\mu_\text{eff}}
+2e^{-\frac{2}{3}\mu_\text{eff}}
+\biggl[\frac{\mu_\text{eff}^2+2\mu_\text{eff}+2}{3\pi^2}
-\frac{121}{9}\biggr]e^{-\mu_\text{eff}}
-2e^{-\frac{4}{3}\mu_\text{eff}}
+{\cal O}(e^{-\frac{5}{3}\mu_\text{eff}}),\\
\widetilde J_{6,(1,3)}^\text{np}
&=-\frac{2}{3}e^{-\frac{1}{3}\mu_\text{eff}}
+2e^{-\frac{2}{3}\mu_\text{eff}}
+\biggl[-\frac{\mu_\text{eff}^2+2\mu_\text{eff}+2}{3\pi^2}
+\frac{148}{9}\biggr]e^{-\mu_\text{eff}}
-2e^{-\frac{4}{3}\mu_\text{eff}}
+{\cal O}(e^{-\frac{5}{3}\mu_\text{eff}}).
\end{align*}

\subsection{Rank deformation $(M_1,M_2)=(k/2,k/2-M)$}

Here we omit the results for $M=k/2$, as we find that the partition functions $|{\widehat Z}_{k,k/2}(N)|$ coincide with $|Z_{k,(k/2,0)}(N)|$ listed in appendix \ref{0121values}, which also implies the coincidence of the reduced grand potential.

\subsubsection{Exact values}
\label{11110121values}

We list the absolute values of the exact values of the partition function
$\widehat Z_{k,M}(N)=Z^{(1,1,1,1)}_k(N,N+M,N+2M,N+M)$ of the $(1,1,1,1)$ model.
\begin{align*}
&
|\widehat Z_{3,1}(0)|=\frac{-3+2 \sqrt{3}}{36},
\quad
|\widehat Z_{3,1}(1)|=\frac{-54-12 \sqrt{3}+55 \pi -18 \sqrt{3} \pi }{5184 \pi },
\nonumber\\&\quad
|\widehat Z_{3,1}(2)|=\frac{-4860+1512 \sqrt{3}-1980
\pi +3240 \sqrt{3} \pi +17649 \pi ^2-10726 \sqrt{3} \pi ^2}{4478976 \pi ^2},
\nonumber\\&\quad
|\widehat Z_{3,1}(3)|=\bigl(-157464-24624 \sqrt{3}+124740 \pi -180792
\sqrt{3} \pi +770310 \pi ^2+107532 \sqrt{3} \pi ^2
\nonumber\\&\qquad
-916913 \pi ^3+368334 \sqrt{3} \pi ^3\bigr)\big/\bigl(1934917632 \pi ^3\bigr),
\nonumber\\&
|\widehat Z_{4,1}(0)|=\frac{-2+\pi }{64 \pi },
\quad
|\widehat Z_{4,1}(1)|=\frac{-24+18 \pi +25 \pi ^2-9 \pi ^3}{18432 \pi ^3},
\nonumber\\&\quad
|\widehat Z_{4,1}(2)|=\frac{-6240+6000 \pi +27200 \pi ^2-8800 \pi ^3-16342
\pi ^4+5175 \pi ^5}{235929600 \pi ^5},
\nonumber\\&\quad
|\widehat Z_{4,1}(3)|=\bigl(-1169280+1340640 \pi +13982640 \pi ^2-2175600 \pi ^3-25862984 \pi ^4+4868542
\pi ^5
\nonumber\\&\qquad
+12528027 \pi ^6-3671325 \pi ^7\bigr)\big/\bigl(3329438515200 \pi ^7\bigr),
\nonumber\\&
|\widehat Z_{6,1}(0)|=\frac{9-\sqrt{3} \pi }{648 \pi },
\quad
|\widehat Z_{6,1}(1)|=\frac{-10692+1620 \sqrt{3} \pi +7641 \pi ^2-1369 \sqrt{3} \pi ^3}{30233088 \pi ^3},
\nonumber\\&\quad
|\widehat Z_{6,1}(2)|=\bigl(104451120-19537200 \sqrt{3} \pi -235272600 \pi ^2+64956600 \sqrt{3} \pi ^3+235542231 \pi ^4
\nonumber\\&\qquad
-45484175 \sqrt{3} \pi ^5\bigr)\big/\bigl(23509249228800
\pi ^5\bigr),
\nonumber\\&\quad
|\widehat Z_{6,1}(3)|=\bigl(-1978094260800+433007104320
\sqrt{3} \pi +7078493363760 \pi ^2
\nonumber\\&\qquad
-3273458151600 \sqrt{3} \pi ^3-24040347545340 \pi ^4+7337186290476 \sqrt{3} \pi ^5+25904976913413 \pi ^6
\nonumber\\&\qquad
-5036313115925
\sqrt{3} \pi ^7\bigr)\big/\bigl(53745529068925747200 \pi ^7\bigr),
\nonumber\\&
|\widehat Z_{6,2}(0)|=\frac{-54+30 \sqrt{3} \pi -11 \pi ^2}{279936 \pi ^2},
\nonumber\\&\quad
|\widehat Z_{6,2}(1)|=\frac{75816-55080 \sqrt{3} \pi -158598 \pi ^2+82290 \sqrt{3} \pi ^3-27001
\pi ^4}{13060694016 \pi ^4},
\nonumber\\&\quad
|\widehat Z_{6,2}(2)|=\bigl(-2610753120+2304223200 \sqrt{3} \pi +18048873600 \pi ^2-9023886000 \sqrt{3} \pi ^3
\nonumber\\&\qquad
-23045154966 \pi ^4+10990565910
\sqrt{3} \pi ^5-3415983775 \pi ^6\bigr)\big/\bigl(30467987000524800 \pi ^6\bigr),
\nonumber\\&\quad
|\widehat Z_{6,2}(3)|=\bigl(19253979884160-19759779446400 \sqrt{3} \pi -295768696920480
\pi ^2
\nonumber\\&\qquad
+129011306709600 \sqrt{3} \pi ^3+1062534604958568 \pi ^4-405384618141720 \sqrt{3} \pi ^5
\nonumber\\&\qquad
-1090402250951574 \pi ^6+484126459239930 \sqrt{3} \pi
^7-145107956077525 \pi ^8\bigr)
\nonumber\\&\qquad
\big/\bigl(23218068557775922790400 \pi ^8\bigr).
\end{align*}

\subsubsection{Reduced grand potential}
\label{1111Jnp}

We list the non-perturbative part of the reduced grand potential $J^\text{np}_{k,(k/2,k/2-M)}(\mu)$.
\begin{align*}
J_{3,(\frac{3}{2},\frac{1}{2})}^\text{np}
&=-\frac{10}{3}e^{-\frac{2}{3}\mu}-8e^{-\frac{4}{3}\mu}
+\biggl[\frac{4\mu^2+2\mu+1}{6\pi^2}-\frac{254}{9}\biggr]e^{-2\mu}
-\frac{473}{9}e^{-\frac{8}{3}\mu}
+{\cal O}(e^{-\frac{10}{3}\mu}),\\
J_{4,(2,1)}^\text{np}&=-2e^{-\frac{1}{2}\mu}
+\biggl[\frac{\mu^2+\mu+1}{\pi^2}-6\biggr]e^{-\mu}
+\frac{16}{3}e^{-\frac{3}{2}\mu}\\
&\quad+\biggl[-\frac{26\mu^2+\mu+9/2}{4\pi^2}+50\biggr]e^{-2\mu}
+{\cal O}(e^{-\frac{5}{2}\mu}),\\
J_{6,(3,2)}^\text{np}&=-\frac{2}{3}e^{-\frac{1}{3}\mu}
+\biggl[-\frac{2\mu^2+2\mu+2}{3\pi^2}+\frac{26}{9}\biggr]e^{-\mu}
-\frac{17}{9}e^{-\frac{4}{3}\mu}
+{\cal O}(e^{-\frac{5}{3}\mu}),\\
J_{6,(3,1)}^\text{np}&=-\frac{26}{3}e^{-\frac{1}{3}\mu}
-32e^{-\frac{2}{3}\mu}
+\biggl[-\frac{2\mu^2+2\mu+2}{3\pi^2}-\frac{1690}{9}\biggr]e^{-\mu}
-\frac{11201}{9}e^{-\frac{4}{3}\mu}
+{\cal O}(e^{-\frac{5}{3}\mu}).
\end{align*}

\subsubsection{Reduced grand potential in effective chemical potential}
\label{1111Jnpeff}

We list the non-perturbative part of the reduced grand potential $\widetilde J_{k,(k/2,k/2-M)}^\text{np}(\mu_\text{eff})$ as a function of the effective chemical potential.
\begin{align*}
\widetilde J_{3,(\frac{3}{2},\frac{1}{2})}^\text{np}
&=-\frac{10}{3}e^{-\frac{2}{3}\mu_\text{eff}}-8e^{-\frac{4}{3}\mu_\text{eff}}
+\biggl[\frac{2\mu_\text{eff}^2+2\mu_\text{eff}+1}{6\pi^2}
-\frac{973}{36}\biggr]e^{-2\mu_\text{eff}}
-57e^{-\frac{8}{3}\mu_\text{eff}}
+{\cal O}(e^{-\frac{10}{3}\mu_\text{eff}}),\\
\widetilde J_{4,(2,1)}^\text{np}
&=-2e^{-\frac{1}{2}\mu_\text{eff}}
+\biggl[\frac{\mu_\text{eff}^2+2\mu_\text{eff}+2}{2\pi^2}
-\frac{7}{2}\biggr]e^{-\mu_\text{eff}}
+\frac{4}{3}e^{-\frac{3}{2}\mu_\text{eff}}\\
&\quad+\biggl[-\frac{9(2\mu_\text{eff}^2+2\mu_\text{eff}+1)}{8\pi^2}
+\frac{99}{4}\biggr]e^{-2\mu_\text{eff}}
+{\cal O}(e^{-\frac{5}{2}\mu_\text{eff}}),\\
\widetilde J_{6,(3,2)}^\text{np}
&=-\frac{2}{3}e^{-\frac{1}{3}\mu_\text{eff}}
+\biggl[-\frac{\mu_\text{eff}^2+2\mu_\text{eff}+2}{3\pi^2}
+\frac{4}{9}\biggr]e^{-\mu_\text{eff}}
-e^{-\frac{4}{3}\mu_\text{eff}}
+{\cal O}(e^{-\frac{5}{3}\mu_\text{eff}}),\\
\widetilde J_{6,(3,1)}^\text{np}
&=-\frac{26}{3}e^{-\frac{1}{3}\mu_\text{eff}}
-32e^{-\frac{2}{3}\mu_\text{eff}}
+\biggl[-\frac{\mu_\text{eff}^2+2\mu_\text{eff}+2}{3\pi^2}
-\frac{1730}{9}\biggr]e^{-\mu_\text{eff}}
-1233e^{-\frac{4}{3}\mu_\text{eff}}
+{\cal O}(e^{-\frac{5}{3}\mu_\text{eff}}).
\end{align*}

\section{Closed string formalism for $(0,M)$}
\label{22closed}

In this appendix we derive the closed string formalism for the $(2,2)$ model with the rank deformation $(M_1,M_2)=(0,M)$.
After establishing the closed string formalism, we show that the Hanany-Witten duality \eqref{HW1111_M0M0} at $M=k/2$ is manifest in this formalism.

The closed string formalism for the ABJM theory was first conjectured in \cite{AHS} and later proved in \cite{HondaABJM}.
The derivation was improved largely later and generalized to the partition function of the orientifold ABJM theory with the orthosymplectic gauge group in \cite{MS2,MN5} and the vacuum expectation values of the half-BPS Wilson loop in the ABJM theory in \cite{KiMo}.
We shall follow these simplified derivations in the following.

We shall prove the closed string formalism
\begin{align}
\frac{Z_{k,(0,M)}(N)}{Z_{k,(0,M)}(0)}=\frac{1}{N!}\int\frac{d^Nx}{(2\pi)^N}\det\langle x_i|{\widehat\rho}_{(0,M)}|x_j\rangle,
\label{22_0M_closed}
\end{align}
where the normalization is
\begin{align}
Z_{k,(0,M)}(0)
=\frac{1}{k^M}\prod_{l<l'}
2\sinh\frac{\pi i(l-l')}{k}
\prod_{a<a'}2\sinh\frac{\pi i(-a+a')}{k},
\label{normalization}
\end{align}
and the density matrix is given as the ``square'' of that of the ABJM theory
\begin{align}
{\widehat\rho}_{(0,M)}={\widehat\rho}_{\text{ABJM}}\,^t{\widehat\rho}_{\text{ABJM}},
\label{square}
\end{align}
with\footnote{
Precisely speaking, ${\widehat\rho}_{\text{ABJM}}$ in \eqref{rhoABJM} is the density matrix for $|Z^\text{ABJM}_k(N,N+M)|$.
Indeed ${\widehat\rho}_\text{ABJM}$ is positive definite which is manifest after the similarity transformation ${\widehat\rho}_\text{ABJM}\rightarrow e^{-\frac{M{\widehat p}}{2k}}{\widehat\rho}_\text{ABJM}e^{\frac{M{\widehat p}}{2k}}$ \cite{HO}.
The density matrix for $Z^\text{ABJM}_k(N,N+M)$ without taking the absolute values is given as $e^{-\frac{\pi iM}{2}}{\widehat\rho}_\text{ABJM}$.
}
\begin{align}
{\widehat\rho}_{\text{ABJM}}&=
e^{\frac{\pi iM}{2}}
\frac{\prod_l2\sinh\frac{{\widehat q}-2\pi il}{2k}}{2\cosh\frac{\widehat q}{2}}
\frac{1}{2\cosh\frac{\widehat p}{2}}
\frac{1}{\prod_l2\cosh\frac{\widehat q-2\pi il}{2k}},\nonumber\\
^t{\widehat\rho}_{\text{ABJM}}&=
e^{\frac{\pi iM}{2}}
\frac{1}{\prod_a2\cosh\frac{\widehat q+2\pi ia}{2k}}
\frac{1}{2\cosh\frac{\widehat p}{2}}
\frac{\prod_a2\sinh\frac{{\widehat q}+2\pi ia}{2k}}{2\cosh\frac{\widehat q}{2}}.
\label{rhoABJM}
\end{align}
Note that the hyperbolic tangent function in \cite{KiMo} appears after we take the trace of ${\widehat\rho}_{\text{ABJM}}$.

First, let us rewrite the partition function \eqref{Z22_2020} into the bra/ket notation as follows
\begin{align}
&Z_{k,(0,M)}=\int
\frac{d^N\kappa}{N!(2\pi)^N}
\frac{d^N\lambda}{N!(2\pi)^N}\nonumber\\
&\times\int\frac{d^{N+M}\mu}{(N+M)!(2\pi)^{N+M}}
\det
\begin{pmatrix}
\langle \lambda_i|\frac{1}{2\cosh\frac{\widehat p}{2}}
e^{\frac{i}{2\hbar}{\widehat q}^2}|\mu_j\rangle\\
\llangle 2\pi il|e^{\frac{i}{2\hbar}{\widehat q}^2}|\mu_j\rangle
\end{pmatrix}
\det
\begin{pmatrix}
\langle\mu_j|\frac{1}{2\cosh\frac{\widehat p}{2}}|\kappa_i\rangle
&\langle\mu_j|{-2\pi ia}\rrangle
\end{pmatrix}\nonumber\\
&\times\int\frac{d^{N+M}\nu}{(N+M)!(2\pi)^{N+M}}
\det
\begin{pmatrix}
\langle\kappa_i|\frac{1}{2\cosh\frac{\widehat p}{2}}
e^{-\frac{i}{2\hbar}{\widehat q}^2}|\nu_j\rangle\\
\llangle 2\pi il|e^{-\frac{i}{2\hbar}{\widehat q}^2}|\nu_j\rangle
\end{pmatrix}
\det
\begin{pmatrix}
\langle\nu_j|\frac{1}{2\cosh\frac{\widehat p}{2}}|\lambda_i\rangle
&\langle\nu_j|{-2\pi ia}\rrangle
\end{pmatrix},
\label{Z22_2020_braket}
\end{align}
with $i=1,\cdots,N$ and $j=1,\cdots,N+M$.
Our strategy to derive the closed string formalism for the partition function of the $(2,2)$ model is completely parallel to that of the ABJM theory in \cite{MS2,MN5,KiMo}, only except that we have to repeat it twice.
Namely, we utilize the following two techniques.
\begin{itemize}
\item
We transform the matrix elements of one of the two determinants in each line of \eqref{Z22_2020_braket} into the delta functions for $\mu_j$ and $\nu_j$.
This is achieved by performing appropriate similarity transformations to each projection operator, which appears in the determinants with the bra and the ket separated
\begin{align}
&|\kappa\rangle\langle\kappa|
\to e^{\frac{i}{2\hbar}{\widehat p}^2}
|\kappa\rangle\langle\kappa|e^{-\frac{i}{2\hbar}{\widehat p}^2},\quad
|\lambda\rangle\langle\lambda|
\to e^{\frac{i}{2\hbar}{\widehat q}^2}e^{\frac{i}{2\hbar}{\widehat p}^2}
|\lambda\rangle\langle\lambda|e^{-\frac{i}{2\hbar}{\widehat p}^2}
e^{-\frac{i}{2\hbar}{\widehat q}^2},\nonumber\\
&|\mu\rangle\langle\mu|
\to e^{\frac{i}{2\hbar}{\widehat p}^2}
|\mu\rangle\langle\mu|e^{-\frac{i}{2\hbar}{\widehat p}^2},\quad
|\nu\rangle\langle\nu|
\to e^{\frac{i}{2\hbar}{\widehat q}^2}e^{\frac{i}{2\hbar}{\widehat p}^2}
|\nu\rangle\langle\nu|
e^{-\frac{i}{2\hbar}{\widehat p}^2}e^{-\frac{i}{2\hbar}{\widehat q}^2}.
\end{align}
This similarity transformation is possible because these operators are simply the identities after the integrations of the variables.
\item
We trivialize one of the two determinants by re-ordering the $N+M$ integration variables ($\rho=\mu$ or $\nu$)
\begin{align}
\int\frac{d^{N+M}\rho}{(N+M)!}
\det[f_j(\rho_{j'})]\det[g_j(\rho_{j'})]
=\int d^{N+M}\rho\Bigl(\prod_j f_j(\rho_j)\Bigr)\det[g_j(\rho_{j'})].
\end{align}
\end{itemize}
After these two rewritings, the $\mu$ and $\nu$ integrations can be performed trivially by substitutions.
After applying these techniques, we find that the $\mu$ integration (the second line in \eqref{Z22_2020_braket}) reduces to the determinant of ${\widehat\rho}_{\text{ABJM}}$ in \eqref{square}, while the $\nu$ integration (the third line in \eqref{Z22_2020_braket}) gives $^t{\widehat\rho}_{\text{ABJM}}$.
The determinant to be trivialized is chosen for the $\mu$ integration in the second line to be the first one, while for $\nu$ integration in the third line to be the second one.

After the similarity transformation, the second line in \eqref{Z22_2020_braket} becomes
\begin{align}
&\int\frac{d^{N+M}\mu}{(N+M)!(2\pi)^{N+M}}
\det
\begin{pmatrix}
\langle\lambda_i|\frac{1}{2\cosh\frac{\widehat q}{2}}|\mu_j\rangle\\
\frac{1}{\sqrt{-i}}e^{\frac{\pi il^2}{k}}\langle 2\pi il|\mu_j\rangle
\end{pmatrix}
\det
\begin{pmatrix}
\langle\mu_j|\frac{1}{2\cosh\frac{\widehat p}{2}}|\kappa_i\rangle
&e^{\frac{\pi ia^2}{k}}\langle\mu_j|{-2\pi ia}\rrangle
\end{pmatrix}
\nonumber\\
&=\frac{e^{-\frac{\pi iMN}{2}}
e^{\frac{\pi i}{k}(\sum_aa^2+\sum_ll^2)}
e^{-\frac{\pi iM}{k}\sum_ll}
e^{\frac{M}{2k}\sum_i(\kappa_i-\lambda_i)}
\prod_{l<l'}
2\sinh\frac{\pi i(l-l')}{k}}{(-ik)^{M/2}}
\det\langle\lambda_i|\widehat\rho_\text{ABJM}|\kappa_{i'}\rangle.
\label{Z22_2020_braket_2ndline_sim}
\end{align}
Here, as explained previously, for the first determinant, since all of the components become the delta functions\footnote{
Notice that we have to deform the integration contour of $\mu_j$ ($j>N$) in the pure-imaginary direction to process the delta functions from $\langle 2\pi il|\mu_j\rangle$.
This deformation is valid only for $M\le k/2$; otherwise the contour crosses the poles of $\langle\mu_j|\frac{1}{2\cosh\frac{\widehat p}{2}}|\kappa_i\rangle$ in the complex plane during the deformation, which we would also have to take into account \cite{MS2}.
}
as $\langle\lambda_i|\frac{1}{2\cosh\frac{\widehat q}{2}}|\mu_j\rangle=\frac{2\pi}{2\cosh\frac{\lambda_i}{2}}\delta(\mu_j-\lambda_i)$ and $\langle 2\pi il|\mu_j\rangle=2\pi\delta(\mu_j-2\pi il)$, we can perform the $\mu$ integrations by trivialization and substitution.
For the second determinant, after extracting the factor $e^{\frac{\pi ia^2}{k}}$ out of the determinant, we can decompose it into the product again by using the Cauchy-Vandermonde determinant in the opposite way.
Similarly, for the third line we find
\begin{align}
&\int\frac{d^{N+M}\nu}{(N+M)!(2\pi)^{N+M}}
\det
\begin{pmatrix}
\langle\kappa_i|\frac{1}{2\cosh\frac{\widehat p}{2}}|\nu_j\rangle\\
e^{-\frac{\pi il^2}{k}}\llangle 2\pi il|\nu_j\rangle
\end{pmatrix}
\det
\begin{pmatrix}
\langle\nu_j|\frac{1}{2\cosh\frac{\widehat q}{2}}|\lambda_i\rangle
&
\frac{1}{\sqrt{i}}
e^{-\frac{\pi ia^2}{k}}\langle\nu_j|{-2\pi ia}\rrangle
\end{pmatrix}\nonumber\\
&=\frac{e^{\frac{\pi iMN}{2}}e^{-\frac{\pi i}{k}(\sum_aa^2+\sum_ll^2)}
e^{-\frac{\pi iM}{k}\sum_aa}e^{\frac{M}{2k}\sum_i(\lambda_i-\kappa_i)}
\prod_{a<a'}2\sinh\frac{\pi i(-a+a')}{k}}{(ik)^{M/2}}
\det\langle\kappa_i|{}^t\widehat\rho_\text{ABJM}|\lambda_{i'}\rangle.
\end{align}
Finally we find that most of the numerical factors cancel, leaving
\begin{align}
Z_{k,(0,M)}(N)=Z_{k,(0,M)}(0)
\int
\frac{d^N\kappa}{N!(2\pi)^N}
\frac{d^N\lambda}{N!(2\pi)^N}
\det\langle\lambda_i|{\widehat\rho}_{\text{ABJM}}|\kappa_j\rangle
\det\langle\kappa_i|\,^t{\widehat\rho}_{\text{ABJM}}|\lambda_j\rangle.
\end{align}
After combining the two determinants by using the formula in appendix A in \cite{MM}, we obtain \eqref{22_0M_closed}.

The Hanany-Witten duality \eqref{HW1111_M0M0} at $M=k/2$ is manifest in the closed string formalism.
Indeed, with the help of the following identity
\begin{align}
\prod_{m=1}^{k/2}2\sinh\frac{x-2\pi i(m-\frac{1}{2})}{k}
=e^{-\frac{\pi ik}{4}}2\cosh\frac{x}{2},
\end{align}
we find that $^t{\widehat\rho}_\text{ABJM}$ coincides with ${\widehat\rho}_{\text{ABJM}}$ for $k\in 2\mathbb{N}$ and $M=k/2$
\begin{align}
{\widehat\rho}_{\text{ABJM}}=\,^t{\widehat\rho}_{\text{ABJM}}=
e^{\frac{\pi iM}{2}}
\sqrt{\frac{
\prod_{m=1}^{k/2}\tanh\frac{{\widehat q}-2\pi i(m-\frac{1}{2})}{2k}
}{2\cosh\frac{\widehat q}{2}}}
\frac{1}{2\cosh\frac{\widehat p}{2}}
\sqrt{\frac{\prod_{m=1}^{k/2}\tanh\frac{{\widehat q}-2\pi i(m-\frac{1}{2})}{2k}
}{2\cosh\frac{\widehat q}{2}}}.
\end{align}
Hence the density matrix is ${\widehat\rho}_{k,(0,M)}={\widehat\rho}_{\text{ABJM}}^2$, which is same as the density matrix of the orbifold ABJM theory \cite{HM}.

\section*{Acknowledgements}
We are grateful to Nadav Drukker, Alba Grassi, Chiung Hwang, Sung-soo Kim, Sungjay Lee, Kazumi Okuyama, Antonio Sciarappa and Futoshi Yagi for valuable discussions.
The work of S.M.\ is supported by JSPS Grant-in-Aid for Scientific Research (C) \# 26400245, Japan-Russia and Japan-Hungary bilateral joint research projects.
We thank Korea Institute for Advanced Study for providing computing resources (KIAS Center for Advanced Computation) for this work.
We would like to thank both Yukawa Institute for Theoretical Physics and Osaka City University for hospitality where part of this work was conducted.


\begin{thebibliography}{99}
\bibitem{KT}
I.~R.~Klebanov and A.~A.~Tseytlin,
``Entropy of near extremal black p-branes,''
Nucl.\ Phys.\ B {\bf 475}, 164 (1996)
[hep-th/9604089].
%
\bibitem{ABJM}
O.~Aharony, O.~Bergman, D.~L.~Jafferis and J.~Maldacena,
``N=6 superconformal Chern-Simons-matter theories, M2-branes and their gravity duals,''
JHEP {\bf 0810}, 091 (2008)
[arXiv:0806.1218 [hep-th]].
%
\bibitem{HLLLP2}
K.~Hosomichi, K.~M.~Lee, S.~Lee, S.~Lee and J.~Park,
``N=5,6 Superconformal Chern-Simons Theories and M2-branes on Orbifolds,''
JHEP {\bf 0809}, 002 (2008)
[arXiv:0806.4977 [hep-th]].
%
\bibitem{ABJ}
O.~Aharony, O.~Bergman and D.~L.~Jafferis,
``Fractional M2-branes,''
JHEP {\bf 0811}, 043 (2008)
[arXiv:0807.4924 [hep-th]].
%
\bibitem{KWY}
A.~Kapustin, B.~Willett and I.~Yaakov,
``Exact Results for Wilson Loops in Superconformal Chern-Simons Theories with Matter,''
JHEP {\bf 1003}, 089 (2010)
[arXiv:0909.4559 [hep-th]].
%
\bibitem{HKPT}
C.~P.~Herzog, I.~R.~Klebanov, S.~S.~Pufu and T.~Tesileanu,
``Multi-Matrix Models and Tri-Sasaki Einstein Spaces,''
Phys.\ Rev.\ D {\bf 83}, 046001 (2011)
[arXiv:1011.5487 [hep-th]].
%
\bibitem{MS}
D.~Martelli and J.~Sparks,
``The large N limit of quiver matrix models and Sasaki-Einstein manifolds,''
Phys.\ Rev.\ D {\bf 84}, 046008 (2011)
[arXiv:1102.5289 [hep-th]].
%
\bibitem{DMP1}
N.~Drukker, M.~Marino and P.~Putrov,
``From weak to strong coupling in ABJM theory,''
Commun.\ Math.\ Phys.\  {\bf 306}, 511 (2011)
[arXiv:1007.3837 [hep-th]].
%
\bibitem{DMP2}
N.~Drukker, M.~Marino and P.~Putrov,
``Nonperturbative aspects of ABJM theory,''
JHEP {\bf 1111}, 141 (2011)
[arXiv:1103.4844 [hep-th]].
%
\bibitem{FHM}
H.~Fuji, S.~Hirano and S.~Moriyama,
``Summing Up All Genus Free Energy of ABJM Matrix Model,''
JHEP {\bf 1108}, 001 (2011)
[arXiv:1106.4631 [hep-th]].
%
\bibitem{MP}
M.~Marino and P.~Putrov,
``ABJM theory as a Fermi gas,''
J.\ Stat.\ Mech.\  {\bf 1203}, P03001 (2012)
[arXiv:1110.4066 [hep-th]].
%
\bibitem{CM}
F.~Calvo and M.~Marino,
``Membrane instantons from a semiclassical TBA,''
JHEP {\bf 1305}, 006 (2013)
[arXiv:1212.5118 [hep-th]].
%
\bibitem{HMO1}
Y.~Hatsuda, S.~Moriyama and K.~Okuyama,
``Exact Results on the ABJM Fermi Gas,''
JHEP {\bf 1210}, 020 (2012)
[arXiv:1207.4283 [hep-th]].
%
\bibitem{PY}
P.~Putrov and M.~Yamazaki,
``Exact ABJM Partition Function from TBA,''
Mod.\ Phys.\ Lett.\ A {\bf 27}, 1250200 (2012)
[arXiv:1207.5066 [hep-th]].
%
\bibitem{HMO2}
Y.~Hatsuda, S.~Moriyama and K.~Okuyama,
``Instanton Effects in ABJM Theory from Fermi Gas Approach,''
JHEP {\bf 1301}, 158 (2013)
[arXiv:1211.1251 [hep-th]].
%
\bibitem{HMO3}
Y.~Hatsuda, S.~Moriyama and K.~Okuyama,
``Instanton Bound States in ABJM Theory,''
JHEP {\bf 1305}, 054 (2013)
[arXiv:1301.5184 [hep-th]].
%
\bibitem{HMMO}
Y.~Hatsuda, M.~Marino, S.~Moriyama and K.~Okuyama,
``Non-perturbative effects and the refined topological string,''
JHEP {\bf 1409}, 168 (2014)
[arXiv:1306.1734 [hep-th]].
%
\bibitem{MM}
S.~Matsumoto and S.~Moriyama,
``ABJ Fractional Brane from ABJM Wilson Loop,''
JHEP {\bf 1403}, 079 (2014)
[arXiv:1310.8051 [hep-th]].
%
\bibitem{HO}
M.~Honda and K.~Okuyama,
``Exact results on ABJ theory and the refined topological string,''
JHEP {\bf 1408}, 148 (2014)
[arXiv:1405.3653 [hep-th]].
%
\bibitem{HM}
M.~Honda and S.~Moriyama,
``Instanton Effects in Orbifold ABJM Theory,''
JHEP {\bf 1408}, 091 (2014)
[arXiv:1404.0676 [hep-th]].
%
\bibitem{MN1}
S.~Moriyama and T.~Nosaka,
``Partition Functions of Superconformal Chern-Simons Theories from Fermi Gas Approach,''
JHEP {\bf 1411}, 164 (2014)
[arXiv:1407.4268 [hep-th]].
%
\bibitem{MN2}
S.~Moriyama and T.~Nosaka,
``ABJM membrane instanton from a pole cancellation mechanism,''
Phys.\ Rev.\ D {\bf 92}, no. 2, 026003 (2015)
[arXiv:1410.4918 [hep-th]].
%
\bibitem{MN3}
S.~Moriyama and T.~Nosaka,
``Exact Instanton Expansion of Superconformal Chern-Simons Theories from Topological Strings,''
JHEP {\bf 1505}, 022 (2015)
[arXiv:1412.6243 [hep-th]].
%
\bibitem{HHO}
Y.~Hatsuda, M.~Honda and K.~Okuyama,
``Large N non-perturbative effects in $\mathcal{N}=4$ superconformal Chern-Simons theories,''
JHEP {\bf 1509}, 046 (2015)
[arXiv:1505.07120 [hep-th]].
%
\bibitem{GAH}
D.~R.~Gulotta, J.~P.~Ang and C.~P.~Herzog,
``Matrix Models for Supersymmetric Chern-Simons Theories with an ADE Classification,''
JHEP {\bf 1201}, 132 (2012)
[arXiv:1111.1744 [hep-th]].
%
\bibitem{CHJ}
P.~M.~Crichigno, C.~P.~Herzog and D.~Jain,
``Free Energy of $D_n$ Quiver Chern-Simons Theories,''
JHEP {\bf 1303}, 039 (2013)
[arXiv:1211.1388 [hep-th]].
%
\bibitem{ADF}
B.~Assel, N.~Drukker and J.~Felix,
``Partition functions of 3d $\hat D$-quivers and their mirror duals from 1d free fermions,''
JHEP {\bf 1508}, 071 (2015)
[arXiv:1504.07636 [hep-th]].
%
\bibitem{MN4}
S.~Moriyama and T.~Nosaka,
``Superconformal Chern-Simons Partition Functions of Affine D-type Quiver from Fermi Gas,''
JHEP {\bf 1509}, 054 (2015)
[arXiv:1504.07710 [hep-th]].
%
\bibitem{CJ}
P.~M.~Crichigno and D.~Jain,
``Non-toric Cones and Chern-Simons Quivers,''
arXiv:1702.05486 [hep-th].
%
\bibitem{GW}
D.~Gaiotto and E.~Witten,
``Janus Configurations, Chern-Simons Couplings, And The theta-Angle in N=4 Super Yang-Mills Theory,''
JHEP {\bf 1006}, 097 (2010)
[arXiv:0804.2907 [hep-th]].
%
\bibitem{HLLLP1}
K.~Hosomichi, K.~M.~Lee, S.~Lee, S.~Lee and J.~Park,
``N=4 Superconformal Chern-Simons Theories with Hyper and Twisted Hyper Multiplets,''
JHEP {\bf 0807}, 091 (2008)
[arXiv:0805.3662 [hep-th]].
%
\bibitem{IK1}
Y.~Imamura and K.~Kimura,
``On the moduli space of elliptic Maxwell-Chern-Simons theories,''
Prog.\ Theor.\ Phys.\  {\bf 120}, 509 (2008)
[arXiv:0806.3727 [hep-th]].
%
\bibitem{TY}
S.~Terashima and F.~Yagi,
``Orbifolding the Membrane Action,''
JHEP {\bf 0812}, 041 (2008)
[arXiv:0807.0368 [hep-th]].
%
\bibitem{IK2}
Y.~Imamura and K.~Kimura,
``N=4 Chern-Simons theories with auxiliary vector multiplets,''
JHEP {\bf 0810}, 040 (2008)
[arXiv:0807.2144 [hep-th]].
%
\bibitem{GHM1}
A.~Grassi, Y.~Hatsuda and M.~Marino,
``Topological Strings from Quantum Mechanics,''
Annales Henri Poincare {\bf 17} (2016) no.11,  3177
[arXiv:1410.3382 [hep-th]].
%
\bibitem{HW}
A.~Hanany and E.~Witten,
``Type IIB superstrings, BPS monopoles, and three-dimensional gauge dynamics,''
Nucl.\ Phys.\ B {\bf 492} (1997) 152
[hep-th/9611230].
%
\bibitem{KKV}
S.~H.~Katz, A.~Klemm and C.~Vafa,
``M theory, topological strings and spinning black holes,''
Adv.\ Theor.\ Math.\ Phys.\  {\bf 3}, 1445 (1999)
[hep-th/9910181].
%
\bibitem{N}
T.~Nosaka,
``Instanton effects in ABJM theory with general R-charge assignments,''
JHEP {\bf 1603}, 059 (2016)
[arXiv:1512.02862 [hep-th]].
%
\bibitem{HKP}
M.~X.~Huang, A.~Klemm and M.~Poretschkin,
``Refined stable pair invariants for E-, M- and $[p, q]$-strings,''
JHEP {\bf 1311}, 112 (2013)
[arXiv:1308.0619 [hep-th]].
%
\bibitem{PTEP}
Y.~Hatsuda, S.~Moriyama and K.~Okuyama,
``Exact instanton expansion of the ABJM partition function,''
PTEP {\bf 2015}, no. 11, 11B104 (2015)
[arXiv:1507.01678 [hep-th]].
%
\bibitem{Marino}
M.~Marino,
``Localization at large N in Chern-Simons-matter theories,''
arXiv:1608.02959 [hep-th].
%
\bibitem{TW}
C.~A.~Tracy and H.~Widom,
``Proofs of two conjectures related to the thermodynamic Bethe ansatz,''
Commun.\ Math.\ Phys.\  {\bf 179} (1996) 667
[solv-int/9509003].
%
\bibitem{KEK}
M.~Hanada, M.~Honda, Y.~Honma, J.~Nishimura, S.~Shiba and Y.~Yoshida,
``Numerical studies of the ABJM theory for arbitrary N at arbitrary coupling constant,''
JHEP {\bf 1205} (2012) 121
[arXiv:1202.5300 [hep-th]].
%
\bibitem{HaOk}
Y.~Hatsuda and K.~Okuyama,
``Probing non-perturbative effects in M-theory,''
JHEP {\bf 1410} (2014) 158
[arXiv:1407.3786 [hep-th]].
%
\bibitem{WZH}
X.~Wang, G.~Zhang and M.~x.~Huang,
``New Exact Quantization Condition for Toric Calabi-Yau Geometries,''
Phys.\ Rev.\ Lett.\  {\bf 115}, 121601 (2015)
[arXiv:1505.05360 [hep-th]].
%
\bibitem{KOO}
T.~Kitao, K.~Ohta and N.~Ohta,
``Three-dimensional gauge dynamics from brane configurations with (p,q) - five-brane,''
Nucl.\ Phys.\ B {\bf 539} (1999) 79
[hep-th/9808111].
%
\bibitem{BHKK}
O.~Bergman, A.~Hanany, A.~Karch and B.~Kol,
``Branes and supersymmetry breaking in three-dimensional gauge theories,''
JHEP {\bf 9910} (1999) 036
[hep-th/9908075].
%
\bibitem{LV}
G.~Lockhart and C.~Vafa,
``Superconformal Partition Functions and Non-perturbative Topological Strings,''
arXiv:1210.5909 [hep-th].
%
\bibitem{MPTY}
V.~Mitev, E.~Pomoni, M.~Taki and F.~Yagi,
``Fiber-Base Duality and Global Symmetry Enhancement,''
JHEP {\bf 1504} (2015) 052
[arXiv:1411.2450 [hep-th]].
%
\bibitem{AKMV}
M.~Aganagic, A.~Klemm, M.~Marino and C.~Vafa,
``The Topological vertex,''
Commun.\ Math.\ Phys.\  {\bf 254} (2005) 425
[hep-th/0305132].
%
\bibitem{IKV}
A.~Iqbal, C.~Kozcaz and C.~Vafa,
``The Refined topological vertex,''
JHEP {\bf 0910} (2009) 069
[hep-th/0701156].
%
\bibitem{KaMa2}
R.~Kashaev, M.~Marino and S.~Zakany,
``Matrix models from operators and topological strings, 2,''
Annales Henri Poincare {\bf 17} (2016) no.10,  2741
[arXiv:1505.02243 [hep-th]].
%
\bibitem{OkuyamaOSp2}
K.~Okuyama,
``Orientifolding of the ABJ Fermi gas,''
JHEP {\bf 1603} (2016) 008
[arXiv:1601.03215 [hep-th]].
%
\bibitem{MePu}
M.~Mezei and S.~S.~Pufu,
``Three-sphere free energy for classical gauge groups,''
JHEP {\bf 1402}, 037 (2014)
[arXiv:1312.0920 [hep-th]].
%
\bibitem{MS1}
S.~Moriyama and T.~Suyama,
``Instanton Effects in Orientifold ABJM Theory,''
JHEP {\bf 1603}, 034 (2016)
[arXiv:1511.01660 [hep-th]].
%
\bibitem{Honda}
M.~Honda,
``Exact relations between M2-brane theories with and without Orientifolds,''
JHEP {\bf 1606} (2016) 123
[arXiv:1512.04335 [hep-th]].
%
\bibitem{MS2}
S.~Moriyama and T.~Suyama,
``Orthosymplectic Chern-Simons Matrix Model and Chirality Projection,''
JHEP {\bf 1604} (2016) 132
[arXiv:1601.03846 [hep-th]].
%
\bibitem{MN5}
S.~Moriyama and T.~Nosaka,
``Orientifold ABJM Matrix Model: Chiral Projections and Worldsheet Instantons,''
JHEP {\bf 1606} (2016) 068
[arXiv:1603.00615 [hep-th]].
%
\bibitem{OWZ}
H.~Ouyang, J.~B.~Wu and J.~j.~Zhang,
``Supersymmetric Wilson loops in $ \mathcal{N}=4 $ super Chern-Simons-matter theory,''
JHEP {\bf 1511}, 213 (2015)
[arXiv:1506.06192 [hep-th]].
%
\bibitem{CDT}
M.~Cooke, N.~Drukker and D.~Trancanelli,
``A profusion of $1/2$ BPS Wilson loops in $\mathcal{N}=4$ Chern-Simons-matter theories,''
JHEP {\bf 1510}, 140 (2015)
[arXiv:1506.07614 [hep-th]].
%
\bibitem{HHMO}
Y.~Hatsuda, M.~Honda, S.~Moriyama and K.~Okuyama,
``ABJM Wilson Loops in Arbitrary Representations,''
JHEP {\bf 1310}, 168 (2013)
[arXiv:1306.4297 [hep-th]].
%
\bibitem{MaMo}
S.~Matsuno and S.~Moriyama,
``Giambelli Identity in Super Chern-Simons Matrix Model,''
arXiv:1603.04124 [hep-th].
%
\bibitem{AHS}
H.~Awata, S.~Hirano and M.~Shigemori,
``The Partition Function of ABJ Theory,''
PTEP {\bf 2013}, 053B04 (2013)
[arXiv:1212.2966 [hep-th]].
%
\bibitem{HondaABJM}
M.~Honda,
``Direct derivation of "mirror" ABJ partition function,''
JHEP {\bf 1312} (2013) 046
[arXiv:1310.3126 [hep-th]].
%
\bibitem{KiMo}
K.~Kiyoshige and S.~Moriyama,
``Dualities in ABJM Matrix Model from Closed String Viewpoint,''
JHEP {\bf 1611} (2016) 096
[arXiv:1607.06414 [hep-th]].
%
\end{thebibliography}
\end{document}